\documentclass[10pt]{article}
    \usepackage{amsmath,amssymb,bm,epsf,epsfig,graphicx}
    \input epsf.sty
    \usepackage[dvipsnames]{xcolor}
    \usepackage[utf8]{inputenc}
    \usepackage[english]{babel}
    \usepackage{amsmath}
    \usepackage{latexsym}
    \usepackage{amsfonts}
    \usepackage{mathtools}
    \usepackage{amssymb}
    \usepackage{scalerel}
    \usepackage{extpfeil}
    \usepackage[left=3cm,right=3cm,top=3.1cm,bottom=3.1cm]{geometry}
    \usepackage{cite}
    \usepackage{tensor}
    \usepackage{tikz}
    \usepackage{tikz-cd}
    \usepackage[mathscr]{euscript}
    \usetikzlibrary{arrows.meta, bending}
    \tikzset{C/.style={circle, minimum size=8mm,
    		node contents={},
    		append after command={\pgfextra{%
    				\draw[-{Straight Barb[flex']}](\tikzlastnode.150) arc (450:110:2.8mm);}
    	}}
    }

    \usepackage{hyperref}
     \usepackage{graphicx}
      \usepackage{subcaption}
    
    \numberwithin{equation}{section}

    \newcommand{\leftpartial}{\mathop{\!\stackrel{\leftarrow}{\partial}}\nolimits}
    \newcommand{\rightpartial}{\mathop{\!\stackrel{\rightarrow}{\partial}}\nolimits}

    \def\p{\partial}

    \def \be {\begin{eqnarray}}
    \def \ee {\end{eqnarray}}
    \def \bal {\begin{align}}
    \def \eal {\end{align}}
    \def \bdm {\begin{displaymath}}
    \def \edm {\end{displaymath}}

    \def\0{\nonumber}

    \def\wc{\omega_\text{c}}
    \def\wo{\omega_\text{o}}

    \def\bfm{\boldsymbol{m}}
    \def\bfl{\boldsymbol{l}}

    \newcommand{\ch}[1]{{\color{black}{#1}}}
    
\begin{document}
     
    \hfill  SISSA 04/2024/FISI

    	\begingroup\allowdisplaybreaks
 
    \vspace*{2.1cm}
    
    \centerline{\Large \bf  Adding stubs to quantum string field theories}
    \vspace {.3cm}

    \begin{center}

    {\large C. Maccaferri$^{(a)}$\footnote{Email: maccafer at unito.it}, R. Poletti$^{(a)}$\footnote{Email: polettiricc at gmail.com}, 
    %\\ \vskip .1 cm
     A. Ruffino$^{(a)}$\footnote{Email: ruffinoalb at gmail.com} and B. Valsesia$^{(b)}$\footnote{Email: bvalsesi at sissa.it}}
    \vskip 1 cm
    $^{(a)}${\it Dipartimento di Fisica, Universit\`a di Torino and  \\INFN  Sezione di Torino \\
    Via Pietro Giuria 1, I-10125 Torino, Italy}
 \vskip .5 cm
    $^{(b)}${\it International School of Advanced Studies (SISSA),\\ via Bonomea 265, 34136 Trieste and\\
    INFN Sezione di Trieste, Italy 
    }
    
    \end{center}
    
    \vspace*{6.0ex}
    
    \centerline{\bf Abstract}
    \bigskip
    Generalizing recent work by Schnabl-Stettinger and Erbin-F\i{}rat, we outline a universal algebraic procedure for  `adding stubs' to string field theories obeying the BV quantum master equation. We apply our results to classical and quantum closed string field theory as well as to open-closed string field theory. We also clarify several aspects of the integration-out process in the co-algebraic formulation of string field theory at the quantum level.
        \baselineskip=16pt
    \newpage
   
   \setcounter{tocdepth}{2}
    \tableofcontents

%%%%%%%%%%%%%%%%%%%%%%%%%%%
%%%%%%%%%%%%%%%%%%%%%%%%%%%
    \section{Introduction and summary}\label{sec:1}
%%%%%%%%%%%%%%%%%%%%%%%%%%%
%%%%%%%%%%%%%%%%%%%%%%%%%%%
String Field Theory (SFT) has witnessed a notable boost in recent times, with advances in concrete physical applications \cite{Cho:2023mhw, large-N, Agmon:2022vdj, Eniceicu:2022xvk, Alexandrov:2022mmy, Sen:2021qdk, Sen:2020eck, Sen:2020cef } as well as  fundamental  progress both at the geometric \cite{ Firat:2023glo, Firat:2023suh, Firat:2023gfn, Erbin:2022rgx, Firat:2021ukc, Cho:2019anu, Costello:2019fuh} and at the algebraic level \cite{Erbin:2020eyc, Koyama:2020qfb, Okawa:2022sjf, Maccaferri:2023gcg}. See \cite{Maccaferri:2023vns, Erler:2019vhl, Erbin:2021smf, Erler:2019loq, deLacroix:2017lif} for recent reviews on the subject.

One noticeable recent progress  has been a better  understanding of the so-called stub deformation \cite{Erbin:2023hcs, Schnabl:2023dbv, Chiaffrino:2021uyd}. This deformation is associated to one of the core aspects of SFT:   the decomposition of the moduli space of Riemann surfaces associated to string amplitudes in `vertex regions', which are encoded into the fundamental interaction vertices of the SFT action and `Feynman' regions which are the portions of moduli space which are covered through the Feynman diagrams, obtained by gluing lower-order vertices with propagators. Essentially by construction, the Feynman regions always contain the part of moduli space where the Riemann surface degenerates and this provides a systematic way of dealing with divergences by suitably interpreting the SFT propagator  \cite{Sen:2019jpm, Larocca:2017pbo, Witten:2013pra}, which is always associated to the degenerating part of the Riemann surface. The distinction between vertex and Feynman regions is arbitrary (up to the obvious requirement that degenerations should be included in the Feynman regions) and it is part of the off-shell data of SFT. Once a choice of consistent vertices is given, there is still the possibility of adjusting the size of the local coordinate patches by rescaling them with canonical `stub' operators $e^{-\lambda L_0}$ (for open strings) or $e^{-\lambda (L_0+\bar L_0)}$  (for closed strings). For $\lambda>0$ this has the effect of increasing the part of moduli space which is covered by the fundamental vertices and reducing the part covered by the propagators in the Feynman diagrams. In particular for very large $\lambda$, almost all the moduli space is covered by the fundamental vertices except the regions where the surface degenerates (which are always covered by the Schwinger parameters of the propagators). This possibility of varying $\lambda$ is the stub deformation. This deformation can be also understood as a smooth way of integrating out off-shell degrees  freedom, in that in the $\lambda\to\infty$ limit the stub operator $e^{-\lambda L_0}$ approaches the projector $P_0$ into the kernel of $L_0$ \footnote{This is certainly true as long as we act on $h\geq 0$ states. When tachyons are present, the stub deformation amplifies them, instead of acting as a suppression factor. It would be interesting to define a different stub operator which can also suppress tachyons. An example would be $1/(1+\lambda L_0)$ which however doesn't work well when there are tachyonic nearly-marginal states which accumulate at $L_0\to 0^-$.} 
\begin{align}
\lim_{\lambda\to+\infty}e^{-\lambda L_0}\sim P_0.
\end{align}
By construction this deformation amounts in a change of the SFT data and thus it falls into the general schemes of field redefinitions. A qualitative `physical' effect of this deformation is to  push (for large $\lambda$) the effect of the higher level states inside the fundamental vertices, leaving the Feynman diagrams to be dominated by  the contribution of the massless fields. However, for any finite $\lambda$, the off-shell degrees of freedom are strictly conserved and therefore one should not expect a qualitative change in the theory, in particular in the vacuum structure \cite{Erler:2023emp}. In this respect the stub deformation appears to be conceptually different from the standard integration-out, where some off-shell (or high energy) degrees of freedom completely  disappear, although the two operations clearly share similarities. 
This similarity with the integrating-out process has been a key-idea in the recent works \cite{Erbin:2023hcs, Schnabl:2023dbv} which gave two related algebraic descriptions of the deformation in the context of (classical) Witten OSFT using the by-now-famous homotopy transfer machinery \cite{Erbin:2020eyc, Koyama:2020qfb, Okawa:2022sjf, Arvanitakis:2020rrk,  Kajiura:2001ng}.

Our work starts from here, in particular from the Erbin-F\i{}rat (EF) formulation where the stub deformation is obtained by integrating-out some auxiliary fields in an appropriately extended theory.
Our first observation consists in a simple reformulation of what Erbin and F\i{}rat did. This simple reformulation is summarized as follows. Given a theory with dynamical field $\rho$, to which we want to add stubs, we add
 a free auxiliary sector with no propagating degrees of freedom, which we denote $\chi$
\begin{align}
\rho&\to (\rho,\chi)\\
S[\rho]=\frac12 \omega(\rho,Q\rho)+S_{int}[\rho]\,&\to\, S'[\rho,\chi] =\frac12 \omega(\rho,Q\rho)+S_{int}[\rho]+\frac12 \omega(\chi,Q\chi).
\end{align}
The field $\chi$ is completely decoupled and, in addition, is taken to live in a subspace where the cohomology of $Q$ is absent. In particular (if we are considering a stub operator of the form $e^{-\lambda L_0}$, having in mind open strings for definitness) the field $\chi$ is taken to live {\it outside} the kernel of $L_0$
\begin{align}
\chi\in (1-P_0 ){\cal H}.
\end{align}
It is clear that integrating out $\chi$ doesn't change the original theory, if not for an unimportant constant (the one-loop determinant of the $\chi$ quadratic term, which can be subtracted away in the definition of the extended action, if one wishes). To perform the stub deformation we
 simply do a linear field redefinition which mixes the free auxiliary field $\chi$ with the physical interacting field $\rho$
\begin{align}
\left(\begin{matrix}\rho\\\chi   \end{matrix}\right)=\left(\begin{matrix} c_\lambda&-\sqrt{1-c_\lambda^2}\\ \sqrt{1-c_\lambda^2}&c_\lambda   \end{matrix}\right)\left(\begin{matrix}\psi\\\Sigma   \end{matrix}\right),
\end{align}
 where $c_\lambda$ is the stub operator, for example $c_\lambda=e^{-\lambda L_0}$. 
Being a  linear and orthogonal  field redefinition this does not change the path integral measure and it also leaves the total kinetic term for the new fields $\psi$ and $\Sigma$ invariant
\begin{align}
S'[\rho,\chi]= S[\psi,\Sigma]=\frac12 \omega(\psi,Q\psi)+\frac12 \omega(\Sigma,Q\Sigma)+S_{int}\left[c_\lambda\psi-\sqrt{1-c_\lambda^2}\Sigma\right].
\end{align}
This action is very closely related to the extended action considered by EF, as we are going to discuss in the next section.
%This, after the trivial renaming  $\sqrt {1-c_\lambda}\Sigma=\Sigma_{\rm EF}$, is the extended action considered by Erbin and F\i{}rat (EF).
It is not difficult to see that the field $\Sigma$ still lives outside of the $Q$ cohomology, in particular
\begin{align}
\Sigma\in (1-P_0 ){\cal H},
\end{align}
therefore it can be completely integrated-out. The stubbed (quantum) theory is the resulting effective action 
\begin{align}
e^{-S_{\rm stub}^{(\lambda)}[\psi]}=\int_L {\cal D}\Sigma\,  e^{- S[\psi,\Sigma]},
\end{align}
where $L$ is a lagrangian submanifold corresponding to an appropriate gauge fixing.
In this paper we will make this simple idea more precise and we will give explicit constructions in the context of classical and quantum closed bosonic SFT as well as  open-closed SFT.

The paper is organized as follows. In section \ref{sec:2} we describe in detail our construction at the classical and quantum level and we apply it to a cubic  theory, assuming the theory makes sense at the (perturbative) quantum level. Then we apply our construction to proper solutions to the BV quantum master equation such as closed SFT and open-closed SFT.  In section \ref{sec:equiv} we give a general proof that tree and loop amplitudes are invariant under the stub deformation. As a non-trivial example we discuss in detail how the stub deformation changes the decomposition of moduli space in the one-loop open string tadpole in open-closed SFT. In the concluding section \ref{sec:6} we put forward some possible explorations that our present work opens up. Two appendices contain important technical material. In appendix \ref{app1} we prove that the quantum homotopy transfer at the co-algebraic level indeed computes the BV path integral. This is done by carefully translating  the work of Doubek, Jur\v{c}o and Pulmann \cite{DJP} from the BV to the co-algebra language. The second appendix \ref{app4} contains an explicit construction of the $SL(2,\mathbb{C})$ vertices which are needed for the one-loop open string tadpole amplitude in open-closed SFT.

 \section{Systematic approach for adding stubs}\label{sec:2}
%%%%%%%%%%%%%%%%%%%%%%%%%%%
%%%%%%%%%%%%%%%%%%%%%%%%%%%
In this section we provide a systematic algebraic procedure for introducing stubs into a generic quantum theory within the BV formalism. Then we will apply it to classical and quantum closed SFT and finally to open-closed SFT.
%    To illustrate this, we will begin by outlining the classical scenario, using Witten's open SFT as a first example. Subsequently, we will transition to the quantum case, initially describing the closed SFT and then delving into the open-closed SFT.
%%%%%%%%%%%%%%%%%%%%%%%%%%%
%%%%%%%%%%%%%%%%%%%%%%%%%%%
    \subsection{Classical stubs}\label{sec:2.1}
    We begin by considering a generic (euclidean for definiteness) classical SFT described by the following action
    \begin{equation}
        S[\rho]=\frac{1}{2}\omega(\rho,Q\rho)+S_{{\rm int}}[\rho],\label{unstubbedtheory}
    \end{equation}
    where $\rho$ is a degree zero string field belonging to the Hilbert space ${\cal H}$, $Q$ is nilpotent and $S_{{\rm int}}[\rho]$ is the interacting part of the starting un-stubbed theory. Moreover, $\omega(\cdot,\cdot)$ is a symplectic  form,  a graded anti-symmetric inner product
    \begin{equation}
        \omega(\rho_{1},\rho_{2})=-(-)^{d(\rho_1)d(\rho_2)} \omega(\rho_{2},\rho_{1}),
    \end{equation}
     in which $d(\cdot)$ is the degree. The symplectic form is non-vanishing if $d(\rho_1)+d(\rho_2)=1$.
%      In the case of open or closed string field theory the symplectic form is related to the BPZ inner product as follows
%    \begin{align}
%        &\omega_{\rm o}(\psi_1,\psi_2)\coloneqq(-)^{d(\psi_1)}\langle \psi_1, \psi_2 \rangle_{\rm o},\\
%        &\omega_{\rm c}(\phi_1,\phi_2)\coloneqq(-)^{d(\phi_1)}\langle \phi_1, c_0^-\phi_2 \rangle_{\rm c},
%    \end{align}
%    in which
%    \begin{equation}
%        c_{0}^{-}=\frac{1}{2}(c_{0}-\bar{c}_{0}),
%    \end{equation}
%    where $c_{0}$ and $\bar{c}_{0}$ are the c-ghost zero modes for the holomorphic and anti-holomorphic sector.\\
    To introduce stubs we consider an operator $c_\lambda:\mathcal{H}\to\mathcal{H}$, which can be written as    \begin{equation}
        c_\lambda=e^{-\lambda L},
    \end{equation}
    where $\lambda$ is the stub parameter and $L:\mathcal{H}\to\mathcal{H}$ is a  Grassmann even operator satisfying $[L,Q]=0$\footnote{In this paper we consider for definiteness $c_\lambda=e^{-\lambda L}$ but it is clear that one can  consider more general stubs of the kind $c(\lambda,L)$, see \cite{Erbin:2023hcs} for explicit examples.}
. In this completely general context, we also require $L$ to be BPZ even, in the sense that \footnote{In \cite{Schnabl:2024fdx} this condition has been relaxed, it would be interesting to see how our formalism can be accordingly extended to this case.}
  
    \begin{equation}
        \omega(\rho_1,L\rho_2)=\omega(L\rho_1,\rho_2).
    \end{equation}
    Furthermore, having in mind the case when $L=L_0$ or $L_0^+$, we require the existence of  a Grassmann odd operator $b$ such that $[Q,b]=L$.  Moreover we also assume that $b=0$ is a `good' gauge fixing, meaning that the cohomology of $Q$ can be systematically searched for inside  ${\rm ker}(b)$ and therefore is contained in ${\rm ker}(L)$. \\
%    Notice that $S$ is a generalization of the common stub operators given by $e^{-\lambda L_{0}}$ and $e^{-\lambda L_{0}^{+}}$ for the open and closed SFT, respectively. 
%    Where $L_{0}^{+}$ is defined as
%    \begin{equation}
%        L_{0}^{+}\coloneqq \frac{1}{2}\left(L_{0}+\bar{L}_{0}\right),
%    \end{equation}
%    in which $L_{0}$ and $\bar{L}_{0}$ are the Virasoro zero modes for the holomorphic and anti-holomorphic sector.\\
    Let us now  introduce a decoupled auxiliary field $\chi$,  without cohomology. This condition can be achieved by requiring that $\chi\in(1-P)\mathcal{H}\coloneqq \bar{P}\mathcal{H}$, where $P$ projects onto ${\rm ker}(L)$.    With this trivial addiction the action \ref{unstubbedtheory} becomes
    \begin{equation}
       S'[\rho,\chi]=\frac{1}{2}\omega(\rho,Q\rho)+\frac{1}{2}\omega(\chi,Q\chi)+S_{{\rm int}}[\rho].\label{step1}
    \end{equation}
    Notice that this is not adding any new propagating degrees of freedom because $\chi\in \bar{P}\mathcal{H}$, where the cohomology is absent. In particular, classically integrating out $\chi$ (i.e. solving its equation of motion) is the same as setting it to zero.

    Now, let us do a trivial field redefinition, a rotation induced by the orthogonal matrix    \begin{equation}\label{orthogonalfieldredef}
        \begin{pmatrix}
            \rho\\\chi
        \end{pmatrix}=O_{\lambda} \begin{pmatrix}
           \psi \\ \Sigma
        \end{pmatrix}=
        \begin{pmatrix}
         c_\lambda   & - \sqrt{1-c_\lambda^{2}}\\ \sqrt{1-c_\lambda^{2}}&c_\lambda
        \end{pmatrix}
        \begin{pmatrix}
           \psi \\ \Sigma
        \end{pmatrix},
    \end{equation}
   which leaves the kinetic terms unchanged but mixes the interactions
   \begin{equation}
      S[\psi,\Sigma]\coloneqq S'[\rho,\chi]= \frac{1}{2}\omega(\psi,Q\psi)+\frac{1}{2}\omega(\Sigma,Q\Sigma)+S_{{\rm int}}\left[c_\lambda\psi- \sqrt{1-c_\lambda^{2}}\Sigma\right].\label{genericstubwithaux}
   \end{equation}
   Notice that since $\chi \in \bar{P}\mathcal{H}$, then  $\Sigma \in \bar{P}\mathcal{H}$ as well, given that $\sqrt{1-c_\lambda^{2}}\, \psi\sim(2\lambda L)^{1/2}\,\psi\in\bar P \mathcal{H} $. Therefore $\Sigma$ can be integrated out completely without loosing physical degrees of freedom. This time however this is not the same   as setting it equal to zero due to the non-trivial couplings between $\Sigma$ and $\psi$. However, since $\Sigma$ doesn't contain physical degrees of freedom, classically integrating it out (i.e. solving the $\Sigma$ equation of motion as a function of $\psi$) should give a theory with the same physical content as the initial one. This new theory is by definition the (classical) stubbed theory
      \begin{equation}\label{genercclassicalstubbedtheory}
       \tilde{S}[\psi]=S[\psi,\Sigma(\psi)]=\frac{1}{2}\omega(\psi,Q\psi)+\tilde{S}_{\rm int}[\psi].
   \end{equation}
   The procedure we have outlined is just a convenient re-writing of what has been done by Erbin and F\i{}rat in \cite{Erbin:2023hcs} in the particular context of Witten theory. Indeed in this case \eqref{genericstubwithaux} reduces to the extended theory described in \cite{Erbin:2023hcs} by renaming $\Sigma_{\rm EF}=\sqrt{1-c_\lambda^2}\Sigma \in \bar{P}{\cal H}$, namely
    \begin{align}
  &       S_{\rm EF}[\psi, \Sigma_{\rm EF}]\coloneqq  S[\psi,\Sigma]\label{cubicstubbeaction}\\
        &=\frac{1}{2}\omega\left(\psi,Q \psi\right)+\frac{1}{2}\omega\left(\Sigma_{\rm EF}, \frac{Q}{1-c_\lambda^2}\Sigma_{\rm EF}\right)-\frac{1}{3}\omega\left((\Sigma_{\rm EF}-c_\lambda\psi),m_2\left((\Sigma_{\rm EF}-c_\lambda\psi),(\Sigma_{\rm EF}-c_\lambda\psi)\right)\right).\0
    \end{align}
%    where $m_2$ is related to the Witten star product $\ast$ as 
%    \begin{equation}
%        m_2(\psi_1,\psi_2)\coloneqq(-)^{d(\psi_1)}\psi_1\ast \psi_2.
%    \end{equation}
  Furthermore in \cite{Erbin:2023hcs} the authors consider a reduction of the theory \eqref{cubicstubbeaction} onto a ``line" defined by the linear relation
    \begin{equation}
        \Sigma_{\rm EF}= \left(c_\lambda-c_\lambda^{-1}\right)\psi \longleftrightarrow \Sigma =-\frac{\sqrt{1-c_\lambda^2}}{c_\lambda}\psi,
    \end{equation} 
    which gives Witten's theory without stubs. In our language this means to undo the field redefinition \ref{orthogonalfieldredef} and then integrate out the free field $\chi$ (which is the same as setting it to zero) thus getting back the original theory without stubs. Thanks to our reformulation, we will be able to extend these results to the quantum theory in a  particularly transparent way.
    %Therefore our approach is fully equivalent to  \cite{Erbin:2023hcs},  although conceptually simpler.
%     Indeed, in this particular configuration, the application of the homological perturbation lemma corresponds to integrate out all those degrees of freedom that do not satisfy the condition imposed by the line equation. Hence, this can be equivalently achieved by performing a field redefinition, which involves rotating the fields in such a way that one of the two degrees of freedom is orthogonal to the line ( $\chi\vert_{\rm line}=0$ whereas $\rho\vert_{\rm line}\neq 0$), and then integrating it out.
%%%%%%%%%%%%%%%%%%%%%%%%%%%
%%%%%%%%%%%%%%%%%%%%%%%%%%%
  
    \subsection{Quantum stubs}\label{sec:2.3}
Let us now repeat the above procedure in the quantum case. This time the integration out of the auxiliary fields really means to perform the path integral over $\chi$ or the `stub-rotated' $\Sigma$. Again we assume to start from a theory without stubs described by the action $S[\rho]$, which we suppose satisfies the quantum BV master equation. Then we add the non-interacting auxiliary field $\chi$.
As a simple warm-up let us review how to perform the BV path integral over the free auxiliary field $\chi$. Remember that $\chi\in\bar P{\cal H}$, which can be decomposed as
\begin{align}
\bar P{\cal H}=hQ {\cal H}\oplus Qh {\cal H},
\end{align}
where we have defined the homotopy operator / propagator
\begin{align}
h\coloneqq \frac b L \,\bar P.
\end{align}
The BV path integral has to be done over a lagrangian submanifold (a maximal subspace where the symplectic form $\omega$ vanishes) of the total space $\bar P{\cal H}$. A gauge fixing is a choice of lagrangian submanifold. In the case at hand we use the obvious gauge fixing $h\chi=0$ and therefore the lagrangian submanifold is the (component-dual of the) vector space $hQ {\cal H}$, on which the symplectic form correctly vanishes.
Let $\{c_i\}$ be a basis of $hQ {\cal H}$ and $\{b_j\}$ a basis of $Qh {\cal H}$. Notice that only the $c_i$'s give a non-trivial contribution to the kinetic term, so that expressing $\chi=\gamma^ic_i+\beta^jb_j$, we remain with
\begin{align}
\omega(\chi,Q \chi)=\gamma^i Q_{ij}\gamma^j,
\end{align}
where $Q_{ij}=\omega(c_i,Q c_j)$.
Doing the BV path integral thus means to perform the (super) gaussian integration over the BV fields $\gamma^i$ according to their grassmanality
 \begin{equation}
      \int_{h\chi=0}\mathcal{D}\chi\,e^{-S[\rho]-\frac{1}{2}\omega(\chi,Q\chi)}=e^{-{\cal A}-S[\rho]},
    \end{equation}
    where ${\cal A}=\frac12\log ({\rm sdet}(Q_{ij}))$.
Therefore, if we start from the extended action 
\begin{equation}
    S'[\rho,\chi]=\frac{1}{2}\omega(\rho,Q\rho)+\frac{1}{2}\omega(\chi,Q\chi)+S_{{\rm int}}[\rho]- \mathcal{A},
\end{equation}
the BV path integral over $\chi$ will precisely reproduce the original action $S(\rho)$.  This sets the difference in vacuum energies between the initial and the extended theory, which are now fully equivalent also at the quantum level.

We proceed now with the second step, namely redefining the fields as described in \eqref{orthogonalfieldredef}, obtaining
\begin{equation}
      S[\psi,\Sigma]\coloneqq S'[\rho(\psi,\Sigma),\chi(\psi,\Sigma)]= \frac{1}{2}\omega(\psi,Q\psi)+\frac{1}{2}\omega(\Sigma,Q\Sigma)+S_{{\rm int}}\left[c_\lambda\psi- \sqrt{1-c_\lambda^{2}}\Sigma \right]-\mathcal{A}.
\end{equation}
The stubbed theory is given by integrating out $\Sigma$ 
\begin{equation}
    \tilde{S}[\psi]=-\log\int_{h\Sigma=0} \mathcal{D}{\Sigma} \,e^{-S'[\psi,\Sigma]}=\Lambda+\frac{1}{2}\omega(\psi, Q\psi)+\tilde{S}_{\rm int}[\psi],\label{stub-action-int}
\end{equation}
where $\Lambda$ is a constant associated with the vacuum bubbles. Notice that the 1-loop bubble that one gets by integrating out $\Sigma$ is just the same as the one obtained by integrating out $\chi$ (which has been subtracted away in the definition of the extended action), so the vacuum energy $\Lambda$ only includes (interacting) vacuum diagrams from two loops onwards. \\
From now on, we will ignore these extra constants (but see the comments in the conclusions) and focus on the dynamical part of the stubbed action, which will obviously  depend on the peculiar interactions of the original theory.

  %%%%%%%%%%%%%%%%%%%%%%%%%%%%%%%%%% 
  \subsection{Example: adding quantum stubs to a cubic theory}\label{sec:cubic}
  %%%%%%%%%%%%%%%%%%%%%%%%%%%%%%%%%
  
    Let us begin with a simple example, a  cubic theory
    \begin{equation}
        S[\rho]=\frac{1}{2}\omega(\rho,Q\rho)+\frac{1}{3}\omega\left(\rho,m_{2}(\rho,\rho)\right),
    \end{equation}
    which we have written starting from a cyclic differential graded associative algebra  $(Q, m_2)$, which is the simplest non-trivial example of a cyclic  $A_\infty$ algebra (see for example \cite{Erbin:2020eyc} for quick review).  The action satisfies the classical BV master equation  thanks to the $A_{\infty}$ relations
    \begin{align}
    Q^2=&0\\
    Qm_2+m_2Q=&0\\
    m_2m_2=&0.
    \end{align}
   But, in general, it does not satisfy the quantum BV master equation.  Decomposing the original string field $\rho=r^a\,v_a$ along a basis $\{v_a\}$, 
the BV bracket $(\cdot,\cdot)$ and the BV laplacian $\Delta$ are defined as follows
\begin{align}
        &\left(\cdot, \cdot\right)= \dfrac{\overleftarrow{\partial}}{\partial r^{a}}(\omega)^{ab}\dfrac{\overrightarrow{\partial}}{\partial r^{b}},\\
        &\Delta=\frac{(-)^{r^{a}}}{2}(\omega)^{ab}\dfrac{\overrightarrow{\partial}}{\partial r^{a}}\dfrac{\overrightarrow{\partial}}{\partial r^{b}}.
    \end{align}
   where 
    \begin{equation}
        (\omega)^{ab}\coloneqq\omega(v^a,v^b),
    \end{equation}
    with  $\{v^{a}\}$ being the $\omega$-canonical dual basis 
    \begin{align}
        \omega(v^a,v_b)=-\omega(v_b,v^a)=\delta^a_b.
    \end{align}
Focusing on the BV laplacian, in the case of the present cubic theory, we get
    \begin{equation}
        \Delta{S}(\rho)=\omega\left(\rho ,m_{2}(U)\right),\label{wittenlaplacian}
    \end{equation}
    where we have introduced the Poisson bi-vector
    \begin{align}
        U=\frac{(-1)^{v^a}}{2}v_a\wedge v^a\in {\cal H}^{\wedge 2},
    \end{align} 
    where $\wedge$ is the (graded) symmetrized tensor product.
%    which is the inverse of the corresponding symplectic form in the sense that
%     \begin{align}
%        (\omega\otimes \boldsymbol{1}_{\mathcal{H}})(1_{\mathcal{H}}\otimes U)=\boldsymbol{1}_{\mathcal{H}}.
%    \end{align}
 Since, thanks to the associativity of $m_2$, we have $(S,S)=0$, the quantum BV master equation reduces to
    \begin{equation}
         \frac{1}{2}(S,S)+\Delta S=\omega\left(\rho,m_{2}(U)\right)
    \end{equation}
and is thus violated unless $m_2(U)=0$.  As it is well-known (see for example \cite{Thorn:1988hm})
 this quantity is non-vanisihing and ill-defined in Witten theory. Formally it receives contributions from the emission of physical closed string states, which appear as poles in the one-loop two point open string amplitude 
 \cite{Freedman:1987fr}. But, for the sake of simplicity (and leaving open the possibility that there could be very peculiar D-branes in very peculiar closed string backgrounds where Witten theory makes sense quantum-mechanically),  we assume here that we have  an associative two-product  $m_2$ such that $m_2(U)=0$. In later examples we will work with proper solutions to the quantum master equation, but the structures we will encounter will be  essentially  the same as in this simpler example, after we assume  $m_2(U)=0$.  That said, let us then introduce the decoupled auxiliary field $\chi$
    \begin{equation}
        S'[\rho,\chi]=\frac{1}{2}\omega(\rho,Q\rho)+\frac{1}{2}\omega(\chi,Q\chi)+\frac{1}{3}\omega\left(\rho,m_{2}(\rho,\rho)\right).
    \end{equation}
    This action can be conveniently packaged as
       \begin{equation}
         S'[R]=\frac{1}{2}\Omega\left(R,m'_{1}(R)\right)+\frac{1}{3}\Omega\left(R,m'_{2}(R,R)\right), \label{cubiccompact0}
    \end{equation}
    by organizing the string field and the auxiliary field in a vector 
    \begin{equation}
        R\coloneqq\begin{pmatrix}\rho \\ \chi \end{pmatrix} \in \mathcal{H}'\coloneqq \mathcal{H} \oplus\bar{P}\mathcal{H},
    \end{equation}
    and redefining the symplectic form and the string products as follows
    \begin{equation}
        \begin{split}
            &m'_1(R)\coloneqq\begin{pmatrix}Q\rho \\Q\chi \end{pmatrix},\\
            &m'_2(R_1,R_2)\coloneqq\begin{pmatrix} m_2(\rho_1,\rho_2)\\ 0\end{pmatrix},\\
            &\Omega(R_1,R_2)\coloneqq {\rm Tr}\begin{pmatrix}\omega(\rho_1,\rho_2)&0 \\0&\omega\left(\chi_1,\chi_2 \right) \end{pmatrix}.
        \end{split}
    \end{equation} 
Let's also define the  Poisson bi-vector $U'$ associated to the symplectic form $\Omega$ on $\cal{H}'$, which satisfies
    \begin{align}
        (\Omega\otimes \boldsymbol{1}_{{\cal H}'})(1_{{\cal H}'}\otimes U')&=\boldsymbol{1}_{{\cal H}'},
    \end{align}
    and is  explicitly written as
    \begin{equation}
       U'=\frac{(-1)^{v'^a}}{2}v'_a\wedge v'^a\in {\cal H}'^{\wedge 2},
    \end{equation}
    where the $\{v'_a\}$ and $\{v'^a\}$ are two canonically normalized basis of $\mathcal{H'}$
        \begin{align}
        \Omega(v'^a,v'_b)=-\Omega(v'_b,v'^a)=\delta^a_b,
    \end{align}
    from which we can write the completeness relation
    \begin{equation}
        1_{{\cal H}'}=v'_a\Omega(v'^{a},\cdot)=-v'^a\Omega(v'_{a},\cdot). \label{completeness}
    \end{equation}
    Now, let us perform the field redefinition \eqref{orthogonalfieldredef}, $R=O_{\lambda}\Psi$, obtaining
    \begin{equation}
         S[\Psi]\coloneqq S'[O_{\lambda}\Psi]=\frac{1}{2}\Omega\left(\Psi,m'_{1}(\Psi)\right)+\frac{1}{3}\Omega\left(\Psi,O^{-1}_{\lambda}m'_{2}(O_{\lambda}\Psi,O_{\lambda}\Psi)\right), \label{actionOlambda}
    \end{equation}
   where we used the fact that we can move $O_{\lambda}$ from the left to the right of $\Omega$ by taking its inverse (since $L$ is bpz even)
   \begin{equation}
       \Omega\left(O_{\lambda}\Psi_1,\Psi_2\right)=\Omega\left(\Psi_1,O^{-1}_{\lambda}\Psi_2\right)\label{OO}
   \end{equation}
   and the fact that this field redefinition leaves the kinetic term unchanged $m'_{1}=O^{-1}_{\lambda}m'_{1}O_{\lambda}$. Therefore, by defining
   \begin{equation}
        \begin{split}
            &M_1\coloneqq  m'_{1}\\
            &M_2(\Psi_1,\Psi_2)\coloneqq O^{-1}_{\lambda}m'_{2}(O_{\lambda}\Psi_1,O_{\lambda}\Psi_2)\\
            &\qquad\qquad\quad =\begin{pmatrix} c_{\lambda}m_2\left(c_\lambda\psi- \sqrt{1-c_\lambda^{2}}\Sigma ,c_\lambda\psi- \sqrt{1-c_\lambda^{2}}\Sigma \right) \\- \sqrt{1-c_\lambda^{2}}m_2\left(c_\lambda\psi- \sqrt{1-c_\lambda^{2}}\Sigma ,c_\lambda\psi- \sqrt{1-c_\lambda^{2}}\Sigma \right) \end{pmatrix},
        \end{split}
    \end{equation} 
    the action \eqref{actionOlambda} becomes
     \begin{equation}
         S[\Psi]=\frac{1}{2}\Omega\left(\Psi,M_{1}(\Psi)\right)+\frac{1}{3}\Omega\left(\Psi,M_2(\Psi,\Psi)\right). \label{cubiccompact}
    \end{equation}
    Let us now uplift this formalism to the level of tensor co-algebras, by extending the structures to the tensor product space $\mathcal{TH'}$, as explained in detail in \cite{Erbin:2020eyc}. In particular, the string products on $\mathcal{H'}$ are promoted to coderivations and the map $O_{\lambda}$ to a cohomorphism. This means that if we introduce the  co-product $\Delta_{\mathcal{TH'}}:\mathcal{TH'}\to \mathcal{TH'}\otimes'\mathcal{TH'}$ we have  the following relations 
    \begin{align}
       &\Delta_{\mathcal{TH'}}\boldsymbol{m}'=\left(\boldsymbol{m}'\otimes'  1_{{\cal T H}'}+  1_{{\cal T H}'}\otimes'\boldsymbol{m}'\right)\Delta_{\mathcal{TH'}},\label{coder}\\
            &\Delta_{\mathcal{TH'}}\boldsymbol{O}_{\lambda}=(\boldsymbol{O}_{\lambda}\otimes'\boldsymbol{O}_{\lambda})\Delta_{\mathcal{TH'}}\label{cohomo},  
    \end{align}
    where we defined $\boldsymbol{m}'=\boldsymbol{m}'_{1}+\boldsymbol{m}'_{2}$
    and
    \begin{align}
    \boldsymbol{m}'_{1}\pi_k=\sum_{i=0}^{k-1} 1^{\otimes i}\otimes M_1\otimes 1^{\otimes(k-i-1)}\\
     \boldsymbol{m}'_{2}\pi_k=\sum_{i=0}^{k-2} 1^{\otimes i}\otimes M_2\otimes 1^{\otimes(k-i-2)},
    \end{align}
    where $\pi_k$ projects onto the $k$-th tensor power of $\mathcal{TH'}$.  Similarly, the cohomomorphism $\boldsymbol{O}_\lambda$ is defined as
    \begin{align}
    \boldsymbol{O}_\lambda\pi_k\coloneqq \left(O_\lambda\right)^{\otimes k}.
    \end{align}
The action \eqref{cubiccompact0} can be rewritten in  Wess-Zumino-Witten (WZW) form
    \begin{equation}
    S'[R]=\int_0^1dt\,\Omega\left(\pi_{1}\boldsymbol{\partial}_t\frac{1}{1-\otimes R(t)}\,,\,\pi_{1} \boldsymbol{m}'\frac{1}{1-\otimes R(t)}\right),
    \end{equation}
    where $$\frac1{1-\otimes x}\coloneqq \sum_{k=0}^\infty x^{\otimes k}$$ is the group element and
where $R(t)$ is an  interpolation such that $R(0)=0$ and $R(1)=R$. 
  Following \cite{ Maccaferri:2023gcg}, the BV master equation can be expressed as
  \begin{align}  
\frac12 {\big (}   S',  S'{\big )}+\Delta  S'=\int_0^1dt\,\Omega\left(\pi_{1}\boldsymbol{\partial}_t\frac{1}{1-\otimes R(t)}\,,\,\pi_{1} (\boldsymbol{m}'^2+\boldsymbol{m}'\boldsymbol{U}')\frac{1}{1-\otimes R(t)}\right).
    \end{align}
 Here $\boldsymbol{U}'$ is the uplift of $U'$ to $\mathcal{TH'}$ (see for example \cite{Okawa:2022sjf, Maccaferri:2023gcg} )
 \begin{align}
 \boldsymbol{U}'=\frac12 (-1)^{v'^a}\boldsymbol{v}'_a\boldsymbol{v}'^a,
 \end{align}
 where $\boldsymbol{v}$ is the zero-product coderivation associated to the vector $v$, which inserts $v$ in all possible ways inside a tensor product, \cite{Erbin:2020eyc}.
 Upon use of the classical $A_\infty$ relations $ (\boldsymbol{m}')^2=0$, the quantum consistency reduces to $\boldsymbol{m}'\boldsymbol{U}'=0$, whose only non-trivial implication is $m_2(U)=0$, which is the quantum consistency of the initial un-stubbed theory we are assuming since the beginning.\\
    In this formalism, the field redefinition \eqref{orthogonalfieldredef} can be written as a map between group like elements, via the cohomomorphism $\boldsymbol{O}_{\lambda}$
    \begin{equation}
        \frac{1}{1-\otimes\Psi}=\boldsymbol{O}_{\lambda}\frac{1}{1-\otimes R},
    \end{equation}
    and thus, using \eqref{OO}, the above action becomes
    \begin{equation}
    S'[O_\lambda\Psi]=S[\Psi]=\int_0^1dt\,\Omega\left(\pi_{1}\boldsymbol{\partial}_t\frac{1}{1-\otimes\Psi(t)}\,,\,\pi_{1} \boldsymbol{M}\frac{1}{1-\otimes\Psi(t)}\right), \label{Maction}
    \end{equation}
    where we consistently defined 
    \begin{equation}
        \boldsymbol{M}\coloneqq \boldsymbol{O}^{-1}_{\lambda}\boldsymbol{m}'\boldsymbol{O}_{\lambda}=\boldsymbol{m}'_1+\boldsymbol{O}^{-1}_{\lambda}\boldsymbol{m}'_{2}\boldsymbol{O}_{\lambda}
    \end{equation}
    that is a coderivation
    \begin{equation}
        \Delta_{\mathcal{TH'}}\boldsymbol{M}=\left(\boldsymbol{M}\otimes'  1_{{\cal TH}'}+  1_{{\cal TH}'}\otimes'\boldsymbol{M}\right)\Delta_{\mathcal{TH'}},
    \end{equation}
    as a consequence of \eqref{coder} and \eqref{cohomo}.\\
    Clearly, the coderivation $\boldsymbol{M}$ satisfies the $A_{\infty}$ relations $\boldsymbol{M}^2=0$, whereas verifying that $\boldsymbol{M}\boldsymbol{U}'=0$ requires just few further steps. The first essential property  is that in complete generality, given an even cohomorphism constructed from a linear operator $F$ on ${\cal H}$,  $\boldsymbol{F}=\oplus_k F^{\otimes k}\pi_k$ , we easily find
        \begin{equation}
        \boldsymbol{F}\boldsymbol{U}=\boldsymbol{U}_{F}\boldsymbol{F},\label{FU=UF}
    \end{equation}
    where $\boldsymbol{U}_{F}$ corresponds to the co-algebra uplifting of the bi-vector 
    \begin{equation}
        U_{F}\coloneqq\boldsymbol{F}U=\frac{(-)^{v^{a}}}{2} Fv_a \wedge F v^a\coloneqq\frac{(-)^{f^{a}}}{2}  f_a \wedge f^a,
    \end{equation}
    namely
    \begin{equation}
        \boldsymbol{U}_{F}=\frac{(-)^{f^{a}}}{2}  \boldsymbol{f}_a\boldsymbol{f}^a.
    \end{equation}
%    Indeed it is easy to check that
%    \begin{equation}
%        \boldsymbol{F}\boldsymbol{U}(\Psi_1\otimes\cdots\otimes\Psi_n)
%        =\boldsymbol{U}_{F}\boldsymbol{F}(\Psi_1\otimes\cdots\otimes\Psi_n).
%    \end{equation}
%      \begin{equation}
%    \begin{split}
%        \boldsymbol{F}\boldsymbol{U}(\Psi_1\otimes\cdots\otimes\Psi_n)&=\frac{(-)^{v^{a}}}{2} \boldsymbol{F}\sum_{i=0}^{n}v^{a}\odot \Psi_1 \odot \cdots \Psi_i \odot v_{a}  \odot \Psi_{i+1} \odot\cdots \odot \Psi_n\\
%        &=\frac{(-)^{v^{a}}}{2} \sum_{i=0}^{n}f^{a}\odot F\Psi_1 \odot \cdots F\Psi_i \odot f_{a}  \odot F\Psi_{i+1} \odot\cdots \odot F\Psi_n\\
%        &=\boldsymbol{U}_{F}\boldsymbol{F}(\Psi_1\otimes\cdots\otimes\Psi_n),
%    \end{split}
%    \end{equation}
%    where $\odot$ is the cyclic tensor product, for more details see [Nilpot].\\
    In our context we are interested in $U'_{O_{\lambda}}$ which can be manipulated using the completeness relation \eqref{completeness} and \eqref{OO}
    \begin{equation}
        \begin{split}
            U'_{O_{\lambda}}&=\frac{(-)^{v'^{a}}}{2} O_{\lambda}v'_a \wedge O_{\lambda} v'^a\\
            &=\frac{(-)^{v'^{a}}}{2} v'_b\Omega(v'^{b},O_{\lambda}v'_a )\wedge O_{\lambda} v'^a\\
            &=\frac{(-)^{v'_{a}}}{2} v'_b\wedge O_{\lambda} v'^a \Omega(v'_a, O^{-1}_{\lambda} v'^{b})\\
            &=\frac{(-)^{v'^{b}}}{2} v'_b\wedge O_{\lambda}  O^{-1}_{\lambda} v'^{b}\\
            &=U'.\label{OU}
        \end{split}
    \end{equation}
    Therefore, from these two results we can state that
    \begin{equation}
        \boldsymbol{O}_{\lambda}\boldsymbol{U}'=\boldsymbol{U}'\boldsymbol{O}_{\lambda},\label{Uinvariance}
    \end{equation}
  and similarly
    \begin{equation}
         \boldsymbol{O}^{-1}_{\lambda}\boldsymbol{U}'=\boldsymbol{U}'\boldsymbol{O}^{-1}_{\lambda}.
    \end{equation}
    Thus, we can easily check that \eqref{Maction} is consistent at quantum level
    \begin{equation}
        \boldsymbol{M}\boldsymbol{U}'=\boldsymbol{O}_{\lambda}^{-1}\boldsymbol{m}'\boldsymbol{O}_{\lambda}\boldsymbol{U}'=\boldsymbol{O}_{\lambda}^{-1}\boldsymbol{m}'\boldsymbol{U}'\boldsymbol{O}_{\lambda}=0,
    \end{equation}
    because $\boldsymbol{m'}\boldsymbol{U'}=0$, as a consequence of the original assumption $m_2(U)=0$.\\
    Now we want to integrate-out the auxiliary field $\Sigma$ as in \eqref{stub-action-int}.
     As discussed in detail in appendix \ref{app1} (see in particular the final equation \eqref{path-integration}), this will generate constant terms (which we ignore here) and effective string products for the remaining string fields $\psi$, which can be conveniently obtained by the (quantum) homotopy transfer, starting from the free strong deformation retract
    \begin{equation}
       \circlearrowright (-h)\left(\mathcal{H}', M_1\right)\xtofrom[\text{$\iota$}]{\text{$\pi$}}\left(\mathcal{H}, Q\right)
    \end{equation}
    where the canonical projector $\pi$ and the canonical inclusion $\iota$ are defined as follows
    \begin{equation}
        \begin{split}
            &\pi(\Psi)=\psi\in \mathcal{H},\\
            &\iota(\psi)=\begin{pmatrix}\psi\\0 \end{pmatrix}\in \mathcal{H}',
        \end{split}
    \end{equation}
    and clearly satisfy $(\iota\pi)^2=\iota\pi$ and $\pi\iota=\boldsymbol{1}$. The needed propagator $h$ is defined as
    \begin{equation}
        h(\Psi)=\begin{pmatrix}0\\ \frac{b}{L}\Sigma \end{pmatrix},\label{prop}
    \end{equation}
   which obeys the annihilation relations and the Hodge-Kodaira decomposition 
    \begin{equation}
        \begin{split}
            &\pi h=h \iota= h^2=0,\\
            & M_{1}h+hM_{1}=1_{\mathcal{H}'}-\iota\pi.
        \end{split}
    \end{equation}
    Notice that in \eqref{prop} the propagator is well defined even without the $\bar{P}$ because $\Sigma\in\bar{P}\cal{H}$.\\
    Now, we can promote all these objects at  the co-algebra level. In doing this  $\pi$ and $\iota$ are promoted to cohomomorphisms  and the propagator is uplifted as  \cite{Erbin:2020eyc}
       \begin{equation}
        \boldsymbol{h}\pi_{n}=\sum_{i=0}^{n}1_{\mathcal{H'}}^{\otimes i} \otimes h \otimes (\iota \pi)^{\otimes n-i-1}\pi_n.
    \end{equation}
This establishes the existence of the strong deformation retract at the level of the tensor co-algebra
    \begin{equation}
       \circlearrowright (-\boldsymbol{h})\left(\mathcal{TH}', \boldsymbol{M}_1\right)\xtofrom[\text{$\boldsymbol{\iota}$}]{\text{$\boldsymbol{\pi}$}}\left(\mathcal{TH}, \boldsymbol{Q}\right).
    \end{equation}

The result of integrating out $\Sigma$ from the interacting theory is obtained by deforming the above strong deformation retract through $\boldsymbol{M}_{2}+\boldsymbol{U}'$ and the quantum homological perturbation lemma gives us the explicit expression for the effective products, that is
    \begin{equation}
        \tilde{\boldsymbol{m}}= \boldsymbol{\pi}(\boldsymbol{M}_1+\boldsymbol{M}_2)\dfrac{1}{\boldsymbol{1}_{\cal TH}+\boldsymbol{h}\left(\boldsymbol{M}_2+\boldsymbol{U}'\right)}\boldsymbol{\iota}=\boldsymbol{Q}+ \boldsymbol{\pi}\boldsymbol{M}_2\dfrac{1}{\boldsymbol{1}_{\cal TH}+\boldsymbol{h}\left(\boldsymbol{M}_2+\boldsymbol{U}'\right)}\boldsymbol{\iota}.
    \end{equation}
As explained in the appendix \ref{app1}, these products define the dynamical part of the  effective action, which can be written as
     \begin{equation}
         \tilde{S}[\psi]=-\log\int_{h\Sigma=0} \mathcal{D}{\Sigma} \,e^{-S[\psi,\Sigma]}=\sum_{n=0}^{\infty}\frac{1}{n+1}\omega\left(\psi, \tilde{m}_{n}\left(\psi^{\otimes n}\right)\right)+\Lambda,
     \end{equation}
     where
     \begin{equation}
         \tilde{m}_{n}=\pi_1 \tilde{\boldsymbol{m}}\pi_n
     \end{equation}
     and $\Lambda$ contains the (connected) vacuum bubbles generated by the path integral. 
     
     Let us now study in detail the effective products by making explicit the dependence on the starting coderivations $\boldsymbol{m'}$, which refers to the theory without stubs (with the addition of the trivial free and auxiliary field $\chi$)
     \begin{equation}
         \begin{split}
              \tilde{\boldsymbol{m}}&= \boldsymbol{Q}+ \boldsymbol{\pi}\boldsymbol{M}_2\dfrac{1}{\boldsymbol{1}_{\cal TH'}+\boldsymbol{h}\left(\boldsymbol{M}_2+\boldsymbol{U}'\right)}\boldsymbol{\iota}\\
              &= \boldsymbol{Q}+ \boldsymbol{\pi}\boldsymbol{O}^{-1}_{\lambda}\boldsymbol{m}'_2\boldsymbol{O}_{\lambda}\dfrac{1}{\boldsymbol{1}_{\cal TH'}+\boldsymbol{h}\left(\boldsymbol{O}^{-1}_{\lambda}\boldsymbol{m}'_2\boldsymbol{O}_{\lambda}+\boldsymbol{U}'\right)}\boldsymbol{\iota}\\
              &= \boldsymbol{Q}+ \boldsymbol{\pi}\boldsymbol{O}^{-1}_{\lambda}\boldsymbol{m}'_2\dfrac{1}{\boldsymbol{1}_{\cal TH'}+\boldsymbol{O}_{\lambda}\boldsymbol{h}\boldsymbol{O}^{-1}_{\lambda}\left(\boldsymbol{m}'_2+\boldsymbol{O}_{\lambda}\boldsymbol{U}'\boldsymbol{O}^{-1}_{\lambda}\right)}\boldsymbol{O}_{\lambda}\boldsymbol{\iota}\\
              &=\boldsymbol{Q}+ \boldsymbol{\pi}\boldsymbol{O}^{-1}_{\lambda}\boldsymbol{m}'_2\dfrac{1}{\boldsymbol{1}_{\cal TH'}+\boldsymbol{O}_{\lambda}\boldsymbol{h}\boldsymbol{O}^{-1}_{\lambda}\left(\boldsymbol{m}'_2+\boldsymbol{U}'\right)}\boldsymbol{O}_{\lambda}\boldsymbol{\iota},
         \end{split}
     \end{equation}
     where in the last step we used \eqref{Uinvariance}. Therefore, we can consistently define 
     \begin{align}
         &\boldsymbol{\pi}_{\lambda}\coloneqq \boldsymbol{\pi}\boldsymbol{O}^{-1}_{\lambda},\label{deflambda1}\\
         &\boldsymbol{\iota}_{\lambda}\coloneqq \boldsymbol{O}_{\lambda}\boldsymbol{\iota},\label{deflambda2}\\
         &\boldsymbol{h}_{\lambda}\coloneqq\boldsymbol{O}_{\lambda}\boldsymbol{h}\boldsymbol{O}^{-1}_{\lambda},\label{deflambda3}
     \end{align}
     obtaining
     \begin{equation}
         \tilde{\boldsymbol{m}}=\boldsymbol{Q}+ \boldsymbol{\pi}_{\lambda}\boldsymbol{m}'_2\dfrac{1}{\boldsymbol{1}_{\cal TH'}+\boldsymbol{h}_{\lambda}\left(\boldsymbol{m}'_2+\boldsymbol{U}'\right)}\boldsymbol{\iota}_{\lambda}.
     \end{equation}
     Notice that $\boldsymbol{\pi}_{\lambda}$ and $\boldsymbol{\iota}_{\lambda}$ are cohomomorphisms because are given by the composition of two cohomomorphisms and they trivially satisfy the retract property, namely $\boldsymbol{\pi}_{\lambda}\boldsymbol{\iota}_{\lambda}=1_{\mathcal{TH}}$ and 
     \begin{equation}
         \boldsymbol{P}_{\lambda}\coloneqq \boldsymbol{\iota}_{\lambda}\boldsymbol{\pi}_{\lambda},
     \end{equation}
     is a projector. In particular, $\boldsymbol{P}_{\lambda}$ is the cohomomorphism associated with the following projector acting on $\mathcal{H}'$
     \begin{equation}
         P_{\lambda}=\begin{pmatrix}
         c^{2}_{\lambda}   & c_{\lambda}\sqrt{1-c_\lambda^{2}}\\ c_{\lambda}\sqrt{1-c_\lambda^{2}}&1-c^{2}_{\lambda}
        \end{pmatrix}.
     \end{equation}
     Let us now focus on the propagator $\boldsymbol{h}_{\lambda}$, which clearly satisfies the annihilation relations $\boldsymbol{h}_{\lambda}^{2}=\boldsymbol{\pi}_{\lambda}\boldsymbol{h}_{\lambda}=\boldsymbol{h}_{\lambda}\boldsymbol{\iota}_{\lambda}=0$, and it acts on a bunch of $n$ states as follows
     \begin{equation}
         \boldsymbol{h}_{\lambda}\pi_{n}=\sum_{i=0}^{n}1_{\mathcal{TH'}}^{\otimes i} \otimes h_{\lambda} \otimes P_{\lambda}^{\otimes n-i-1}\pi_n,\label{propcoalgebre}
     \end{equation}     that is the usual structure of a propagator lifted to the co-algebras formalism. Moreover, the explicit expression of the propagator is
     \begin{equation}
         h_{\lambda}=\begin{pmatrix}
         \frac{b}{L}(1-c^{2}_{\lambda})   & -\frac{b}{L}c_{\lambda}\sqrt{1-c_\lambda^{2}}\\ -\frac{b}{L}c_{\lambda}\sqrt{1-c_\lambda^{2}}&\frac{b}{L}c^{2}_{\lambda}
        \end{pmatrix}.
     \end{equation}
     Therefore, adding stubs through the procedure described above can be equivalently implemented by applying this peculiar homotopy transfer referred to the action  \eqref{cubiccompact0}, namely considering  the following strong deformation retract 
     \begin{equation}
       \circlearrowright (-\boldsymbol{h}_{\lambda})\left(\mathcal{TH}', \boldsymbol{m_1}' \right)\xtofrom[\text{$\boldsymbol{\iota}_{\lambda}$}]{\text{$\boldsymbol{\pi}_{\lambda}$}}\left(\mathcal{TH}, \boldsymbol{Q}\right)
    \end{equation}\label{SDRforstubp}
  and deforming it with $   \boldsymbol{m_1}' \to \boldsymbol{m}'+\boldsymbol{U}'$.
  This describes the transfer between the two nilpotent structures
    \begin{align}
    \tilde{\boldsymbol{m}}+\boldsymbol{U}= \,&\boldsymbol{\pi}_{\lambda}(\boldsymbol{m}'+\boldsymbol{U}')\dfrac{1}{\boldsymbol{1}_{\cal TH'}+\boldsymbol{h}_{\lambda}\left(\boldsymbol{m}'_2+\boldsymbol{U}'\right)}\boldsymbol{\iota}_{\lambda}\\
    =\,&\boldsymbol{\pi}(\boldsymbol{M}+\boldsymbol{U}')\dfrac{1}{\boldsymbol{1}_{\cal TH'}+\boldsymbol{h}\left(\boldsymbol{M}_2+\boldsymbol{U}'\right)}\boldsymbol{\iota}.
    \end{align}
    
   Notice that, as explained in appendix \ref{app1} for a generic integration-out, the transfer of $\boldsymbol{U}'$ just gives $\boldsymbol{U}$
   \begin{align}
   \boldsymbol{U}=\boldsymbol{\pi}_{\lambda}(\boldsymbol{U}')\dfrac{1}{\boldsymbol{1}_{\cal TH'}+\boldsymbol{h}_{\lambda}\left(\boldsymbol{m}'_2+\boldsymbol{U}'\right)}\boldsymbol{\iota}_{\lambda},
   \end{align}
   without contaminations from the interactions.
  
It is important to realize that, although the quantum cubic action is supposed to be the same as the classical one, the resulting  quantum action with stubs  is different from its classical counterpart \cite{Erbin:2023hcs, Schnabl:2023dbv}. This is  because it has to satisfy the quantum BV master equation (under the assumption that $m_2(U)=0$) and not the classical.
   For instance, we can notice a tadpole term that is not present in the classical treatment but which is necessary for the final theory to satisfy the BV quantum master equation. In particular if we consider the action weighted by the string coupling constant $\kappa$ according to the topological expansion we get
     \begin{equation}
    \begin{split}
        \tilde{m}_{0}&=-\kappa \boldsymbol{\pi}_{\lambda}\boldsymbol{m}'_{2}\boldsymbol{h}_{\lambda}\boldsymbol{U}'\boldsymbol{1}_{\cal TH'}+ O(\kappa^2)\\
        &=-\frac{1}{2}\kappa\boldsymbol{\pi}_{\lambda}\begin{pmatrix}m_{2}\left(v^{i}\wedge \frac{b}{L}(1-c_{\lambda}^{2})v_i\right)\\0\end{pmatrix}+O(\kappa^2)\\
        &=-\kappa m_2\left(v^i,\frac{b}{L}\left(1-c_\lambda^2\right)v_i\right)+O(\kappa^2),
        \end{split}
    \end{equation}
where in the last step we used the completeness relation to move $\sqrt{1-c_\lambda^2}$ from the left to the right in the Poisson bi-vector and the fact that $c_\lambda$ commutes with the Siegel propagator. This correction is expected: in the quantum Witten theory the one-loop tadpole is fully constructed from a Feynman diagram, but after adding stubs a part of moduli space (far from open string degeneration) is eliminated in the Feynman diagram and the elementary vertex  $\tilde{m}_{0}$ is added to compensate. This vertex explicitly contains the closed string degeneration region but, just like the original Feynman diagram, it doesn't tell us how to deal with it. This appears to be general: adding open string stubs does not help in dealing with closed string degeneration. This problem can be fully dealt with in open-closed SFT, as we will explicitly see in the following.
%%%%%%%%%%%%%%%%%%%%%%%%%%%%%%%%
%%%%%%%%%%%%%%%%%%%%%%%%%%%%%%%%
    
    \subsection{Closed string field theory with stubs}\label{sec:2.4}
    In this section we will apply the prescription for adding stubs described above to the case of closed SFT, which is a theory described by a truly quantum BV master action \cite{Zwiebach:1992ie}. The action can be written as
    \begin{equation}
        S_{c}[\rho_{\rm c}]=\sum_{g=0}^{\infty}\kappa^{2g-2}\sum_{k=0}^{\infty}\frac{1}{k!} \omega_{c}\left(\rho_{\rm c},l_{k}^{(g)}(\rho_{\rm c}^{\wedge k})\right)=\frac{1}{2\kappa^{2}}\omega_{c}(\rho_{\rm c}, Q_{\rm c}\rho_{\rm c})+S_{\rm int }[\rho_{\rm c}],\label{closedwithoutstubs}
    \end{equation}
    where $\rho_{\rm c}=r^{i}_{\rm c}c_i\in \cal{H}_{\rm c}$ is the closed string field, $c_i$ the basis vectors and $Q_{\rm c} $ the closed BRST charge. As discussed in the previous section we introduce an auxiliary field $\chi_{c}\in \bar{P}_{\rm c}\cal{H}_{\rm c}$, where $P_{\rm c}$ projects onto ${\rm ker}(L_{\rm c})$ (the standard choice being $L_{\rm c}=L_0+\bar L_0$), as described in \eqref{step1}, obtaining
    \begin{equation}
        S'_{\rm c}[\rho_{\rm c},\chi_{\rm c}]=\frac{1}{2\kappa^2}\omega_{\rm c}\left(\chi_{\rm c},Q_{\rm c}\chi_{\rm c}\right)+\frac{1}{2\kappa^{2}}\omega_{c}(\rho_{\rm c}, Q_{\rm c}\rho_{\rm c})+S_{\rm int }[\rho_{\rm c}].\label{closedprimedaction}
    \end{equation}
Again we organize the two fields in a vector 
    \begin{equation}
        R_{\rm c}\coloneqq\begin{pmatrix}\rho_{\rm c} \\\chi_{\rm c } \end{pmatrix} \in \mathcal{H}'_{\rm c }\coloneqq \mathcal{H}_{\rm c } \oplus\bar{P}_{\rm c }\mathcal{H}_{\rm c },
    \end{equation}
   and introduce a collective symplectic form and its corresponding Poisson bi-vector
    
    \begin{equation}
        \Omega_{\rm c}(R_{1},R_{2})\coloneqq {\rm Tr}\begin{pmatrix}\omega_{\rm c}(\rho_{1},\rho_{2})&0 \\0&\omega_{\rm c}\left(\chi_1,\chi_2\right) \end{pmatrix},
    \end{equation}
    
    \begin{equation}
       U'_{\rm c}=\left[\dfrac{(-)^{c^i}}{2}\begin{pmatrix}c_i\\ 0 \end{pmatrix}\wedge \begin{pmatrix}c^i\\ 0 \end{pmatrix}+\dfrac{(-)^{\bar c^j}}{2}\begin{pmatrix}0\\ \bar c_j\end{pmatrix}\wedge \begin{pmatrix} 0\\ \bar c^{j} \end{pmatrix}\right]\in {\cal H}_{\rm c}'^{\wedge 2},
    \end{equation}
    such that
    \begin{align}
    (\Omega_{\rm c}\otimes \boldsymbol{1})(1\otimes U'_{\rm c})&=\boldsymbol{1}_{\cal {H'}_{\rm c}}.
    \end{align}
    Let us then extend the multi string products on the new Hilbert space $\mathcal{H}'_{\rm c}$
   \begin{equation}
       l'^{(g)}_{n}(R_1,\cdots , R_{n})\coloneqq \begin{pmatrix}
            l'^{(g)}_{n}(\rho_1,\cdots,\rho_n)\\0
       \end{pmatrix},
   \end{equation}
   which can be uplifted to the co-algebras formalism as coderivations acting on the symmetrized closed tensor algebra \cite{Erbin:2020eyc}
   \begin{align}
        \bfl'\coloneqq  \sum_{g}\kappa^{2g}\sum_{n}\boldsymbol{l}'^{(g)}_{n} :\mathcal{SH}'_{\rm c}\to\mathcal{SH}'_{\rm c}.
    \end{align}
    Therefore, the action \eqref{closedprimedaction} can be rewritten in the WZW form by introducing the interpolated field $R_{\rm c}(t)$ such that $R_{\rm c}(0)=0$ and $R_{\rm c}(1)=R_{\rm c}$
    \begin{align}
    S'[R_{\rm c}]=\int_0^1dt\,\frac{\Omega_{\rm c}}{\kappa^2}\left(\pi_{1}\boldsymbol{\partial}_t\,e^{\wedge R_{\rm c}(t)}\,,\,\pi_{1} \boldsymbol{l}'e^{\wedge R_{\rm c}(t)}\right),\label{WZWclosedprimedaction}
    \end{align}
    where $\pi_{1}$ projects onto the single closed string state sector.\\
    Moreover, the multi-string products satisfy the quantum $L_{\infty}$ homotopy algebra, which in this formalism reads
    \begin{equation}
        (\boldsymbol{l}'+\boldsymbol{U}'_{\rm c})^2=0,
    \end{equation}
    and this guarantees the validity of the quantum BV master equation \cite{Markl:1997bj}.
    Let us now perform the field redefinition \eqref{orthogonalfieldredef} adapted to this case
     \begin{equation}
        R_{\rm c}=\begin{pmatrix}
            \rho_{\rm c}\\\chi_{\rm c}
        \end{pmatrix}=O_{\lambda}^{({\rm c})} \begin{pmatrix}
           \phi \\ \Sigma_{\rm c}
        \end{pmatrix}=
        \begin{pmatrix}
         c_\lambda^{({\rm c})}   & - \sqrt{1-c_\lambda^{({\rm c})2}}\\ \sqrt{1-c_\lambda^{({\rm c})2}}&c_\lambda^{({\rm c})}
        \end{pmatrix}
        \begin{pmatrix}
           \phi \\ \Sigma_{\rm c}
        \end{pmatrix}\coloneqq \Phi,
    \end{equation}
    where we can standardly choose $c_\lambda^{({\rm c})} =e^{\lambda L_{\rm c}}\to e^{\lambda L_{0}^+}$.
    This rotation uplifts to a cohomomorphism between two group-like elements
    \begin{equation}
        e^{\wedge R_{\rm c}}=\boldsymbol{O}_{\lambda}^{({\rm c})}e^{\wedge \Phi}.
    \end{equation}
    Thus, by using the property \eqref{OO} the action can be rewritten as follows
    \begin{align}
    S'[\Phi]=\int_0^1dt\,\frac{\Omega_{\rm c}}{\kappa^2}\left(\pi_{1}\boldsymbol{\partial}_t\,e^{\Phi(t)}\,,\,\pi_{1} \boldsymbol{L}e^{\Phi(t)}\right),\label{C-cubic-stubbed-WZW}
    \end{align}
    where we defined
    \begin{equation}
        \boldsymbol{L}\coloneqq \boldsymbol{O}_{\lambda}^{({\rm c})-1}\boldsymbol{l}'\boldsymbol{O}_{\lambda}^{({\rm c})},
    \end{equation}
    that is clearly a coderivation that satisfies the quantum $L_{\infty}$ relations, indeed
    \begin{equation}
        (\boldsymbol{L}+\boldsymbol{U}'_{\rm c})^{2}=\boldsymbol{O}_{\lambda}^{({\rm c})-1}\left(\boldsymbol{l}'+\boldsymbol{O}_{\lambda}^{({\rm c})}\boldsymbol{U}'_{\rm c}\boldsymbol{O}_{\lambda}^{({\rm c})-1}\right)^2\boldsymbol{O}_{\lambda}^{({\rm c})}=\boldsymbol{O}_{\lambda}^{({\rm c})-1}(\boldsymbol{l}'+\boldsymbol{U}'_{\rm c})^2\boldsymbol{O}_{\lambda}^{({\rm c})}=0,\label{homotopyproof}
    \end{equation}
    where in the last step we used \eqref{Uinvariance}.\\    
    Since the products satisfy the quantum homotopy relations we can consistently apply the homological perturbation lemma to integrate out the auxiliary field $\Sigma$ ending up on a quantum stubbed closed SFT action. We therefore introduce the objects which define the free strong deformation retract, namely the canonical inclusion $\iota_{\rm c}$, the canonical projector $\pi_{\rm c}$ and the propagator $h_{\rm  c}$ as follows
    \begin{align}
        &\pi_{\rm c}(\Phi)=\phi,\\
        &\iota_{\rm c} (\phi)=
        \begin{pmatrix}
            \phi \\ 0
        \end{pmatrix},\\
        &h_{\rm c}(\Phi)=\begin{pmatrix}
            0\\ \frac{b_{c}}{L_{c}}\Sigma_{\rm c}
        \end{pmatrix},
    \end{align}
    which clearly satisfy the retract properties $\pi_{\rm c}\iota_{\rm c}=1_{{\cal H}_{\rm c}}$ and $(\iota_{\rm c}\pi_{\rm c})^2=\iota_{\rm c}\pi_{\rm c}$, the annihilation relations $h_{\rm c}\iota_{\rm c}=\pi_{\rm c}h_{\rm c}=h_{\rm c}^2=0$ and the Hodge-Kodaira decomposition
    \begin{equation}
        L_{1}^{(0)}h_{\rm c}+h_{\rm c} L_{1}^{(0)}=1_{{\cal H}'_{\rm c}}-\iota_{\rm c}\pi_{\rm c}.
    \end{equation}
    We can then promote such objects to the language of co-algebras by following the well-known definitions of propagator and projectors on the symmetrized tensor space, as described for example in appendix B of \cite{Erbin:2020eyc}.
%    Notice that if we apply $\boldsymbol{\pi}_{\rm c}$ to the group like element of $\mathcal{SH}'_{\rm c}$ we obtain the group-like element of $\mathcal{SH}_{\rm c}$, namely 
%    \begin{equation}
%        \boldsymbol{\iota}_{\rm c}e^{\wedge \Phi}=e^{\wedge \phi}.
%    \end{equation}
     We can then apply the homological perturbation lemma, which tells us that the effective products that will describe the transferred theory take the following form
          \begin{align}
              \tilde{\boldsymbol{l}}&=\boldsymbol{Q}_{\rm c}+\boldsymbol{\pi}_{\rm c}\boldsymbol{\delta L}\frac{1}{\boldsymbol{1}_{\mathcal{SH'}_{\rm c}}+\boldsymbol{h}_{\rm c}\left(\boldsymbol{\delta L}+\kappa^{2}\boldsymbol{U'}_{\rm c}\right)}\boldsymbol{\iota}_{\rm c}=\\
              &=\boldsymbol{Q}_{\rm c}+ \boldsymbol{\pi}_{\rm c}\boldsymbol{O}^{({\rm c})-1}_{\lambda}\boldsymbol{\delta l}'\dfrac{1}{\boldsymbol{1}_{\cal SH'}+\boldsymbol{O}_{\lambda}^{({\rm c})}\boldsymbol{h}_{\rm c}\boldsymbol{O}^{({\rm c})-1}_{\lambda}\left(\boldsymbol{\delta l}'+\kappa^{2}\boldsymbol{U}'_{\rm c}\right)}\boldsymbol{O}_{\lambda}^{({\rm c})}\boldsymbol{\iota}_{\rm c}\\
              &=\boldsymbol{Q}_{\rm c}+ \boldsymbol{\pi}_{\lambda}^{({\rm c})}\boldsymbol{\delta l}'\dfrac{1}{\boldsymbol{1}_{\cal SH'}+\boldsymbol{h}_{\lambda}^{({\rm c})}\left(\boldsymbol{\delta l}'+\kappa^{2}\boldsymbol{U}'_{\rm c}\right)}\boldsymbol{\iota}_{\lambda}^{({\rm c})},
          \end{align}
          in which we consistently defined 
          \begin{align}
         &\boldsymbol{\pi}_{\lambda}^{({\rm c})}\coloneqq \boldsymbol{\pi}_{\rm c}\boldsymbol{O}^{({\rm c})-1}_{\lambda},\\
         &\boldsymbol{\iota}_{\lambda}^{({\rm c})}\coloneqq \boldsymbol{O}_{\lambda}^{({\rm c})}\boldsymbol{\iota}_{\rm c},\\
         &\boldsymbol{h}_{\lambda}^{({\rm c})}\coloneqq\boldsymbol{O}_{\lambda}^{({\rm c})}\boldsymbol{h}_{\rm c}\boldsymbol{O}^{({\rm c})-1}_{\lambda}.\label{closedstubbedproducts}
     \end{align}
    Therefore, the action for the subbed closed string field (ignoring the vacuum bubbles, which are not captured by the homotopy transfer) reads
    \begin{align}
    \tilde{S}_{\rm c}\ch{[}\phi\ch{]}=\int_0^1dt\,\frac{\omega_{\rm c}}{\kappa^2}\left(\pi_{1}\boldsymbol{\partial}_t\,e^{\wedge \phi(t)}\,,\,\pi_{10} \tilde{\boldsymbol{l}}\,e^{\wedge \phi(t)}\right) ,\label{closedwithstubs}
    \end{align}
    where $ \tilde{\boldsymbol{l}}=\sum_{g=0}^\infty \kappa^{2g}  \tilde{\boldsymbol{l}}^{(g)}$. In particular the truncation to  $g=0$ is the classical stubbed closed SFT, which obeys the classical master equation.
%%%%%%%%%%%%%%%%%%%%%%%%%%%%%%%%
%%%%%%%%%%%%%%%%%%%%%%%%%%%%%%%%
    
    \subsection{Open-closed string field theory with stubs}\label{sec:2.5}
    
  %%%%%%%%%%%%%%%%%%%%%%  
    
    In this section we will apply our method to open-closed string field theory \cite{Zwiebach:1997fe}. We will follow the recent formulation of OC-SFT presented in \cite{Maccaferri:2023gcg}, to which we refer the reader for further details. The BV master action is given by
    \begin{equation}
        S_{{\rm oc}}[\rho_{\rm o},\rho_{\rm c}]=\sum_{g,b=0}^{\infty}\kappa^{2g+b-2}\sum_{k=0}^{\infty}\sum_{\{p_1,...,p_b\}}^{\infty}\frac{1}{b!k!(p_1)\cdots(p_b)}\mathcal{A}^{g,b}_{k;p_1,...,p_b}\left(\rho_{\rm c}^{\wedge k}  \otimes' \rho_{\rm o}^{\odot p_1}\wedge'\cdots\wedge' \rho_{\rm o}^{\odot p_b} \right),
    \end{equation}
    where $\rho_{\rm o}=r^{i}_{\rm o}o_i\in \cal{H}_{\rm o}$ is the open string field,  $\rho_{\rm c}=r^{i}_{\rm c}c_i\in \cal{H}_{\rm c}$ the closed string field, $o_i$ the basis vectors of $\mathcal{H}_{\rm o}$ and $c_i$ the basis vectors of $\mathcal{H}_{\rm c}$. The $\mathcal{A}^{g,b}_{k;p_1,...,p_b}$ represent all the possible couplings on all possible Riemann surfaces, including the kinetic terms for closed strings (sphere) and open strings (disk). The action can be written in terms of open-closed multi-string products as \cite{Maccaferri:2023gcg}
       \begin{align}
    &S_{\rm oc}[\rho_{\rm c},\rho_{\rm o}]=\sum_{g,b=0}^{\infty}\kappa^{2g+b-2}\sum_{k=0}^{\infty}\frac{1}{b!}{\ch{\Bigg [}}\frac1{(k+1)!}
    \wc\left(\rho_{\rm c},l^{(g,b)}_{k,0}\left(\rho_{\rm c}^{\wedge k}\right)\right)\label{OCact}\\
    &+\ch{\sum_{\substack{\{p_1,...,p_{b-1}\}=0\\ {p_b=1}}}^{\infty}}\frac{\ch{C({p_1,\ldots, p_b})}}{k!\,(p_1)\cdots(p_b)}
    \wo\left(\rho_{\rm o},m^{(g,b)}_{k\left[p_1,...,p_{b-1}\right]p_b-1}\left(\rho_{\rm c}^{\wedge k}\otimes'\rho_{\rm o}^{\odot p_1}\cdots\rho_{\rm o}^{\odot p_{b-1}}\otimes''\rho_{\rm o}^{\otimes (p_b-1)}\right)\right){\ch{\Bigg ]}}\0
    \end{align}
with the important identification
   \begin{align}
    &\wo\left(\Psi_{b,1}, m^{(g,b)}_{k\left[p_1,...,p_{b-1}\right]p_b-1}\left((\Phi)_k\otimes' [\Psi]_{ p_1}\wedge' \cdots\wedge'  [\Psi]_{ p_{b-1}}\otimes'' \Psi_{b,2}\otimes\cdots\otimes\Psi_{b,p_b}\right)\right)\0\\
    \coloneqq&(-1)^\epsilon \wc\left(\Phi_1,l^{(g,b)}_{k-1,[p_1,\cdots, p_b]}\left(\Phi_2\wedge\cdots\wedge\Phi_k\otimes'[\Psi]_{p_1}\wedge'\cdots\wedge'[\Psi]_{p_b}\right)\right)\label{dual}\\
    =&\mathcal{A}^{g,b}_{k;\{p_1,...,p_b\}}\left((\Phi)_k\otimes'[\Psi]_{p_1}\wedge'...\wedge' [\Psi]_{p_b}\right)\0
    \end{align}
and other more standard cyclicity relations.
The constrained $l$ and $m$ products can then be upgraded to coderivation-like operators $\boldsymbol{l}$ and $\boldsymbol{m}$ on the open-closed tensor algebra $\mathcal{SH}_{\rm c}\otimes' \mathcal{SCH}_{\rm o}$ (consisting of symmetrized closed strings ($\mathcal{SH}_{\rm c}$) and cyclized open strings ($\mathcal{CH}_{\rm o}$) on symmetrized boundaries ($\mathcal{SCH}_{\rm o}$)) in a way which is described in detail in \cite{Maccaferri:2023gcg}.

     Following the general strategy, we isolate the two kinetic terms and we introduce an auxiliary field for each sector  
    \begin{equation}
    \begin{split}
        S'_{{\rm oc}}[\rho_{\rm o},\chi_{\rm o},\rho_{\rm c},\chi_{\rm c}]&=\frac{1}{2\kappa}\wo(\rho_{\rm o},Q_{\rm o}\rho_{\rm o})+\frac{1}{2\kappa}\wo\left(\chi_{\rm o},Q_{\rm o}\chi_{\rm o}\right)\\
        &+\frac{1}{2\kappa^2}\wc(\rho_{\rm c},Q_{\rm c}\rho_{\rm c})+ \frac{1}{2\kappa^2}\wc\left(\chi_{\rm c},Q_{\rm c}\chi_{\rm c }\right)\\ &+ S_{{\rm int}}[\rho_{\rm o},\rho_{\rm c}].
    \end{split}
        \end{equation}
        in which $\chi_{\rm o}\in \bar{P}_{\rm o}$ and $\chi_{\rm c}\in \bar{P}_{\rm c}$ where $P_{\rm o}$ and $P_{\rm c}$ project onto ${\rm ker}(L_{\rm o})$ and ${\rm ker}(L_{\rm c})$, respectively. The standard  choice of stubs is $L_{\rm c}=L_0^+$ and $L_{\rm o}=L_0$. \\
    We then define 
\begin{equation}
        R_{\rm c}\coloneqq\begin{pmatrix}\rho_{\rm c} \\\chi_{\rm c } \end{pmatrix} \in \mathcal{H}'_{\rm c }\coloneqq \mathcal{H}_{\rm c } \oplus\bar{P}_{\rm c }\mathcal{H}_{\rm c },
    \end{equation}
    \begin{equation}
        R_{\rm o}\coloneqq\begin{pmatrix}\rho_{\rm o} \\\chi_{\rm o} \end{pmatrix} \in \mathcal{H}'_{\rm o }\coloneqq \mathcal{H}_{\rm o} \oplus\bar{P}_{\rm o}\mathcal{H}_{\rm o},
    \end{equation}
     and
    \begin{equation}
        \Omega_{\rm o}(R_{{\rm o}(1)},R_{{\rm o}(2)})\coloneqq {\rm Tr}\begin{pmatrix}\omega_{\rm o}(\rho_{{\rm o}(1)},\rho_{{\rm o}(2)})&0 \\0&\omega_{\rm o}\left(\chi_{{\rm o}(1)},\chi_{{\rm o}(2)}\right) \end{pmatrix},
    \end{equation}
    \begin{equation}
        \Omega_{\rm c}(R_{{\rm c}(1)},R_{{\rm c}(2)})\coloneqq {\rm Tr}\begin{pmatrix}\omega_{\rm c}(\rho_{{\rm c}(1)},\rho_{{\rm c}(2)})&0 \\0&\omega_{\rm c}\left(\chi_{{\rm c}(1)},\chi_{{\rm c}(2)}\right) \end{pmatrix},
    \end{equation}
    \begin{equation}
       U'_{\rm o}=\left[\dfrac{(-)^{o^i}}{2}\begin{pmatrix}o_i\\ 0 \end{pmatrix}\wedge \begin{pmatrix}o^i\\ 0 \end{pmatrix}+\dfrac{(-)^{\bar o^j}}{2}\begin{pmatrix}0\\ \bar o_j\end{pmatrix}\wedge \begin{pmatrix} 0\\ \bar o^{j} \end{pmatrix}\right]\in {\cal H}_{\rm o}'^{\wedge 2},
    \end{equation}
    \begin{equation}
       U'_{\rm c}=\left[\dfrac{(-)^{c^i}}{2}\begin{pmatrix}c_i\\ 0 \end{pmatrix}\wedge \begin{pmatrix}c^i\\ 0 \end{pmatrix}+\dfrac{(-)^{\bar c^j}}{2}\begin{pmatrix}0\\ \bar c_j\end{pmatrix}\wedge \begin{pmatrix} 0\\\bar c^{j} \end{pmatrix}\right]\in {\cal H}_{\rm c}'^{\wedge 2},
    \end{equation}
    such that
    \begin{align}
    (\Omega_{\rm o}\otimes \boldsymbol{1})(1\otimes U'_{\rm o})&=\boldsymbol{1}_{\cal {H'}_{\rm o}},
    \end{align}
    \begin{align}
    (\Omega_{\rm c}\otimes \boldsymbol{1})(1\otimes U'_{\rm c})&=\boldsymbol{1}_{\cal {H'}_{\rm c}}.
    \end{align}
   Now, we promote the multi string products on the new Hilbert spaces
     \begin{align}
        &l'^{(g,b)}_{k[p_1,\cdots, p_{b}]}(R_{\rm c}^{\wedge k}\otimes' R_{\rm o}^{\odot p_b}\wedge' \cdots\wedge' R_{\rm o}^{\odot p_1} )\coloneqq\begin{pmatrix}
            l^{(g,b)}_{k[p_1,\cdots, p_{b}]}(\rho_{\rm c}^{\wedge k}\otimes' \rho_{\rm o}^{\odot p_b}\wedge' \cdots\wedge' \rho_{\rm o}^{\odot p_1} )\\0
        \end{pmatrix},\\
        &m'^{(g,b)}_{k[p_1,\cdots, p_{b-1}]p_b}(R_{\rm c}^{\wedge k}\otimes' R_{\rm o}^{\odot p_1}\wedge' \cdots\wedge' R_{\rm o}^{\odot p_{b-1}}\otimes'' R_{\rm o}^{\otimes p_{b}} )\coloneqq\notag\\
        &\qquad\qquad\qquad\qquad\qquad\qquad\begin{pmatrix}
           m^{(g,b)}_{k[p_1,\cdots, p_{b-1}]p_b}(\rho_{\rm c}^{\wedge k}\otimes' \rho_{\rm o}^{\odot p_1}\wedge' \cdots\wedge' \rho_{\rm o}^{\odot p_{b-1}}\otimes'' \rho_{\rm o}^{\otimes p_{b}} )\\0
        \end{pmatrix},
    \end{align}
    which can be uplifted to the co-algebras formalism in the sense described in \cite{Maccaferri:2023gcg}.
    \begin{align}
        &\bfl'= \sum_{g,b}\kappa^{2g+b}\sum_{k,\left\{p_1,...,p_b\right\}}\boldsymbol{l}^{(g,b)}_{k\left[p_1,...,p_b\right]}: \mathcal{SH'}_{\rm c}\otimes' \mathcal{SCH'}_{\rm o}\to\mathcal{SH'}_{\rm c}\otimes' \mathcal{SCH'}_{\rm o}\\
        &\bfm'= \sum_{g,b}\kappa^{2g+b-1}\sum_{k,\left\{p_1,...,p_b\right\}}\boldsymbol{m}^{(g,b)}_{k\left[p_1,...,p_{b-1}\right]p_b}: \mathcal{SH'}_{\rm c}\otimes' \mathcal{SCH'}_{\rm o}\to\mathcal{SH'}_{\rm c}\otimes' \mathcal{SCH'}_{\rm o}.
    \end{align}
Therefore, by defining an interpolation for each field, namely $R_{\rm o}(t)$ and  $R_{\rm c}(t)$ such that for $t=0$ they return $0$ and for $t=1$ $R_{\rm o}$ and $R_{\rm c}$, respectively, we can rewrite the open-closed SFT action with free auxiliary fields in the WZW form
        \begin{align}
    S'[R_{\rm c},R_{\rm o}]=\int_0^1dt\,&\frac{\Omega_{\rm c}}{\kappa^2}\left(\pi_{10}\boldsymbol{\partial}_t\,{\cal G}(R_{\rm c}(t),R_{\rm o}(t)))\,,\,\pi_{10} \bfl'\,{\cal G}(R_{\rm c}(t),R_{\rm o}(t))\right)+\notag\\
    &\frac{\Omega_{\rm o}}{\kappa}\left(\pi_{01}\,\boldsymbol{\partial}_t\,{\cal G}(R_{\rm c}(t),R_{\rm o}(t))\,,\,\pi_{01}\bfm'\,{\cal G}(R_{\rm c}(t),R_{\rm o}(t))\right),\label{ocactionRR}
    \end{align}
    where $\pi_{10}$ and $\pi_{01}$ returns a single copy of $\mathcal{H}'_{\rm c}$ and  $\mathcal{H}'_{\rm o}$, respectively. In addition, the group like element is defined as in \cite{Maccaferri:2023gcg}
    \begin{equation}
       {\cal G}(R_{\rm c},R_{\rm o}) = \sum_{k,b,p_1,...,p_b\geq 0}\frac{1}{k!b!(p_1)\cdots(p_b)}\left(R_{\rm c}^{\wedge k}\otimes'R_{\rm o}^{\odot p_1}\wedge' \cdots \wedge' R_{\rm o}^{\odot p_b}\right),
    \end{equation}
       where $(p)=p+\delta_{p,0}$.
    Furthermore, the products satisfy the quantum open-closed homotopy algebra, namely
 \begin{align}
    &\pi_{10}\left\{\dfrac{1}{2}[\boldsymbol{l}',\boldsymbol{l}']+\boldsymbol{l}'\boldsymbol{m}'+\left[\boldsymbol{l}',\kappa\boldsymbol{U}'_{o}+\kappa^{2}\boldsymbol{U}'_{c}\right]\right\}=0,\\
    &\kappa\pi_{01}\left\{\dfrac{1}{2}[\boldsymbol{m}',\boldsymbol{m}']+\boldsymbol{m}'\boldsymbol{l}'+\left[\boldsymbol{m}',\kappa\boldsymbol{U}'_{o}+\kappa^{2}\boldsymbol{U}'_{c}\right]\right\}=0,
    \end{align}
     which can be succinctly written as $$( \boldsymbol{l}'+\boldsymbol{m}'+\kappa\boldsymbol{U}'_{o}+\kappa^{2}\boldsymbol{U}'_{c})^2=0.$$

    Let us now implement the field redefinition \eqref{orthogonalfieldredef}, with two independent stub parameters $\lambda_{\rm c}$ and $\lambda_{\rm o}$
    \begin{equation}
        R_{\rm c}=\begin{pmatrix}
            \rho_{\rm c}\\\chi_{\rm c}
        \end{pmatrix}=O_{\lambda_{\rm c}}^{({\rm c})} \begin{pmatrix}
           \phi \\ \Sigma_{\rm c}
        \end{pmatrix}=
        \begin{pmatrix}
         c_{\lambda_{\rm c}}^{({\rm c})}   & - \sqrt{1-c_{\lambda_{\rm c}}^{({\rm c})2}}\\ \sqrt{1-c_{\lambda_{\rm c}}^{({\rm c})2}}&c_{\lambda_{\rm c}}^{({\rm c})}
        \end{pmatrix}
        \begin{pmatrix}
           \phi \\ \Sigma_{\rm c}
        \end{pmatrix}\coloneqq \Phi,
    \end{equation}
    \begin{equation}
        R_{\rm o}=\begin{pmatrix}
            \rho_{\rm o}\\\chi_{\rm o}
        \end{pmatrix}=O_{\lambda_{\rm o}}^{({\rm o})} \begin{pmatrix}
           \psi \\ \Sigma_{\rm o}
        \end{pmatrix}=
        \begin{pmatrix}
         c_{\lambda_{\rm o}}^{({\rm o})}   & - \sqrt{1-c_{\lambda_{\rm o}}^{({\rm o})2}}\\ \sqrt{1-c_{\lambda_{\rm o}}^{({\rm o})2}}&c_{\lambda_{\rm o}}^{({\rm o})}
        \end{pmatrix}
        \begin{pmatrix}
           \psi \\ \Sigma_{\rm o}
        \end{pmatrix}\coloneqq \Psi.
    \end{equation}
   As discussed in the previous sections the orthogonal matrices which define the above field redefinitions can be upgraded to the co-algebras formalism as an open-closed cohomomorphism-like object acting on the group-like element
   \begin{equation}
       {\cal G}(\Phi,\Psi) =\boldsymbol{O}_{\lambda_{\rm o},\lambda_{\rm c}} {\cal G}(R_{\rm c},R_{\rm o}),
   \end{equation}
    where
    \begin{equation}
    \boldsymbol{O}_{\lambda_{\rm o},\lambda_{\rm c}}=\boldsymbol{O}^{({\rm c}) }_{\lambda_{\rm c}}\otimes' \boldsymbol{O}^{({\rm o}) }_{\lambda_{\rm o}}:\mathcal{SH'}_{\rm c}\otimes' \mathcal{SCH'}_{\rm o}\to\mathcal{SH}_{\rm c}\otimes' \mathcal{SCH}_{\rm o}.
    \end{equation}
   Therefore, thanks to \eqref{OO} the action \eqref{ocactionRR} can be rewritten as
   \begin{align}
    S'[\Phi,\Psi]=\int_0^1dt\,&\frac{\Omega_{\rm c}}{\kappa^2}\left(\pi_{10}\boldsymbol{\partial}_t\,{\cal G}(\Phi(t),\Psi(t)))\,,\,\pi_{10} \boldsymbol{L}\,{\cal G}(\Phi(t),\Psi(t))\right)+\notag\\
    &\frac{\Omega_{\rm o}}{\kappa}\left(\pi_{01}\,\boldsymbol{\partial}_t\,{\cal G}(\Phi(t),\Psi(t))\,,\,\pi_{01}\boldsymbol{M}\,{\cal G}(\Phi(t),\Psi(t))\right),\label{ocactionPhiPsi}
    \end{align}
    where we defined 
    \begin{align}
        &\boldsymbol{L}\coloneqq \boldsymbol{O}^{-1 }_{\lambda_{\rm o},\lambda_{\rm c}}\boldsymbol{l}'\boldsymbol{O}_{\lambda_{\rm o},\lambda_{\rm c}},\\
        &\boldsymbol{M}\coloneqq \boldsymbol{O}^{-1}_{\lambda_{\rm o},\lambda_{\rm c}}\boldsymbol{m}'\boldsymbol{O}_{\lambda_{\rm o},\lambda_{\rm c}},
    \end{align}
       Moreover, the new multi-string products satisfy the quantum homotopy algebra 
     \begin{align}
    &\pi_{10}\left\{\dfrac{1}{2}[\boldsymbol{L},\boldsymbol{L}]+\boldsymbol{L}\boldsymbol{M}+\left[\boldsymbol{L},\kappa\boldsymbol{U}'_{o}+\kappa^{2}\boldsymbol{U}'_{c}\right]\right\}=0,\\
    &\kappa\pi_{01}\left\{\dfrac{1}{2}[\boldsymbol{M},\boldsymbol{M}]+\boldsymbol{M}\boldsymbol{L}+\left[\boldsymbol{M},\kappa\boldsymbol{U}'_{o}+\kappa^{2}\boldsymbol{U}'_{c}\right]\right\}=0,
    \end{align}
    and this can be seen by following the same steps as in \eqref{homotopyproof}. Therefore, \eqref{ocactionPhiPsi} satisfies the quantum master equation, and we can  consistently apply the quantum homotopy transfer with the goal of integrating out the two auxiliary fields $\Sigma_{\rm o}$ and $\Sigma_{\rm c} $. Furthermore, precisely because the products $\boldsymbol{L}$ and $\boldsymbol{M}$ define a quantum nilpotent structure
    $$(\boldsymbol{L}+\boldsymbol{M}+\kappa\boldsymbol{U}'_{o}+\kappa^{2}\boldsymbol{U}'_{c})^2=0,$$
     we can assert that the effective theory will also be a solution of the corresponding quantum master equation.\\
        Let us then define the desired strong deformation retract by introducing a canonical projector and a canonical inclusion promoted to the co-algebras formalism as cohomorphisms
        \begin{align}
            &\boldsymbol{\pi}=\boldsymbol{\pi}_{\rm c}\otimes'\boldsymbol{\pi}_{\rm o},  \\
            &\boldsymbol{\iota}=\boldsymbol{\iota}_{\rm c}\otimes'\boldsymbol{\iota}_{\rm o},
        \end{align}
      where
      \begin{align}
          &\pi_{\rm c}(\Phi)=\phi,\\ 
          &\iota_{\rm c}(\phi)=\begin{pmatrix}\phi \\  0\end{pmatrix},
      \end{align}
      the same for $\pi_{\rm o}$ and $\iota_{\rm o}$. Hence, if we apply $\boldsymbol{\pi}$ to the group like element we get
      \begin{equation}
          \boldsymbol{\pi}{\cal G}(\Phi,\Psi) =\sum_{k,b,p_1,...,p_b\geq 0}\frac{1}{k!b!(p_1)\cdots(p_b)}\left(\phi^{\wedge k}\otimes'\psi^{\odot p_1}\wedge' \cdots \wedge' \psi^{\odot p_b}\right)=\mathcal{G}(\phi,\psi)\coloneqq\mathcal{G}.
      \end{equation}
      Let us now define the propagator
      \begin{equation}
          \boldsymbol{h}=\boldsymbol{h}_{c}\otimes' \boldsymbol{\iota}_{\rm o}\boldsymbol{\pi}_{\rm o}+\boldsymbol{1}_{\mathcal{SH'}_{\rm c}}\otimes'\boldsymbol{h}_{\rm o},
      \end{equation}
      where
      \begin{align}
          &h_{\rm c}(\Phi)=\begin{pmatrix} 0\\ \frac{b_{\rm c}}{L_{\rm c}} \Sigma_{\rm c}    \end{pmatrix},\\
          &h_{\rm o}(\Psi)=\begin{pmatrix} 0\\ \frac{b_{\rm o}}{L_{\rm o}} \Sigma_{\rm o}    \end{pmatrix}.
          \end{align}
          Finally, by applying the quantum homological perturbation lemma we get the following  action for the stubbed open-closed theory
          \begin{align}
    \tilde{S}_{\rm oc}\ch{[}\phi,\psi\ch{]}=\int_0^1dt\left(\frac{\omega_{\rm c}}{\kappa^2}\left(\pi_{10}\boldsymbol{\partial}_t\,{\cal G}\,,\,\pi_{10} \tilde{\boldsymbol{l}}\,{\cal G}\right)+ \frac{\omega_{\rm o}}{\kappa}\left(\pi_{01}\,\boldsymbol{\partial}_t\,{\cal G}\,,\,\pi_{01}\tilde{\boldsymbol{m}}\,{\cal G}\right)\right).
    \end{align}
    where the effective products are
     \begin{equation}
            \tilde{\boldsymbol{l}}+\tilde{\boldsymbol{m}}=\boldsymbol{Q}_{\rm c}+\boldsymbol{Q}_{\rm o}+
              \boldsymbol{\pi}_{\lambda_{\rm o},\lambda_{\rm c}}(\boldsymbol{\delta m}'+\boldsymbol{\delta l}')\frac{1}{\boldsymbol{1}_{\mathcal{SH}_{\rm c}\otimes'\mathcal{SCH}_{\rm o}}+\boldsymbol{h}_{ \lambda_{\rm o},\lambda_{\rm c}}\left(\boldsymbol{\delta m }'+\boldsymbol{\delta l  }'+\kappa \boldsymbol{U}'_{\rm o}+\kappa^{2}\boldsymbol{U}'_{\rm c}\right)}\boldsymbol{\iota}_{\lambda_{\rm o},\lambda_{\rm c}}\label{openclosedstubbedproducts}
        \end{equation}
       and we have introduced 
         \begin{align}
             &\boldsymbol{\pi}_{\lambda_{\rm o},\lambda_{\rm c}}\coloneqq \boldsymbol{\pi}\boldsymbol{O}^{-1}_{\lambda_{\rm o},\lambda_{\rm c}},\\
             &\boldsymbol{\iota}_{\lambda_{\rm o},\lambda_{\rm c}}\coloneqq \boldsymbol{O}_{\lambda_{\rm o},\lambda_{\rm c}}\boldsymbol{\iota},\\
             &\boldsymbol{h}_{ \lambda_{\rm o},\lambda_{\rm c}}\coloneqq \boldsymbol{O}_{\lambda_{\rm o},\lambda_{\rm c}}\boldsymbol{h}\boldsymbol{O}^{-1}_{\lambda_{\rm o},\lambda_{\rm c}}.
         \end{align}
%%%%%%%%%%%%%%%%%%%%%%%%%%%%%%%%
\section{Equivalence of amplitudes}\label{sec:equiv}
%%%%%%%%%%%%%%%%%%%%%%%%%%%%%%%%
  In this section we will show that the on-shell amplitudes for a theory with and without stubs are equivalent at the quantum level. For the sake of clarity, we will consider `quantum' $A_{\infty}$ theories (of the kind analyzed in section \ref{sec:cubic}), but the proof can be straightforwardly generalized to quantum $L_{\infty}$ theories and to open-closed SFT. \\ Our proof will rely on showing that the two theories give rise to the same `almost' minimal model, namely the effective action for fields in the image of $P$,  properly containing the cohomology. Then amplitudes will trivially be the same, by further restricting the external legs to the cohomology. %\footnote{The equivalence of the two almost minimal models is at the end of the day a consequence of the fact that we use the same $L$ for the stub operator $e^{-\lambda L}$ and for the propagator $\frac{b}{L}$.}\\
 % that can be written as the product of a canonical inclusion and a canonical projection, $P=\iota_0 \pi_0$\\ 
  Let's start with the original theory without stubs (trivially extended by the free auxiliary field $\chi$, which certainly does not change the amplitudes). In this case the minimal model effective products (the amplitudes) are given by  the following perturbed SDR
    \begin{align}
       \circlearrowright (-\boldsymbol{h}'_{0})\left(\mathcal{TH}', \boldsymbol{m_1}'\right)&\xtofrom[\text{$\boldsymbol{\iota}'_{0}$}]{\text{$\boldsymbol{\pi}'_{0}$}}\left(\mathcal{T}P\mathcal{H},\boldsymbol{\pi}_0 \boldsymbol{Q} \boldsymbol{\iota}_0\right)\0\\
       \downarrow\quad\quad\quad&\quad\quad\quad\quad\quad\downarrow\\
          \circlearrowright (-\hat{\boldsymbol{h}}'_{0})\left(\mathcal{TH}', \boldsymbol{m}'+\boldsymbol{U}'\right)&\xtofrom[\text{$\hat{\boldsymbol{\iota}}'_{0}$}]{\text{$\hat{\boldsymbol{\pi}}'_{0}$}}\left(\mathcal{T}P\mathcal{H}, \hat{\boldsymbol{m}}+\boldsymbol{U}_{\pi_{0}}\right),\0
    \end{align}
    where the (un-perturbed) projectors are defined as follows
    \begin{align}
        &\boldsymbol{\pi}'_{0}\coloneqq \boldsymbol{\pi}_0\boldsymbol{\pi}:\mathcal{TH}'\to \mathcal{T}P\mathcal{H}, \\
        &\boldsymbol{\iota}'_{0}\coloneqq \boldsymbol{\iota}\boldsymbol{\iota}_0:\mathcal{T}P\mathcal{H}\to \mathcal{TH}'
    \end{align}
    and the propagator $\boldsymbol{h}'_{0}$ is the co-algebraic uplifting \eqref{propcoalgebre} of $h'_{0}$ which acts on an element of $\mathcal{H}'$ as
    \begin{equation}
        h'_{0}\coloneqq\begin{pmatrix}
            h_0 & 0\\ 0&\frac{b}{L}
        \end{pmatrix}\coloneqq\begin{pmatrix}
            \frac{b}{L}\bar{P} & 0\\ 0&\frac{b}{L}
        \end{pmatrix}.
    \end{equation}
   Notice that  $\bar{P}$ is not needed in the second slot because there are no states belonging to ${\rm ker}(L)$.\\
   Using the homotopy transfer, the effective products read
   \begin{equation}
       \hat{\boldsymbol{m}}=\boldsymbol{\pi}'_{0}\boldsymbol{m}'\frac{1}{\boldsymbol{1}_{\mathcal{TH}'}+\boldsymbol{h}'_{0}(\boldsymbol{\delta m}'+\boldsymbol{U}')}\boldsymbol{\iota}'_{0}.\label{directtransfer}
   \end{equation}
  Let's now discuss the amplitudes of the stubbed theory, defined by the homotopy transfer \eqref{SDRforstubp}. To obtain the almost minimal model in this scenario, we must apply a further integration-out. Therefore, computing amplitudes for the stubbed theory corresponds to applying two consecutive homotopy transfers to the starting theory (with the non-propagating free field $\chi$), without stubs
      \begin{eqnarray}
       \circlearrowright (-\boldsymbol{h}_{\lambda})\left(\mathcal{TH}', \boldsymbol{m_1}'\right)\xtofrom[\text{$\boldsymbol{\iota}_{\lambda}$}]{\text{$\boldsymbol{\pi}_{\lambda}$}} &\circlearrowright (-\boldsymbol{h}_{0})\left(\mathcal{TH}, \boldsymbol{Q}\right)&\!   \xtofrom[\text{$\boldsymbol{\iota}_{0}$}]{\text{$\boldsymbol{\pi}_{0}$}}\left(\mathcal{T}P\mathcal{H}, \boldsymbol{\pi}_0 \boldsymbol{Q} \boldsymbol{\iota}_0\right)\0\\
       \downarrow&&\downarrow\\
          \circlearrowright (-\tilde{\boldsymbol{h}}_{\lambda})\left(\mathcal{TH}', \boldsymbol{m}'+\boldsymbol{U}'\right)\xtofrom[\text{$\tilde{\boldsymbol{\iota}}_{\lambda}$}]{\text{$\tilde{\boldsymbol{\pi}}_{\lambda}$}} &\circlearrowright (-\hat{\boldsymbol{h}}_{0})\left(\mathcal{TH}, \tilde{\boldsymbol{m}}+\boldsymbol{U}\right)&\!   \xtofrom[\text{$\hat{\boldsymbol{\iota}}_{0}$}]{\text{$\hat{\boldsymbol{\pi}}_{0}$}}\left(\mathcal{T}P\mathcal{H}, \hat{\boldsymbol{m}}_{\lambda}+\boldsymbol{U}_{\pi_{0}}\right)\0
            \end{eqnarray}
%The two almost minimal models are the same iff $\hat{\boldsymbol{m}}_{\lambda}= \hat{\boldsymbol{m}}$.
   This is an occurrence of the {\it horizontal composition} described in detail in \cite{Erbin:2020eyc}, which tells us that we can write 
       \begin{equation}
         \hat{\boldsymbol{m}}_{\lambda}=\boldsymbol{\pi}_{0}\boldsymbol{\pi}_{\lambda}\boldsymbol{m}'\frac{1}{\boldsymbol{1}_{\mathcal{TH}'}+\left(\boldsymbol{h}_{\lambda}+\boldsymbol{\iota}_{\lambda}\boldsymbol{h}_{0}\boldsymbol{\pi}_{\lambda}\right)\left(\boldsymbol{\delta m}'+\boldsymbol{U}'\right)}\boldsymbol{\iota}_{\lambda}\boldsymbol{\iota}_{0}.\label{compositetransfer}
    \end{equation}
    This expression \eqref{compositetransfer} has to be compared with \eqref{directtransfer}. 
    The two expressions apparently differ because of the $\pi$ projectors, the $\iota$ inclusions and the propagators. 
    Let's start by analyzing the apparent differences in the propagators. 
      First of all we have to understand that, as already pointed out in \cite{Erbin:2020eyc}, the composite co-algebraic propagator $\boldsymbol{h}_{\lambda}+\boldsymbol{\iota}_{\lambda}\boldsymbol{h}_{0}\boldsymbol{\pi}_{\lambda}$ does not appear to be of the typical form \eqref{propcoalgebre}, indeed its action is given by
    \begin{equation}
    \begin{split}
       &\left( \boldsymbol{h}_{\lambda}+\boldsymbol{\iota}_{\lambda}\boldsymbol{h}_{0}\boldsymbol{\pi}_{\lambda}\right) \pi_n= 
       \sum_{k=0}^{n-1}\left[1_{\mathcal{H}'}^{\otimes k}\otimes h_{\lambda}\otimes P_{\lambda}^{\otimes (n-k-1)}+P_{\lambda}^{\otimes k}\otimes \iota_{\lambda}h_{0}\pi_{\lambda}\otimes \left(\iota_{\lambda}P\pi_{\lambda}\right)^{\otimes(n-k-1)}\right]\pi_n. \label{compositeprop}
    \end{split}
       \end{equation}
       Nevertheless, this composite propagator satisfies (by construction) the Hodge-Kodaira decomposition and thus \eqref{compositeprop} is just an alternative equivalent  way to uplift the propagator $h_{\lambda}+\iota_{\lambda}h_{0}\pi_{\lambda}$ from $\mathcal{H}'$ to $\mathcal{TH}'$. Additionally, it is possible to show that after unpackaging  \eqref{compositetransfer} the effective products are given in terms of $h_{\lambda}+\iota_{\lambda}h_{0}\pi_{\lambda}$ exactly as if we had upgraded the `propagator' $h_{\lambda}+\iota_{\lambda}h_{0}\pi_{\lambda}$ according to the prescription \eqref{propcoalgebre}. Therefore, we can directly compare the propagators in \eqref{directtransfer} and \eqref{compositetransfer} as operators on $\mathcal{H}'$.\\
       Let us then compute explicitly the composite propagator using the definitions \eqref{deflambda1}, \eqref{deflambda2} and \eqref{deflambda3}
       \begin{equation}
           \begin{split}
              h_{\lambda}+\iota_{\lambda}h_{0}\pi_{\lambda}&= O_{\lambda}\left(h_0+\iota h_{0}\pi\right) O_{\lambda }^{-1}\\
              &= O_{\lambda}\begin{pmatrix}
                  \frac{b}{L}\bar{P} & 0\\ 0&\frac{b}{L}
              \end{pmatrix} O_{\lambda }^{-1}\\
              &=\begin{pmatrix}
                  \frac{b}{L}\bar{P}c_{\lambda}^{2}+\frac{b}{L}(1-c_{\lambda}^{2})&\frac{b}{L}\bar{P}c_{\lambda}\sqrt{1-c_{\lambda}^{2}}-\frac{b}{L}c_{\lambda}\sqrt{1-c_{\lambda}^{2}}\\
                  \frac{b}{L}\bar{P}c_{\lambda}\sqrt{1-c_{\lambda}^{2}}-\frac{b}{L}c_{\lambda}\sqrt{1-c_{\lambda}^{2}}& \frac{b}{L}\bar{P}(1-c_{\lambda}^{2})+\frac{b}{L}c_{\lambda}^{2}
              \end{pmatrix}\\
              &= \begin{pmatrix}
            \frac{b}{L}\bar{P} & 0\\ 0&\frac{b}{L}
        \end{pmatrix}=h'_{0},
           \end{split}
       \end{equation}
       where in the last step we used the fact that $\bar{P}(1-c_{\lambda}^{2})=(1-c_{\lambda}^{2})$ or equivalently $\bar{P}\sqrt{1-c_{\lambda}^{2}}=\sqrt{1-c_{\lambda}^{2}}$, because $\sqrt{1-c_{\lambda}^{2}}\rho\in \bar{P}\mathcal{H}\quad\forall\rho\in\mathcal{H}$, as explained in section \ref{sec:2.1}. \\
       Let us now study the composite canonical projection $\pi_0\pi_{\lambda}$ by applying it on a test state $R\in \mathcal{H}'$ 
       \begin{equation}
           \begin{split}
              \pi_0\pi_{\lambda} R&=\pi_0\pi O_{\lambda}^{-1} \begin{pmatrix}
                  \rho\\\chi
              \end{pmatrix}\\
              &=\pi_{0}\left(c_{\lambda}\rho+\sqrt{1-c_{\lambda}^{2}}\chi\right)\\
              &=\pi_{0}c_{\lambda}\rho\\&=\pi_0\rho,
           \end{split}
       \end{equation}
       where the third line is justified by the fact that $\sqrt{1-c_{\lambda}^{2}}\,\chi$ does not belong to ${\rm ker}(L)$ and thus it is annihilated by $\pi_0$. Moreover, in the last step we noticed that $\pi_{0}c_{\lambda}\rho=\pi_{0}\rho$ because $\pi_0$ returns a non-vanishing result when $\rho\in {\rm ker}(L)$ and clearly in such a situation $c_{\lambda}\rho=e^{-\lambda L}\rho=\rho$. Therefore, the composite canonical projector acts effectively as 
       \begin{equation}
           \pi_0\pi_{\lambda}=\pi_0\pi=\pi'_{0}.
       \end{equation}
       Similarly, it is possible to show that the composite canonical inclusion $\iota_{\lambda}\iota_0=\iota'_0$ and thus we can conclude that  $$\hat{\boldsymbol{m}}_{\lambda}= \hat{\boldsymbol{m}}.$$
%%%%%%%%%%%%%%%%%%%%%%%%%%%%%%%%
    \subsection{Example: open string one-loop tadpole in OC-SFT}\label{sec:3}
    %%%%%%%%%
   After the general proof we have just given we can concretely test the equivalence between amplitudes in open-closed SFT with and without stubs. To compute an amplitude in OC-SFT, as in the previous section, we apply the homological perturbation lemma to the following strong deformation retract
    \begin{equation}
       \circlearrowright (-\boldsymbol{H}_{0})\left(\mathcal{SH}_{\rm c}\otimes'\mathcal{SCH}_{\rm o}, \boldsymbol{Q}_{\rm c}+\boldsymbol{Q}_{\rm o}\right)\xtofrom[\text{$\boldsymbol{\iota}_0$}]{\text{$\boldsymbol{\pi}_0$}}\left(\mathcal{S}P_{0}^{+}\mathcal{H}_{\rm c}\otimes'\mathcal{SC}P_{0}\mathcal{H}_{\rm o}, \boldsymbol{\pi}_{0}\left(\boldsymbol{Q}_{\rm c}+\boldsymbol{Q}_{\rm o}\right)\boldsymbol{\iota}_{0}\right)
    \end{equation}
    in which the projector $\boldsymbol{\iota_{0}}\boldsymbol{\pi_{0}}$ is defined as
    \begin{equation}
        \boldsymbol{\iota_{0}}\boldsymbol{\pi_{0}}=\boldsymbol{P}_{0}^{+}\otimes' \boldsymbol{P}_{0},
    \end{equation}
    where $P_{0}$ and $P_{0}^{+}$ project onto the ${\rm ker}(L_{0})$ and ${\rm ker}(L_{0}^{+})$, respectively.
  As discussed in \cite{large-N},  the open-closed co-algebraic  propagator $\boldsymbol{H}_{0}$ can be chosen as
    \begin{equation}
          \boldsymbol{H}_{0}=\boldsymbol{h}_{0}^{+}\otimes' \boldsymbol{P}_{0}+\boldsymbol{1}_{\mathcal{SH'}_{\rm c}}\otimes'\boldsymbol{h}_{0},
      \end{equation}
    where $\boldsymbol{h}_{0}^{+}$ and $\boldsymbol{h}_{0}$ are the co-algebraic promotion of the Siegel gauge propagators
    \begin{align}
    &h_{0}^{+}=\frac{b_{0}^{+}}{L_{0}^{+}}\bar{P}_{0}^{+},\\    
    &h_{0}=\frac{b_{0}}{L_{0}}\bar{P}_{0},
    \end{align}
    see appendix A of \cite{large-N} for more details.
    Therefore, the effective products of the almost minimal model, which define the amplitudes are
    \begin{equation}
        \left(\hat{\boldsymbol{l}}+\hat{\boldsymbol{m}}\right)=\boldsymbol{\pi}_{0}\left(\boldsymbol{l}+\boldsymbol{m}\right)\frac{1}{\boldsymbol{1}_{\mathcal{SH}_{\rm c}\otimes'\mathcal{SCH}_{\rm o}}+\boldsymbol{H}_{0}\left(\boldsymbol{\delta m }+\boldsymbol{\delta l  }+\kappa \boldsymbol{U}_{\rm o}+\kappa^{2}\boldsymbol{U}_{\rm c}\right)}\boldsymbol{\iota}_{0}.\label{minimalmodelproducts}
    \end{equation}
    \\

It is particularly illuminating, as a concrete example, to specifically discuss the open string one-loop tadpole, which is simple enough and yet non-trivially tests the quantum structure of our construction.\footnote{Another simple (but classical) example we have studied is the closed string two-point amplitude on the disk, which was analyzed in detail in \cite{cosmo}. In this case  the stub deformation  creates an explicit fundamental vertex in the middle of moduli space, which was absent (with a specific choice of $SL(2,\mathbb{C})$ vertices) in \cite{cosmo}.} We denote this amplitude as $A_{0;1}^{(0,2)}(\varphi_{\rm o})$. In this particular case, the relevant homotopy relation is 
    \begin{equation}
        [\boldsymbol{m}^{(0,2)},\boldsymbol{m}^{(0,1)}]+\pi_{0,1}\boldsymbol{m}^{(0,1)}\boldsymbol{l}^{(0,1)}+[\boldsymbol{m}^{(0,2)},\boldsymbol{l}^{(0,0)}]+[\boldsymbol{m}^{(0,1)}, \boldsymbol{U}_{\rm o}]=0,
    \end{equation}
    and if we apply it to the identity we get
    \begin{equation}\label{annulushomotopyrelation}
        m_{1,0}^{(0,1)}\left(l_{0,0}^{(01)}\right)+Q_{\rm o}m_{0,0}^{(0,2)}+m_{0,2}^{(0,1)}(U_{\rm o})=0.
    \end{equation}
Using \ref{minimalmodelproducts} we can find the expression for the desired amplitudes without and with stubs
\begin{equation}
    A_{0;1}^{0,2}(\varphi_{\rm o})=\wo\left(\varphi_{\rm o},{m}_{0,0}^{0,2}\right)-\wo\left(\varphi_{\rm o},m_{1,0}^{(0,1)}\left(\frac{b_{0}^{+}}{L_{0}^{+}}\bar{P}_{0}^{+}l_{0,0}^{(0,1)}\right) \right)-\wo\left(\varphi_{\rm o},m_{0,2}^{(0,1)}\left(\boldsymbol{h}_{0}U_{\rm o}\right) \right),
\end{equation}
and
\begin{equation}
 A_{\lambda\,0;1}^{0,2}(\varphi_{\rm o})=\wo\left(\varphi_{\rm o},\tilde{m}_{0,0}^{0,2}\right)-\wo\left(\varphi_{\rm o},\tilde{m}_{1,0}^{(0,1)}\left(\frac{b_{0}^{+}}{L_{0}^{+}}\bar{P}_{0}^{+}\tilde{l}_{0,0}^{(0,1)}\right) \right)-\wo\left(\varphi_{\rm o},\tilde{m}_{0,2}^{(0,1)}\left(\boldsymbol{h}_{0}U_{\rm o}\right) \right).
\end{equation}
As before, let us proceed by extracting the vertices with stubs that appear in the previous expression from equation \ref{openclosedstubbedproducts} which we list below
\begin{align}
    &\tilde{l}_{0,0}^{(0,1)}= c_{\lambda_{\rm c}}^{({\rm c})}l_{0,0}^{(0,1)},\\
    &\tilde{m}_{1,0}^{(0,1)}(\cdot)= c_{\lambda_{\rm c}}^{({\rm c})}m_{1,0}^{(0,1)}( c_{\lambda_{\rm o}}^{({\rm o})}\cdot),\\
   &\tilde{m}_{0,2}^{(0,1)}(\cdot,\cdot)= c_{\lambda_{\rm o}}^{({\rm o})}m_{0,2}^{(0,1)}( c_{\lambda_{\rm o}}^{({\rm o})}\cdot, c_{\lambda_{\rm o}}^{({\rm o})}\cdot),\\
   &\tilde{m}_{0,0}^{(0,2)}= c_{\lambda_{\rm o}}^{({\rm o})}m_{0,0}^{(0,2)}- c_{\lambda_{\rm o}}^{({\rm o})}m_{1,0}^{(0,1)}\left(H_{\rm c}l_{0,0}^{(0,1)}\right)- c_{\lambda_{\rm o}}^{({\rm o})}m_{0,2}^{(0,1)}\left(\boldsymbol{H}_{\rm o}U_{\rm o}\right).
\end{align}
Therefore, the stubbed amplitude becomes
\begin{equation}
\begin{split}
      A_{\lambda\,0;1}^{0,2}(\varphi_{\rm o})&=\wo\left(\varphi_{\rm o},m_{0,0}^{(0,2)}\right)-\wo\left(\varphi_{\rm o},m_{1,0}^{(0,1)}\left(\left[H_{\rm c}+\frac{b_{0}^{+}}{L_{0}^{+}}\bar{P}_{0}^{+} c_{\lambda_{\rm c}}^{({\rm c})2}\right]l_{0,0}^{(0,1)}\right) \right)\\&\qquad-\wo\left(\varphi_{\rm o},m_{0,2}^{(0,1)}\left(\left[\boldsymbol{H}_{\rm o}+\boldsymbol{c}^{(\rm o)}_{\lambda_{\rm o}}\boldsymbol{h}_{0}\right]U_{\rm o}\right) \right)\\
    &=\wo\left(\varphi_{\rm o},\tilde{m}_{0,0}^{(0,2)}\right)-\wo\left(\varphi_{\rm o},m_{1,0}^{(0,1)}\left(\frac{b_{0}^{+}}{L_{0}^{+}}\bar{P}_{0}^{+}l_{0,0}^{(0,1)}\right) \right)\\&\qquad-\wo\left(\varphi_{\rm o},m_{0,2}^{(0,1)}\left(\left[\boldsymbol{H}_{\rm o}+\boldsymbol{c}^{(\rm o)}_{\lambda_{\rm o}}\boldsymbol{h}_{0}\right]U_{\rm o}\right) \right).
    \end{split}
\end{equation}
 Thus, following the general picture we have previously discussed, the two amplitudes are equivalent if
\begin{equation}
    \left[\boldsymbol{H}_{\rm o}+\boldsymbol{c}^{(\rm o)}_{\lambda_{\rm o}}\boldsymbol{h}_{0}\right]U_{\rm o}=\boldsymbol{h}_{0}U_{\rm o},
\end{equation}
which is simply demonstrated as
\begin{equation}
\begin{split}
    \left[\boldsymbol{H}_{\rm o}+\boldsymbol{c}^{(\rm o)}_{\lambda_{\rm o}}\boldsymbol{h}_{0}\right]U_{\rm o}&=\frac{1}{2}o^{i}\wedge H_{o}o_{i}+\boldsymbol{c}^{(\rm o)}_{\lambda_{\rm o}}\left(\frac{1}{2}o^{i}\wedge h_{o}o_{i}\right)\\
    &=\frac{1}{2}o^{i}\wedge\left[H_{\rm o}+\frac{b_{0}}{L_{0}}\bar{P}_{0} c_{\lambda_{\rm o}}^{({\rm o})2}\right]o_{i}\\
    &=\frac{1}{2}o^{i}\wedge \frac{b_{0}}{L_{0}}\bar{P}_{0}o_{i}=\boldsymbol{h}_{0}U_{\rm o},
\end{split}
\end{equation}
where we inserted the completeness relation to move operators from the basis vector on the left to the basis vectors on the right.\\ %and we used \ref{Hh}. \\
It is also instructive to provide a geometric picture of the stub deformation in this example. This is possible by explicitly constructing the OC-SFT vertices that realize this amplitude. We have done this is in appendix \ref{app4}, by explicitly solving the homotopy relation \eqref{annulushomotopyrelation}. Using these results, we can rewrite the stubbed and the un-stubbed amplitude in terms of surface states
\begin{equation}
\begin{split}
   A_{0;1}^{0,2}(\varphi_{\rm o})=&-\frac{1}{2\pi i}\int_{0}^{\infty}ds\langle\Sigma_{1,1}^{(0,1)}\vert\left(1_{\mathcal{H}_{\rm o}}\otimes b_{0}^{+}e^{-sL_{0}^{+}}\right)\left(\varphi_{\rm o} \otimes l_{0,0}^{(0,1)}\right)\\
    &+\int_{t_{\rm min}}^{t_{\rm max}}dt \langle \Sigma_{0,1}^{(0,2)}(t)\vert B_{t}(V_t)\varphi_{\rm o}\\
    &-\int_{0}^{\infty}ds\langle \Sigma_{0,3}^{(0,1)}\vert \left(1_{\mathcal{H}_{\rm o}}\otimes 1_{\mathcal{H}_{\rm o}}\otimes b_{0}e^{-sL_0}\right)\left(o^i\otimes \varphi_{\rm o} \otimes o_i\right)
\end{split}
\end{equation}
and
 \begin{equation}
 \begin{split}
     A_{\lambda \,0;1}^{0,2}(\varphi_{\rm o})=&-\frac{1}{2\pi i}\langle\Sigma_{1,1}^{(0,1)}\vert\left(1_{\mathcal{H}_{\rm o}}\otimes h_{0}^{+} c_{\lambda_{\rm c}}^{({\rm c})2}\right)\left(\varphi_{\rm o} \otimes l_{0,0}^{(0,1)}\right)\\
     &-\frac{1}{2\pi i}\langle\Sigma_{1,1}^{(0,1)}\vert\left(1_{\mathcal{H}_{\rm o}}\otimes H_{\rm c}\right)\left(\varphi_{\rm o} \otimes l_{0,0}^{(0,1)}\right)\\
    &+\int_{t_{\rm min}}^{t_{\rm max}}dt \langle \Sigma_{0,1}^{(0,2)}(t)\vert B_{t}(V_t)\varphi_{\rm o}\\
    &-\langle \Sigma_{0,3}^{(0,1)}\vert \left(1_{\mathcal{H}_{\rm o}}\otimes 1_{\mathcal{H}_{\rm o}}\otimes h_{0} c_{\lambda_{\rm o}}^{({\rm o})2}\right)\left(o^i\otimes \varphi_{\rm o} \otimes o_i\right)\\
    &-\langle \Sigma_{0,3}^{(0,1)}\vert \left(1_{\mathcal{H}_{\rm o}}\otimes 1_{\mathcal{H}_{\rm o}}\otimes H_{\rm o}\right)\left(o^i\otimes \varphi_{\rm o} \otimes o_i\right).\\
 \end{split}
 \end{equation}
 Now, using the standard Schwinger representation for the  propagators, we get
 \begin{equation}
     \begin{split}
          A_{\lambda \,0;1}^{0,2}(\varphi_{\rm o})= &-\frac{1}{2\pi i}\int_{2\lambda_{\rm c}}^{\infty}ds\langle\Sigma_{1,1}^{(0,1)}\vert\left(1_{\mathcal{H}_{\rm o}}\otimes b_{0}^{+}e^{-sL_{0}^{+}}\right)\left(\varphi_{\rm o} \otimes l_{0,0}^{(0,1)}\right)\\
    &-\frac{1}{2\pi i}\int_{0}^{2\lambda_{\rm c}}ds\langle\Sigma_{1,1}^{(0,1)}\vert\left(1_{\mathcal{H}_{\rm o}}\otimes b_{0}^{+}e^{-sL_{0}^{+}}\right)\left(\varphi_{\rm o} \otimes l_{0,0}^{(0,1)}\right)\\
    &+\int_{t_{\rm min}}^{t_{\rm max}}dt \langle \Sigma_{0,1}^{(0,2)}(t)\vert B_{t}(V_t)\varphi_{\rm o}\\
    &-\int_{0}^{2\lambda_{\rm o}}ds\langle \Sigma_{0,3}^{(0,1)}\vert \left(1_{\mathcal{H}_{\rm o}}\otimes 1_{\mathcal{H}_{\rm o}}\otimes b_{0}e^{-sL_0}\right)\left(o^i\otimes \varphi_{\rm o} \otimes o_i\right)\\
   & -\int_{2\lambda_{\rm o}}^{\infty}ds\langle \Sigma_{0,3}^{(0,1)}\vert \left(1_{\mathcal{H}_{\rm o}}\otimes 1_{\mathcal{H}_{\rm o}}\otimes b_{0}e^{-sL_0}\right)\left(o^i\otimes \varphi_{\rm o} \otimes o_i\right),
     \end{split}\label{brughu}
 \end{equation}
 that is clearly equivalent to the amplitude without stubs. Furthermore, we can easily interpret the reorganization of the terms which build-up the amplitude after having introduced the stubs. In particular, we have a new fundamental vertex given by the sum of the second, third and fourth line of the above expression \eqref{brughu} and thus now the fundamental vertex involves integration over a bigger region of moduli space. Conversely, the portion of moduli space covered by Feynman diagrams is reduced. Indeed the amplitude can be rewritten as
 \begin{equation}
     \begin{split}
         A_{\lambda \,0;1}^{0,2}(\varphi_{\rm o})= & -\int_{2\lambda_{\rm o}}^{\infty}ds\langle \Sigma_{0,3}^{(0,1)}\vert \left(1_{\mathcal{H}_{\rm o}}\otimes 1_{\mathcal{H}_{\rm o}}\otimes b_{0}e^{-sL_0}\right)\left(o^i\otimes \varphi_{\rm o} \otimes o_i\right)\\
    &+\int_{t_{\rm c}(e^{-\lambda_{\rm c}})}^{t_{\rm o}(e^{-\lambda_{\rm o}})}dt \langle \Sigma_{0,1}^{(0,2)}(t)\vert B_{t}(V_t)\varphi_{\rm o}\\
  &-\frac{1}{2\pi i}\int_{2\lambda_{\rm c}}^{\infty}ds\langle\Sigma_{1,1}^{(0,1)}\vert\left(1_{\mathcal{H}_{\rm o}}\otimes b_{0}^{+}e^{-sL_{0}^{+}}\right)\left(\varphi_{\rm o} \otimes l_{0,0}^{(0,1)}\right),
     \end{split}
 \end{equation}
 where $t_{\rm o}(q)$ and $t_{\rm c}(q)$ are defined in \eqref{topen} and \eqref{tclosed}. In particular the local coordinate map associated with the surface state  $\langle \Sigma_{0,1}^{(0,2)}(t)\vert$ is defined as follows
 \begin{equation}
 \hat{F}_{t}(w)=
 \begin{cases}
    &G_{\rm c}(w)\qquad \qquad t<t_{\rm min}\\
    &F_{t}(w)\qquad\qquad t_{\rm min}<t<t_{\rm max }\\
    &G_{\rm o}(w) \qquad\qquad t>t_{\rm max}
 \end{cases}
 \end{equation}
where the maps $G_{\rm c},\,  G_{\rm c},\,F_{t}$ are defined in appendix \ref{app4}.
    \section{Conclusions}\label{sec:6}
%%%%%%%%%%%%%%%%%%%%%%%%%%%%%%%%
%%%%%%%%%%%%%%%%%%%%%%%%%%%%%%%%    
After having clarified the quantum algebraic structure of the stub deformation in bosonic SFT,  we would like to give some possible directions for future explorations
\begin{itemize}
\item  According to our picture, adding stubs also generates constant terms in the action, from the path integral over $\Sigma$. In this paper we have ignored these constant terms, but it is important to understand why they are there. The first of these terms is the one-loop determinant of the $\Sigma$ kinetic term. Since this is the same as the one-loop determinant of the $\chi$ kinetic term, this term has to be canceled by a counterterm in the original extended action \eqref{step1}, if we want that integrating out $\chi$ reproduces {\it exactly} the initial theory. The other vacuum terms are generated by the interactions of $\Sigma$ with $\psi$ and they start at two loops.
The reason why they are present is that the stubbed and the original action should give rise to the same vacuum energy when one reduces them to the (almost) minimal model, i.e. after integrating out the whole $\bar P H$. This is also expected from (quantum) background independence and the field-redefinition invariance of the path-integral measure \cite{Sen:1993kb}.\footnote{We thank A. H. F\i{}rat for a discussion on this point.}
% To understand why they are present, we have to consider that, in general, the quantum theory (stubbed or not)  will have tadpoles starting at one-loop.  In a 1-PI repackaging of the quantum BV master action the tadpoles have to be canceled by shifting the string field \cite{Sen:2014dqa}. This shift will generate  constant terms in the action which will start with a two-loop vacuum diagram. The portion of the total moduli space that will be covered by the generated vacuum diagram after the shift will clearly change by adding stubs. However if the stubbed theory is really equivalent to the original one, the total vacuum energy after the vacuum shift should be the same. 
 We expect this to be true only after accounting for the extra constant terms that are generated by integrating-out $\Sigma$ (which, after discarding the one-loop determinant, also start at two loops).   It would be worth to systematically check this, which  we expect to be a generalization of the arguments presented in section \ref{sec:equiv}. On the technical side, we would like to understand how the vacuum bubbles that are generated by the path integral can be formulated using the same co-algebraic homotopy transfer that generates amplitudes with external legs.

\item It could be a useful exercise to repeat our analysis in the context of superstring field theory. In particular our addition of a free  non-propagating field seems to share some similarity with Sen's method of introducing the R sector via a free (but propagating) field. It would be interesting to see these two structures at work simultaneously.
\item Although the stub deformation is a field redefinition, it is not invertible in a strict sense. This is geometrically clear since the inverse stub operator $e^{+\lambda L_0}$ has the interpretation of `removing' a strip of worldsheet, which may not be a well-defined operation in general. Nonetheless there could be a well-defined domain of definition for inverse stubs: for sure we expect to be possible to remove stubs from a theory where we have just added them, if the length of the inverse stub is smaller than the length of the added ones. The use of inverse stubs in the kinetic operator of closed SFT could be a way of regulating  the one-loop determinant, which produces the torus vacuum amplitude an infinite number of times on infinite equivalent fundamental domains  \cite{Zwiebach:1992ie}. Similarly, the vacuum energy which is generated in open-closed SFT after open string integration-out and closed string vacuum-shift as discussed in \cite{large-N} seems to overcount the annulus moduli space and inverse open-string stubs could play a role in fixing this problem. More generally, the introduction of quantum stubs could help in approaching the elusive nature of the cosmological constant in SFT.

\item Open-Closed SFT has a natural `interpolating' form  \cite{Zwiebach:1992bw} where, at the extreme of the interpolation, the moduli space is mostly covered by open or closed string propagators \cite{Firat:2023gfn, Cho:2019anu}. It would be  interesting to relate these general interpolations with open and closed stub deformations. Again, this seems to require the possibility of increasing the closed string stubs and reducing at the same time the open string stubs (and viceversa). 
\end{itemize}
In general, we hope the technical results in this paper will be useful for better understanding the quantum structure of string (field) theory.

%%%%%%%%%%%%%%%%%%%%%%%%%%%%%%%%
%%%%%%%%%%%%%%%%%%%%%%%%%%%%%%%%   
    \section*{Acknowledgments}
%%%%%%%%%%%%%%%%%%%%%%%%%%%%%%%%
%%%%%%%%%%%%%%%%%%%%%%%%%%%%%%%%
 We thank H. Erbin, A. H. F\i{}rat  and J. Vo\v{s}mera for useful discussions.
  CM thanks the Kavli Institute for Theoretical Physics (KITP), Santa Barbara, for hospitality during the workshop {\it ``What is String Theory? Weaving Perspectives Together"}, where part of this work was completed.  This research was supported in part by grant NSF PHY-2309135 to the KITP.  The work of CM  and AR  is partially supported by the MUR PRIN contract 2020KR4KN2 “String Theory as a bridge between Gauge Theories and Quantum Gravity”.  The work of CM, AR and BV is also partially supported by the INFN project STeFI “String Theory and Fundamental Interactions”.
    \appendix

\section{Path integral and co-algebraic homotopy transfer}\label{app1}

In this section we will prove explicitly that the homotopy transfer at the level of co-algebras  (that we heavily use in the main text) performs the path integral over the degrees of freedom which we want to integrate out. This amounts in showing that the transferred products given by the homotopy transfer precisely reproduce the effective vertices  obtained by standard QFT manipulations. The equivalence of the homotopy transfer to the path integral has been discussed in detail in \cite{DJP}, at the level of Batalin-Vilkovisky formalism. Here we will translate this language to the `dual' co-algebra language, on which our SFT formalism is  based. In order to be as self-contained as possible we will start from the very beginning. 

Let us then consider a generic SFT action $S[\Phi]$ that satisfies the BV quantum master equation, where $\Phi\in \mathcal{H}$ can be decomposed as $\Phi=\phi^{i}f_{i}$ where the $f_{i}$'s are the basis vectors of a graded vector space ${\cal H}$, the space of string fields.\\
Let's furthermore distinguish between the free kinetic term and the interactions
\begin{equation}
    S[\Phi]=S_{\rm free}[\Phi]+S_{\rm int}[\Phi],
\end{equation}
and assume  a kinetic term of the form
\begin{equation}
    S_{\rm free}[\Phi]=\frac{1}{2}\omega\left(\Phi,Q \Phi\right)=\frac{1}{2}\omega(f_{i},Qf_{j})\phi^{i}\phi^{j}=\frac{1}{2}Q_{ij}\phi^i\phi^j,
\end{equation}
where $\omega$ is a symplectic form on ${\cal H}$,   $Q$ is a BRST operator (an odd nilpotent linear operator on ${\cal H}$) and $Q_{ij}$ is defined as
\begin{equation}
    Q_{ij}\coloneqq\frac{1}{2}\omega(f_{i},Qf_{j}),
\end{equation}
with
\begin{equation}
    Q_{ij}=(-)^{f_{j}}Q_{ji},
\end{equation}
where $(-)^{v}$ accounts for the degree of the vector $v$.
Let us then introduce a projector $P$ and define $\varphi=P\Phi$ and $R=\bar{P}\Phi$. Our aim is to find the Wilsonian effective action $\tilde{S}[\varphi]$ by integrating out $R$ in the BV path integral, namely
\begin{equation}
    e^{-\tilde{S}[\phi]}=\int_{L} \mathcal{D}R\, e^{-S[\varphi+R]},
\end{equation}
where $L$ is a Lagrangian submanifold  corresponding to some gauge fixing. To better understand this point let us introduce a `propagator' $h$ giving rise to  the Hodge-Kodaira decomposition
\begin{align}
P+hQ+Qh=1,
\end{align}
together with $h^2=Ph=0$, so that $(P, hQ, Qh)$ are orthogonal projectors. 
 This partitions the total space ${\cal H}$ into
 \begin{equation}
    \mathcal{H}=A\oplus B\oplus C,
\end{equation} 
where we have defined 
\begin{align}
    &A\coloneqq P\mathcal{H} ={\rm Span}\{a_{i}\},\\
    &B \coloneqq Qh\mathcal{H}= {\rm Span}\{b_{i}\},\\
    &C\coloneqq hQ\mathcal{H}={\rm Span}\{c_{i}\}.
\end{align}
We also define dual basis vectors by
\begin{equation}
    \omega(a^i,a_j)=\omega(b^i,c_j)=\omega(c^i,b_j)=\delta^i_j,
\end{equation}
from which we can write down the completeness relation 
\begin{equation}
    1_{\mathcal{H}}=a_i\omega(a^i,\cdot)+b_i\omega(c^i,\cdot)+c_i\omega(b^i,\cdot).
\end{equation}
Under this decomposition the symplectic form is given by
\begin{equation}
   \omega_{ij}= \begin{pmatrix}
        \omega_{ij}^{A} & 0 & 0\\
        0 & 0 & \omega_{ij}^{BC}\\
        0 & -\omega_{ij}^{BC} & 0
    \end{pmatrix},
\end{equation}
where we defined 
\begin{align}
    &\omega_{ij}^{A}\coloneqq \omega(a_i,a_j),\\
    &\omega_{ij}^{BC}\coloneqq \omega(b_i,c_j).
\end{align}
Therefore, the string field can be written as
\begin{align}
     \Phi=\varphi+R=\alpha^{i}a_{i}+\beta^{i}b_{i}+\gamma^{i}c_{i}.
\end{align}
The distinction between fields and anti-fields in $(1-P){\cal H}\equiv\bar P{\cal H}$ depends on a choice of lagrangian submanifold (a maximal subspace of $\bar P{\cal H}$ where $\omega$ vanishes). In particular it is possible to arrange all anti-fields of $\bar{P}\mathcal{H}$ into $B$. Consequently, the Lagrangian submanifold that defines the gauge fixing will be given by the condition $\beta^{i}=0$ or equivalently $h R=h\Phi=0$. Thus, the path integral becomes 
\begin{equation}
    \int_{L}\mathcal{D}R=\int_{L}\prod_{ij}d\beta^{i}d\gamma^{j}=\int \prod_{ij}d\beta^{i}d\gamma^{j}\delta(\beta^{i}).
\end{equation}
By defining $R'=\gamma^{i}c_{i}$, the effective action reads
\begin{equation}
    e^{-\tilde{S}[\phi]}=\int \prod_{i}d\gamma^{i} e^{-S[\varphi+R']}.
\end{equation}
To perform the above integral we introduce a source term $J_{i}\gamma^{i}$, with $d(J_{i})=d(\gamma^{i})$, and we isolate the kinetic term of the $C$ sector 
\begin{equation}
\begin{split}
    e^{-\tilde{S}[\phi]}&=  \int \prod_{i}d\gamma^{i} e^{-\frac{1}{2}Q_{ij}\gamma^{i}\gamma^{j}-S_{\rm free}[\varphi]-S_{\rm int}[\varphi+R']+J_i\gamma^i}\bigr\vert_{J_{i}=0}\\
    &=e^{-S_{\rm free}[\varphi]-S_{\rm int}[\varphi+\p_{J_{i}}c_{i}]}\int \prod_{i}d\gamma^{i}e^{-\frac{1}{2}Q_{ij}\gamma^{i}\gamma^{j}+J_i\gamma^i}\bigr\vert_{J_{i}=0}.
    \end{split}
\end{equation}
The problem then reduces to the computation of a Gaussian integral. Assuming positivity of $Q_{ij}$ (also by  deforming possible `tachyonic'  contours in the complex plane so that, for grassmann  even variables, the integral is done on a steepest path), the integral converges when $Q_{ij}$ is non-degenerate. This condition is satisfied if ${\rm ker}(Q)$ is contained in ${\rm Im}(P)$. Under this assumption we can readily compute the path integral by completing the square. To do this, we define
\begin{equation}
    h^{ij}\coloneqq (-)^{b_i}\omega(b^{i},hb^{j}),
\end{equation}
that satisfies the following properties
\begin{align}
    &h^{ij}=(-)^{b^i}h^{ji},\\
    &Q_{ji}h^{jk}=\delta_{i}^{k},\\
    &Q_{ij}h^{jk}=(-)^{c_{i}}\delta_{i}^{k}.
\end{align}
Then we can easily verify
\begin{equation}
    -\frac{1}{2}Q_{ij}\gamma^{i}\gamma^{j}+J_i\gamma^i=-\frac{1}{2}\left[Q_{ij}(\gamma^{i}-h^{ik}J_{k})(\gamma^{j}-h^{jl}J_{l})\right]+\frac{1}{2}h^{ij}J_{i}J_{j},
\end{equation}
and thus, by making the change of variables $x^{i}=\gamma^{i}-h^{ik}J_{k}$ we get
\begin{equation}
\begin{split}
    e^{-\tilde{S}[\phi]}&=e^{-S_{\rm free}[\varphi]-S_{\rm int}[\varphi+\p_{J_{i}}c_{i}]}e^{\frac{1}{2}h^{ij}J_iJ_j}\int \prod_{i}dx^{i}e^{-\frac{1}{2}Q_{ij}x^{i}x^{j}}\bigr\vert_{J_{i}=0}\\
    &={\rm sdet}(Q_{ij})^{-\frac{1}{2}}e^{-S_{\rm free}[\varphi]-S_{\rm int}[\varphi+\p_{J_{i}}c_{i}]}e^{\frac{1}{2}h^{ij}J_i J_j}\bigr\vert_{J_{i}=0},
\end{split}
    \end{equation}
%The constant term arising from the super determinant in the above expression is related to the exponential of the annulus partition function, as described by A. Sen in \cite{Sen:2021qdk}. The sole distinction lies in the fact that, in our scenario, the indices exclusively run on the basis vectors within sector $C$. Particularly, the constant ${\cal A}$ introduced by A. Sen in this context is given by
%\begin{equation}
%    e^{-{\cal A}}={\rm sdet}(Q_{ij})^{-\frac{1}{2}}\longrightarrow {\cal A}=\frac{1}{2}\log\left(\frac{{\rm det}(Q_{I_f,J_f})}{{\rm det}(Q_{I_b,J_b})}\right),
%\end{equation}
%where the indices $I_f$ runs over the fermionic degrees of freedom whereas $I_b$ over the bosonic ones. 
Therefore, the effective action can be written as
\begin{equation}
    \tilde{S}[\varphi]={\cal A}-\log\left(e^{-S_{\rm free}[\varphi]-S_{\rm int}[\varphi+\p_{J_{i}}c_{i}]}e^{\frac{1}{2}h^{ij}J_i J_j}\right)\bigr\vert_{J_{i}=0}.
\end{equation}
Additionally, we can introduce a further source term $J_{i}\gamma^{i}$, getting
\begin{equation}
    \begin{split}
      \tilde{S}[\varphi]&={\cal A}-\log\left(e^{-S_{\rm free}[\varphi]-S_{\rm int}[\varphi+\p_{J_{i}}c_{i}]}e^{\frac{1}{2}h^{ij}J_i J_j}e^{J_{i}\gamma^{i}}\right)\bigr\vert_{J_{i}=0,\gamma^{i}=0}   \\
      &={\cal A}-\log\left(e^{-S_{\rm free}[\varphi]-S_{\rm int}[\varphi+\p_{J_{i}}c_{i}]}e^{\frac{1}{2}h^{ij}\p_{\gamma^{i}} \p_{\gamma^{j}}}e^{J_{i}\gamma^{i}}\right)\bigr\vert_{J_{i}=0,\gamma^{i}=0}\\
      &={\cal A}-\log\left(e^{\frac{1}{2}h^{ij}\p_{\gamma^{i}} \p_{\gamma^{j}}}e^{-S_{\rm free}[\varphi]-S_{\rm int}[\varphi+\p_{J_{i}}c_{i}]}e^{J_{i}\gamma^{i}}\right)\bigr\vert_{J_{i}=0,\gamma^{i}=0}\\
      &={\cal A}-\log\left(e^{\frac{1}{2}h^{ij}\p_{\gamma^{i}} \p_{\gamma^{j}}}e^{-S_{\rm free}[\varphi]-S_{\rm int}[\varphi+\gamma^ic_{i}]}\right)\bigr\vert_{\gamma^{i}=0}.\label{path-int}
    \end{split}
\end{equation}
Let us then introduce a canonical projector $\Pi:{\cal F(H)}\to{\cal F}(A)$, where ${\cal F(H)}$ is the functional space over ${\cal H}$ and ${\cal F}(A)$ the same over $A$. This projector is defined such that
\begin{align}
    &\Pi(\alpha^{i})=\alpha^{i},\\
    &\Pi(\beta^{i})=\Pi(\gamma^{i})=0,
\end{align}
and it is generalized on a generic functional $f\in{\cal F(H)}$ by Taylor expanding the functional in powers of $\alpha$, $\beta$ and $\gamma$. With this understanding, the condition $\gamma^{i}=0$ in the effective action can be implemented through the introduction of $\Pi$
\begin{equation}
    \tilde{S}[\varphi]={\cal A}-\log\left(\Pi e^{\frac{1}{2}h^{ij}\p_{\gamma^{i}} \p_{\gamma^{j}}}e^{-S_{\rm free}[\varphi]-S_{\rm int}[\varphi+\gamma^ic_{i}]}\right).
\end{equation}
Furthermore let us define
\begin{equation}
    \tilde{\Pi}\coloneqq \Pi e^{\frac{1}{2}h^{ij}\p_{\gamma^{i}} \p_{\gamma^{j}}},\label{Ptilde}
\end{equation}
thus obtaining
\begin{equation}
    \tilde{S}[\varphi]={\cal A}-\log\left(\tilde{\Pi} e^{-S_{\rm free}[\varphi]-S_{\rm int}[\varphi+\gamma^ic_{i}]}\right).
    \label{standard_path_integral}
\end{equation}
We introduced $\tilde{\Pi}$ to establish a connection between this standard QFT procedure and \cite{DJP}, where the authors demonstrate that $\tilde{\Pi}$ represents the transferred projector derived from the homological perturbation lemma (HPL) at the functional level, applied to the following free SDR
\begin{equation}
       \circlearrowright (-H)\left({\cal F(H)}, d\right)\xtofrom[\text{$I$}]{\text{$\Pi$}}\left({\cal F}(A), \Pi dI\right),\label{freeSDR}
       \end{equation}
       in which the starting differential and the propagator are defined as
\begin{align}
    &d\coloneqq (S_{\rm free}[\varphi+R'],\cdot)=\gamma^{i}\omega(c^{j},Qc_{i})\frac{\overrightarrow{\p}}{\p\beta^{j}}+\alpha^{i}\omega(a^{j},Qa_{i})\frac{\overrightarrow{\p}}{\p\alpha^{j}},\\
    &H\coloneqq \frac{1}{\#_{\beta+\gamma}}\beta^{i}\omega(b^{j},hb_{i})\frac{\overrightarrow{\p}}{\p\gamma^{j}},
\end{align}
where $\#_{\beta+\gamma}$ counts the number of occurrence of the variables $\beta$ and $\gamma$ in each monomial on which $H$ acts. The inclusion $I$ trivially embeds a function of $\alpha$ into the space of functions of $(\alpha, \beta, \gamma)$. It is easy to see that the propagator satisfies the Hodge-Kodaira decomposition 
\begin{equation}
    dH+Hd=1-I\Pi,
\end{equation}
Next, we consider adding a deformation given by the BV Laplacian $\delta_{1}=\Delta$
\begin{equation}
       \circlearrowright (-\tilde{H})\left({\cal F(H)}, d+\Delta\right)\xtofrom[\text{$\tilde{I}$}]{\text{$\tilde{\Pi}$}}\left({\cal F}(A), \tilde{d}\right),
\end{equation}
then by the homological perturbation lemma the transferred canonical projector will be given by
\begin{equation}
    \tilde{\Pi}=\Pi \frac{1}{1+\delta_{1}H}.\label{Ptilde2}
\end{equation}
An important result of \cite{DJP}  (section 4.1.1) is that \ref{Ptilde} and \ref{Ptilde2} are in fact the same projector, when applied to a generic functional $f\in {\cal F(H)}$, namely
\begin{equation}
    \Pi e^{\frac{1}{2}h^{ij}\p_{\gamma^{i}} \p_{\gamma^{j}}}f=\Pi \frac{1}{1+\delta_{1}H}f.
\end{equation}
 \\ 
This result allows us to draw a connection between the more standard path integral approach and the application of the HPL to the free strong deformation retract \ref{freeSDR}. Moreover it ensures that $\tilde{\Pi}$ defined as \ref{Ptilde} satisfies all the properties required by the HPL.\\
%The significance of establishing this equivalence between the "conventional" approach to the path integral and %the approach through the HPL lies in ensuring that $\tilde{\Pi}$ has all the properties outlined in the theorem, %including for instance the chain relations \ref{chainrel}. This enables us to deduce the explicit expression of %the effective products, thereby demonstrating, as desired, their alignment with those determined by the homotopy %transfer at the level of Hilbert space. \\
Notice that, as it is evident from \ref{path-int}, this path integral will also compute the vacuum bubbles, which will appear as constant terms in the effective action. In particular if we set $\varphi$ to zero in the effective action \eqref{standard_path_integral} we get
\begin{equation}
    \tilde{S}[0]=\mathcal{A}-\log\left(\tilde{\Pi}e^{-S_{\rm int}[\gamma^ic_{i}]}\right),
\end{equation}
and the second term will produce the (interacting) vacuum bubbles, in addition to the one-loop determinant $\mathcal{A}$. We can isolate the bubbles by adding and subtracting them from \ref{standard_path_integral}
\begin{equation}
    \tilde{S}[\varphi]=\mathcal{A}-\log\left(\tilde{\Pi}e^{-S_{\rm int}[\gamma^ic_{i}]}\right)-\log\left(\frac{\tilde{\Pi} e^{-S_{\rm free}[\varphi]-S_{\rm int}[\varphi+\gamma^ic_{i}]}}{\tilde{\Pi}e^{-S_{\rm int}[\gamma^ic_{i}]}}\right).
\end{equation}
The last term
\begin{equation}
    \tilde{S}'[\varphi]=-\log\left(\frac{\tilde{\Pi} e^{-S_{\rm free}[\varphi]-S_{\rm int}[\varphi+\gamma^ic_{i}]}}{\tilde{\Pi}e^{-S_{\rm int}[\gamma^ic_{i}]}}\right)
\end{equation}
 is then the effective action without the vacuum bubbles, which correctly vanishes when $\varphi=0$. 
In order to extract this normalized action and concretely link it to the products defining the initial UV action, we follow again \cite{DJP}, who applied the HPL to the free SDR \ref{freeSDR}, but with a different perturbation $\delta_{2}$ defined as
\begin{equation}
    \delta_2=(S_{\rm int}[\varphi+R'],\cdot)+\Delta.
\end{equation}
This gives the following perturbed SDR
\begin{equation}
       \circlearrowright (-\tilde{H}_2)\left({\cal F(H)}, d+\delta_2\right)\xtofrom[\text{$\tilde{I}_2$}]{\text{$\tilde{\Pi}_2$}}\left({\cal F}(A), \tilde{d}_2\right)
\end{equation}
and the HPL dictates that
\begin{align}
 \tilde{\Pi}_2=\Pi \frac{1}{1+\delta_{2}H}\label{Ptilde-2}.
\end{align}
Focusing now on the transferred differential
\begin{equation}
    \tilde{d_{2}}=\Pi d I+ \Pi \delta_2\frac{1}{1+H\delta_2}I, \label{d2a}
\end{equation}
it is possible to show (section 4.2 of \cite{DJP}) that $\tilde{d}_{2}$ can be independently written as
\begin{equation}
    \tilde{d}_{2}=(\tilde{S}'[\varphi], \cdot)_{\alpha\alpha}+\Delta_{\alpha\alpha},\label{d2b}
\end{equation}
 where the subscript $(\cdot)_{\alpha\alpha}$ means that  the BV bracket and BV laplacian only contain derivatives in $\alpha$, namely
\begin{align}
   &\left(\cdot, \cdot\right)_{\alpha\alpha}=\dfrac{\overleftarrow{\partial}}{\partial \alpha^{i}}(\omega_{A})^{ij}\dfrac{\overrightarrow{\partial}}{\partial \alpha^{j}},\\
        &\Delta_{\alpha\alpha}=\frac{1}{2}(-)^{\alpha^{i}}(\omega_{A})^{ij}\dfrac{\overrightarrow{\partial}}{\partial \alpha^{i}}\dfrac{\overrightarrow{\partial}}{\partial \alpha^{j}}.
\end{align}
Notice that any possible constant part of $\tilde{S}'[\varphi]$  in \eqref{d2b} would be eliminated by the BV bracket. We can then determine the effective products of $\tilde{S}'[\varphi]$ by imposing the equality of \ref{d2a} and \ref{d2b}
\begin{equation}
    \begin{split}
  (\tilde{S}'[\varphi], \cdot)_{\alpha\alpha}+\Delta_{\alpha\alpha}&=\Pi d I+ \Pi \delta_2\frac{1}{1+H\delta_2}I      \\
  &=\Pi d I+\Pi \delta_2I- \Pi \delta_2\frac{1}{1+H\delta_2}H\delta_2 I\\
  &=\Pi (S[\varphi+R],\cdot)I+\Pi\Delta I- \Pi \delta_2\frac{1}{1+H\delta_2}H\delta_2 I\\
  &=\Pi (S[\Phi],\cdot)I+\Delta_{\alpha\alpha}-\Pi \frac{1}{1+\delta_{2}H}\delta_2 H\delta_2 I,
    \end{split}
\end{equation}
where we named $\Phi=\varphi+R$. The above expression allows us to establish a relation between the BV bracket of the effective theory and the starting action, that can be further simplified as
\begin{equation}
    (\tilde{S}'[\varphi], \cdot)_{\alpha\alpha}=\Pi (S[\Phi],\cdot)I-\Pi \frac{1}{1+\delta_{2}H}\delta_2 H\delta_2 I .\label{effectivebracket}
\end{equation}
By applying this expression to $\varphi$ we will now show that we precisely recover the effective vertices as obtained by the HPL at the level of the Hilbert space $\mathcal{H}$ and the co-algebraic description used in the main text.\\ 
In order to find the effective products we assume that the starting action and the effective action can be written in WZW form as follows:
\begin{equation}
    S[\Phi]=\int_{0}^{1}dt\,\omega(\dot{\Phi},\pi_{1}\boldsymbol{l}e^{\wedge\Phi}),
\end{equation}
\begin{equation}
    \tilde{S}'[\varphi]=\int_{0}^{1}dt\,\omega(\dot{\varphi},\pi_{1}\tilde{\boldsymbol{l}}e^{\wedge\varphi}),
\end{equation}
where both $\boldsymbol{l}$ and $\tilde{\boldsymbol{l}}$ are assumed to be cyclic. 
Then, if we apply \ref{effectivebracket} to $\varphi$ we get
%\begin{equation}
%    \pi_{1}\tilde{\boldsymbol{l}}e^{\wedge \varphi}=\pi_{1}\boldsymbol{\pi}\boldsymbol{l}\boldsymbol{\iota}e^{\wedge \varphi}-\Pi \frac{1}{1+\delta_{2}H}\delta_2 H\delta_2 I\varphi.\label{stepltilde}
%\end{equation}
\begin{equation}
   (\tilde{S}'[\varphi], \varphi)_{\alpha\alpha}=   \Pi(S[\Phi],I\varphi)-\Pi \frac{1}{1+\delta_{2}H}\delta_2 H\delta_2 I\varphi.\label{stepltilde}
\end{equation}
The left hand side can be obtained by a direct computation
\begin{equation}
    \begin{split}
    (\tilde{S}'[\varphi], \varphi)_{\alpha\alpha}=\tilde{S}'[\varphi]\frac{\leftpartial}{\partial\alpha^i}\omega_A^{ij}\frac{\rightpartial}{\partial\alpha^j}(\alpha^k a_k)=(-)^{\alpha^{i}}\frac{\rightpartial}{\partial\alpha^i}\tilde{S}'[\varphi]\omega_A^{ij}a_j.
    \end{split}
    \label{LHS}
\end{equation}
The derivative acts on the action as:
\begin{equation}
    \begin{split}
    \frac{\rightpartial}{\partial\alpha^j}\tilde{S}'[\varphi]
    &=\frac{\rightpartial}{\partial\alpha^j}\int_0^1 dt\,\omega(\pi_1\boldsymbol{\partial}_t e^{\wedge \varphi},\pi_1\tilde{\boldsymbol{l}} e^{\wedge \varphi})\\
    &=(-)^{\alpha^i}\int_0^1 dt\,\omega(\pi_1\boldsymbol{\partial}_t\boldsymbol{\partial}_{\alpha_i} e^{\wedge \varphi},\pi_1\tilde{\boldsymbol{l}} e^{\wedge \varphi})+\int_0^1 dt\,\omega(\pi_1\boldsymbol{\partial}_t e^{\wedge \varphi},\pi_1\tilde{\boldsymbol{l}}\boldsymbol{\partial}_{\alpha_i} e^{\wedge \varphi})
    \end{split}
    \label{partial_S_step}
\end{equation}
where $\boldsymbol{\partial}_{\alpha_i}$ should be considered as a coderivation on the same footing as $\boldsymbol{\partial}_t$. Then using the cyclic property of $\tilde{\boldsymbol{l}}$ we can rewrite the second term as:
\begin{equation}
    \omega(\pi_1\boldsymbol{\partial}_t e^{\wedge \varphi},\pi_1\tilde{\boldsymbol{l}}\boldsymbol{\partial}_{\alpha_i} e^{\wedge \varphi})=
    (-)^{\alpha_i}\omega(\pi_1\boldsymbol{\partial}_{\alpha^i}e^{\wedge \varphi},\pi_1\boldsymbol{\partial}_{t}\tilde{\boldsymbol{l}} e^{\wedge \varphi}),
\end{equation}
so that the two terms in \ref{partial_S_step} can be resummed using the Leibniz rule:
\begin{equation}
    \begin{split}
    \frac{\rightpartial}{\partial\alpha^j}\tilde{S}'[\varphi]
    &=(-)^{\alpha^i}\int_0^1dt\,\partial_t\,\omega(\pi_1\boldsymbol{\partial}_{\alpha^i} e^{\wedge \varphi},\pi_1\tilde{\boldsymbol{l}} e^{\wedge \varphi})\\
    &=(-)^{\alpha^i}\omega(\pi_1\boldsymbol{\partial}_{\alpha^i} e^{\wedge \varphi},\pi_1\tilde{\boldsymbol{l}} e^{\wedge \varphi})=(-)^{\alpha^i}\omega(a_i,\pi_1\tilde{\boldsymbol{l}} e^{\wedge \varphi}).
    \end{split}
\end{equation}
Plugging this back into \ref{LHS} we can use the completeness relation to build
\begin{equation}
    (\tilde{S}'[\varphi], \varphi)_{\alpha\alpha}=\omega(a_i,\pi_1\tilde{\boldsymbol{l}} e^{\wedge \varphi})\omega_A^{ij}a_j=\pi_1\tilde{\boldsymbol{l}} e^{\wedge \varphi}.
    \label{tildeSvarphi}
\end{equation}
Let's now turn to the first term on the right hand side of \ref{stepltilde}. Since the $\beta,\gamma$ part of the Poisson bracket vanishes when acting on $\varphi$, we are left with
\begin{equation}
    \Pi(S[\Phi],I\varphi)=\Pi\left[S[\Phi]\frac{\leftpartial}{\partial\alpha^i}\omega_A^{ij}\frac{\rightpartial}{\partial\alpha^j}I\varphi\right]
\end{equation}
where $I\varphi$ must be understood now as an element of $\cal F(H)$. Following analogous steps as in \ref{tildeSvarphi} we get
\begin{equation}
    \Pi(S[\Phi],I\varphi)=\Pi\left[\omega(a_i,\pi_1\boldsymbol{l}e^{\wedge \Phi})\right]\omega_A^{ij}a_j.
\end{equation}
The symplectic form gives a non zero result only if $\pi_1\boldsymbol{l}e^{\wedge \Phi}\in A$ so the previous expression is equivalent to
\begin{equation}
    \Pi(S[\Phi],I\varphi)=\Pi\left[\omega(a_i,\pi_1\boldsymbol{\pi}\boldsymbol{l}e^{\wedge \Phi})\right]\omega_A^{ij}a_j=\Pi[\pi_1\boldsymbol{\pi}\boldsymbol{l}e^{\wedge \Phi}].
\end{equation}
As $\Pi$ kills all $\beta$'s  and $\gamma$'s we can trade it with the $\boldsymbol{\iota}$ immersion acting on $\varphi$ and write
\begin{equation}
    \Pi(S[\Phi],I\varphi)=\pi_1\boldsymbol{\pi}\boldsymbol{l}\boldsymbol{\iota} e^{\wedge \varphi}.
    \label{traiding_Pi_with_iota}
\end{equation}
The second term in the RHS of \ref{tildeSvarphi} can be computed trough similar steps, giving  the following chain of relations
\begin{equation}
    \begin{split}
     \delta_{2}I\varphi&=\pi_{1}\boldsymbol{\delta l}e^{\wedge \Phi},\\
    H\delta_{2}I\varphi&=(-)^{\gamma^{j}}\beta^{j}\pi_{1}\boldsymbol{\delta l}\boldsymbol{(}hb_{j}\boldsymbol{)}e^{\wedge \Phi},\\
     \delta_{2}H\delta_{2}I\varphi&=\pi_1 \boldsymbol{\delta l}(\boldsymbol{h}\boldsymbol{\delta l}+\boldsymbol{h}\boldsymbol{U})e^{\wedge \Phi},
    \end{split}
\end{equation}
where $\boldsymbol{(}hb_{j}\boldsymbol{)}$ is understood as the zero-coderivation associated with the state $hb_{j}$. Additionally, we can notice that each application of a $\delta_{2}H$ on $\delta_2 H\delta_2 \varphi$ consists in inserting a $(\boldsymbol{h}\boldsymbol{\delta l}+\boldsymbol{h}\boldsymbol{U})$ in front of $e^{\wedge\Phi}$, namely
\begin{equation}
    (\delta_{2}H)^{n}\delta_2 H\delta_2 \varphi=\pi_1 \boldsymbol{\delta l}(\boldsymbol{h}\boldsymbol{\delta l}+\boldsymbol{h}\boldsymbol{U})^{n+1}e^{\wedge \Phi},
\end{equation}
Therefore \ref{stepltilde} becomes
\begin{equation}
    \begin{split}
         \pi_{1}\tilde{\boldsymbol{l}}e^{\wedge \varphi}&=\pi_{1}\boldsymbol{\pi}\boldsymbol{l}\boldsymbol{\iota}e^{\wedge \varphi}-\pi_{1}\boldsymbol{\pi}\boldsymbol{\delta l}\frac{1}{1+\boldsymbol{h}(\boldsymbol{\delta l}+\boldsymbol{U})}\boldsymbol{h}(\boldsymbol{\delta l}+\boldsymbol{U})\boldsymbol{\iota}e^{\wedge\varphi}\\
         &=\pi_1 \boldsymbol{\pi}\boldsymbol{l}\frac{1}{1+\boldsymbol{h}(\boldsymbol{\delta l}+\boldsymbol{U})}\boldsymbol{\iota}e^{\wedge\varphi},\label{HPL-coa}
    \end{split}
\end{equation}
where we have also traded the $\Pi$ in the second term of \ref{stepltilde} with a $\boldsymbol{\iota}$ acting on $e^{\wedge\varphi}$, following the same reasoning adopted in \ref{traiding_Pi_with_iota}.  This establishes that the HPL at the co-algebraic level which we use in the main text is the result of the path integral.

Moreover, we can remove the $\pi_1$ from \eqref{HPL-coa} and add the projected Poisson bi-vector $\boldsymbol{U}_{\pi}$ (defined on $A$)  on both sides of the above expression, obtaining
\begin{equation}
     \tilde{\boldsymbol{l}}+\boldsymbol{U}_{\pi}=\boldsymbol{\pi}\boldsymbol{l}\frac{1}{1+\boldsymbol{h}(\boldsymbol{\delta l}+\boldsymbol{U})}\boldsymbol{\iota}+\boldsymbol{U}_{\pi}.\label{Utransf}
\end{equation}
 Remarkably this is equivalent to
what we would get by applying the homotopy transfer to the nilpotent structure $(\boldsymbol{l}+\boldsymbol{U})$, that is
\begin{equation}
    \tilde{\boldsymbol{l}}+\boldsymbol{U}_{\pi}=\boldsymbol{\pi}(\boldsymbol{l}+\boldsymbol{U})\frac{1}{1+\boldsymbol{h}(\boldsymbol{\delta l}+\boldsymbol{U})}\boldsymbol{\iota}.
\end{equation}
To see this equivalence, let us manipulate 
\begin{equation}
    \begin{split}
   \boldsymbol{\pi}(\boldsymbol{l}+\boldsymbol{U})\frac{1}{1+\boldsymbol{h}(\boldsymbol{\delta l}+\boldsymbol{U})}\boldsymbol{\iota}&=\boldsymbol{\pi}\boldsymbol{l}\frac{1}{1+\boldsymbol{h}(\boldsymbol{\delta l}+\boldsymbol{U})}\boldsymbol{\iota}+\boldsymbol{\pi}\boldsymbol{U}\frac{1}{1+\boldsymbol{h}(\boldsymbol{\delta l}+\boldsymbol{U})}\boldsymbol{\iota}\\
        &=\boldsymbol{\pi}\boldsymbol{l}\frac{1}{1+\boldsymbol{h}(\boldsymbol{\delta l}+\boldsymbol{U})}\boldsymbol{\iota}+\boldsymbol{U}_{\pi}\boldsymbol{\pi}\frac{1}{1+\boldsymbol{h}(\boldsymbol{\delta l}+\boldsymbol{U})}\boldsymbol{\iota}\\
        &=\boldsymbol{\pi}\boldsymbol{l}\frac{1}{1+\boldsymbol{h}(\boldsymbol{\delta l}+\boldsymbol{U})}\boldsymbol{\iota}+\boldsymbol{U}_{\pi}\boldsymbol{\pi}\boldsymbol{\iota}-\boldsymbol{U}_{\pi}\boldsymbol{\pi}\boldsymbol{h}(\boldsymbol{\delta l}+\boldsymbol{U})\frac{1}{1+\boldsymbol{h}(\boldsymbol{\delta l}+\boldsymbol{U})}\boldsymbol{\iota}\\
        &=\boldsymbol{\pi}\boldsymbol{l}\frac{1}{1+\boldsymbol{h}(\boldsymbol{\delta l}+\boldsymbol{U})}\boldsymbol{\iota}+\boldsymbol{U}_{\pi}=  \tilde{\boldsymbol{l}}+\boldsymbol{U}_{\pi},
    \end{split}
\end{equation}
where in the second line we have used \eqref{FU=UF} and in the third line $\boldsymbol{\pi}\boldsymbol{\iota}=1$ and $\boldsymbol{\pi h}=0$.\\
So, in conclusion, we have proven the following general formula for the Wilsonian effective action
\begin{equation}
    \tilde{S}[\varphi]={\cal A}-\log\left(e^{\frac{1}{2}h^{ij}\p_{\gamma^{i}} \p_{\gamma^{j}}}e^{-S_{\rm int}[\gamma^ic_{i}]}\right){\Big |}_{\gamma^{i}=0}+\int_{0}^{1}dt\,\omega(\dot{\varphi},\pi_{1}\tilde{\boldsymbol{l}}e^{\varphi}),\label{path-integration}
\end{equation}
which describes the transfer between the nilpotent structures
\begin{align}
(\boldsymbol{l}+\boldsymbol{U})^2=0\quad\to\quad (\tilde{\boldsymbol{l}}+\boldsymbol{U}_\pi)^2=0,
\end{align}
which, as expected, guarantees that the effective action is a solution to the BV quantum master equation, provided this is true for the initial UV action. 
The first two terms in \eqref{path-integration} give the expression of the (connected) vacuum bubbles, that are the additive constants in the effective action. These constants do not enter in the BV quantum master equation, yet they are important for the correct computation of observables. The dynamical part of the action, on the right, is given by the homotopy transfer at the co-algebra level. 
It would be interesting and useful to understand if also the constant part of the action could be written at the co-algebra level, encoding the vacuum amplitudes that one can construct using the interaction vertices and the propagators.

\section{ \texorpdfstring{$SL(2,\mathbb{C})$}{} vertices for the open string one-loop tadpole}\label{app4}
In this appendix, we will construct the OC-SFT vertices needed to compute the open string one-loop tadpole. The moduli space associated with this amplitude is one dimensional. For the sake of simplicity we will use $SL(2,\mathbb{C})$ vertices as much as we can.\footnote{It would be also interesting to use hyperbolic vertices \cite{Firat:2023gfn, Cho:2019anu}.}  A similar construction has been discussed in \cite{Sen:2020eck}, but the closed string degeneration region was not addressed.
%and we will parameterize it by using the length of the annulus. 
%Notice that by doing so we cover the entire moduli space, as it is possible to move the open string puncture along one boundary through a translation and flip the two boundaries through an $SL(2, \mathbb{C})$ transformation.\\
Specifically, we can characterize the annulus as a cylinder, $C_{(\pi,2\pi t)}$, with height $\pi$ and circumference $2\pi t$, where $t\in[0,\infty)$ parameterizes the moduli space.
Let's denote $z$  the complex coordinate on the cylinder, which is subject to the identification 
\begin{equation}
    z\simeq z+2\pi t. \label{annulusidentification}
\end{equation}
Within this parameterization, we have the open string degeneration in the limit $t\to \infty$, which corresponds to an extremely wide and flattened cylinder that leads to the collision of boundaries.  On the other hand, the limit $t\to 0$ is the closed string degeneration, represented by a very narrow tube.\\
As anticipated in the main text, this amplitude receives contribution from three Feynman diagrams, the first one containing  the open string degeneration, the second one containing the closed string degeneration and the third one which is a fundamental vertex covering the interior of the moduli space
\begin{equation}\label{Oannulisamplitude}
A_{0;1}^{0,2}(\varphi_{\rm o})=-\wo\left(\varphi_{\rm o},m_{0,2}^{(0,1)}\left(\boldsymbol{h}_{0}U_{\rm o}\right) \right)-\wo\left(\varphi_{\rm o},m_{1,0}^{(0,1)}\left(\frac{b_{0}^{+}}{L_{0}^{+}}\bar{P}_{0}^{+}l_{0,0}^{(0,1)}\right) \right)+\wo\left(\varphi_{\rm o},m_{0,0}^{(0,2)} \right).
\end{equation}

%%%%%%%%%%%%%%%%%%%%%%%%%%%%%%% 
%%%%%%%%%%%%%%%%%%%%%%%%%%%%%%%%
\subsection{Open string non-separating plumbing fixture}
Let us begin by describing the open channel
$$-\wo\left(\varphi_{\rm o},m_{0,2}^{(0,1)}\left(\boldsymbol{h}_{0}U_{\rm o}\right) \right).$$
To start with we  give an explicit expression to the product $m_{0,2}^{(0,1)}$ 
\begin{equation}
    \wo\left(\Psi_1,m_{0,2}^{(0,1)}(\Psi_2,\Psi_3)\right)=(-)^{\Psi_1+\Psi_2}\langle f_{-\xi}\circ\Psi_1(0)f_{0}\circ\Psi_2(0)f_{\xi}\circ\Psi_3(0)\rangle_{\rm UHP}
\end{equation}
and we define the $f$-maps as generic $SL(2,\mathbb{R})$ transformations which insert the punctures on the real axis of the UHP in $-\xi$, $0$ and $\xi>0$ respectively,  namely $f_\xi(0)=\xi$, $f_0(0)=0$ and $f_{-\xi}(0)=-\xi$.  We impose cyclicity by considering the $SL(2,\mathbb{R})$ transformation $g$ that cyclically permutes the location of the punctures, $g(0)=-\xi$, $g(\xi)=0$ and $g(-\xi)=\xi$ and enforcing the following conditions
\begin{equation}
    g\circ f_{-\xi}=f_{\xi},\qquad\qquad  g\circ f_{0}=f_{-\xi}, \qquad\qquad  g\circ f_{\xi}=f_{0}.
\end{equation}
Additionally we also impose twist symmetry $f_{\xi}(-w)=- f_{-\xi}(w)$ and $f_0(-w)=-f_0(w)$.
This gives a one-parameter ($\lambda$) family of  maps
\begin{align}
    &f_{-\xi}(w_1)=\frac{\xi ^2+\lambda \xi w_1}{\xi -3 \lambda w_1},\label{f-xi}\\
    &f_{0}(w_2)=\lambda w_2,\\
    &f_{\xi}(w_3)=\frac{\lambda \xi w_3-\xi ^2}{\xi +3 \lambda w_3}.\label{fxi}
\end{align}
%where $w_1$, $w_2$, $w_3$ are the local coordinates associated respectively to $\Psi_1$, $\Psi_2$ and $\Psi_3$. 
The two positive real parameters $\lambda$ and $\xi$ must satisfy $\xi>3\lambda$ in order to avoid the overlap of the local coordinate patches.\\
Presently, our aim is to implement the plumbing fixture procedure, identifying the local coordinates associated with the punctures located in $\xi$ and $-\xi$ as follows
\begin{equation}
    w_1 \simeq-\frac{q}{w_3} \label{plumbinfixtureopen},
\end{equation}
where $q\in[0,1]$.
%and the minus sign is necessary to ensure that when $q$ is set to $1$, the two semi-disks in the respective local coordinates are mapped to each other without any flipping. \\
By inverting \eqref{f-xi} and \eqref{fxi} we can rewrite \eqref{plumbinfixtureopen} getting an identification on the UHP (with coordinate $u$) which defines an annulus
\begin{equation}
    u \simeq f_{-\xi}\circ I_{-q}\circ f^{-1}_{\xi}(u),\label{identification}
\end{equation}
where we have denoted $I_{-q}(w)=-q/w$. In the following, we will use the symbol $\Sigma_{\rm open}^{\rm p.f.}(q)$ to denote this surface, which is reported in figure \ref{fig:open1}.
\begin{figure}[ht]
    \centering
    \includegraphics[scale=0.35]{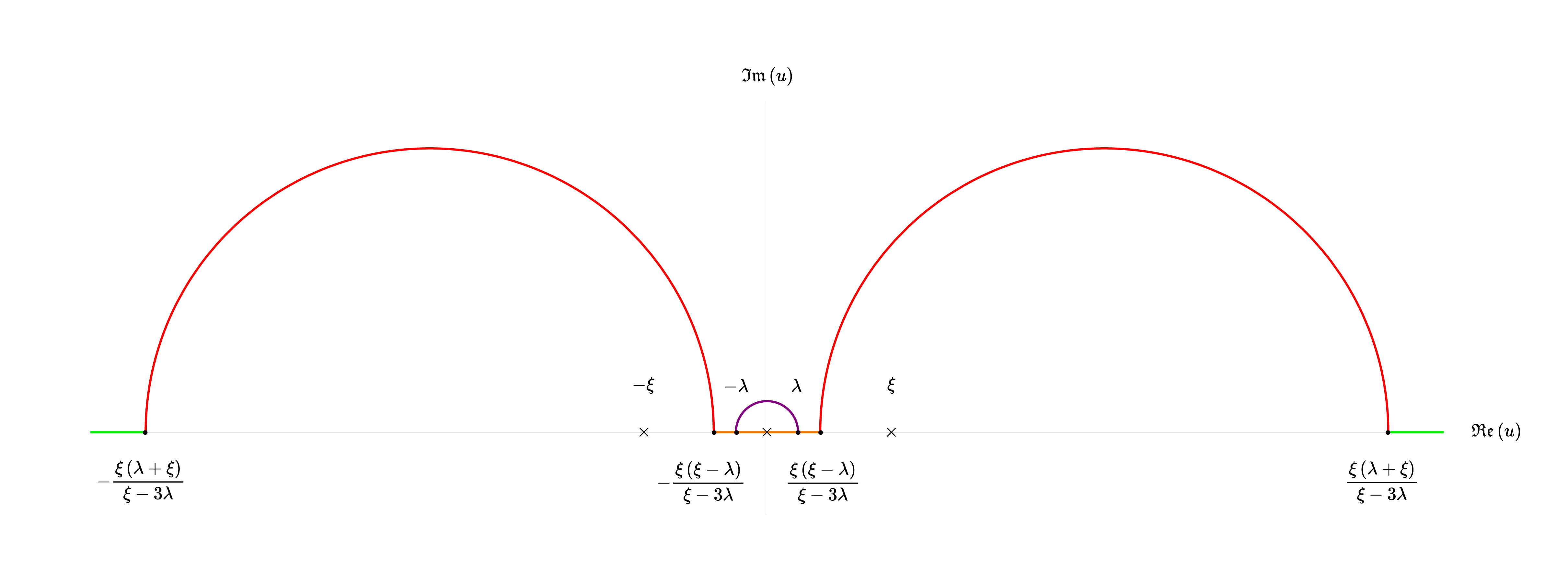}
    \caption{Annulus given by the plumbing fixture procedure with boundaries in green and orange. The red curves, identified through the equivalence relation \eqref{plumbinfixtureopen}, delineate the contours of the local patches associated with maps $f_{-\xi}$ and $f_{\xi}$. The $\times$  denote the open string punctures and the purple half-circle the contour of the local patch related to $f_{0}$. }
    \label{fig:open1}
\end{figure}
Note that identification \eqref{identification} corresponds to composition of $SL(2,\mathbb{R})$ maps,  and therefore, following \cite{Sen:2020eck}, we can find the similarity transformation that `diagonalizes' the previous expression. In other words, we introduce another $SL(2,\mathbb{R})$ map 
\begin{equation}
u=p(z')=\xi \sqrt{\frac{\xi^2-\lambda^2}{\xi^2-9\lambda^2}}\left(\frac{1-z'}{1+z'}\right),
\end{equation}
 such that
\begin{equation}
    z'\simeq p^{-1}\circ f_{-\xi}\circ I_{-q}\circ f^{-1}_{\xi} \circ p(z')  =\Lambda z',
\end{equation}
where the real number $\Lambda$ is the ratio of the two eigenvalues of the diagonal matrix associated with 
\begin{equation}
    p^{-1}\circ f_{-\xi}\circ I_{-q}\circ f^{-1}_{\xi} \circ p.
\end{equation}
Its explicit expression is
\begin{equation}
    \Lambda(\lambda,\xi,q)=\frac{\xi ^2+3 \lambda ^2 q-\sqrt{\xi ^4+9 \lambda ^4 q^2-10 \lambda ^2 \xi ^2 q}}{\xi ^2+3 \lambda ^2 q+\sqrt{\xi ^4+9 \lambda ^4 q^2-10 \lambda ^2 \xi ^2 q}},
\end{equation}
which is real thanks to the non-overlap condition  $\xi>3\lambda$.  The resulting surface is shown in figure \ref{fig:open2}.\\
Now it is easy to map to the frame where the annulus is described by \eqref{annulusidentification}. Indeed, defining $$z=-\log(z')$$ we get
\begin{equation}
    z\simeq z+\log\left(\frac{1}{\Lambda}\right).
\end{equation}
%in which the minus is needed because $\Lambda<1$. 
\begin{figure}[ht]
    \centering
    \includegraphics[scale=0.4]{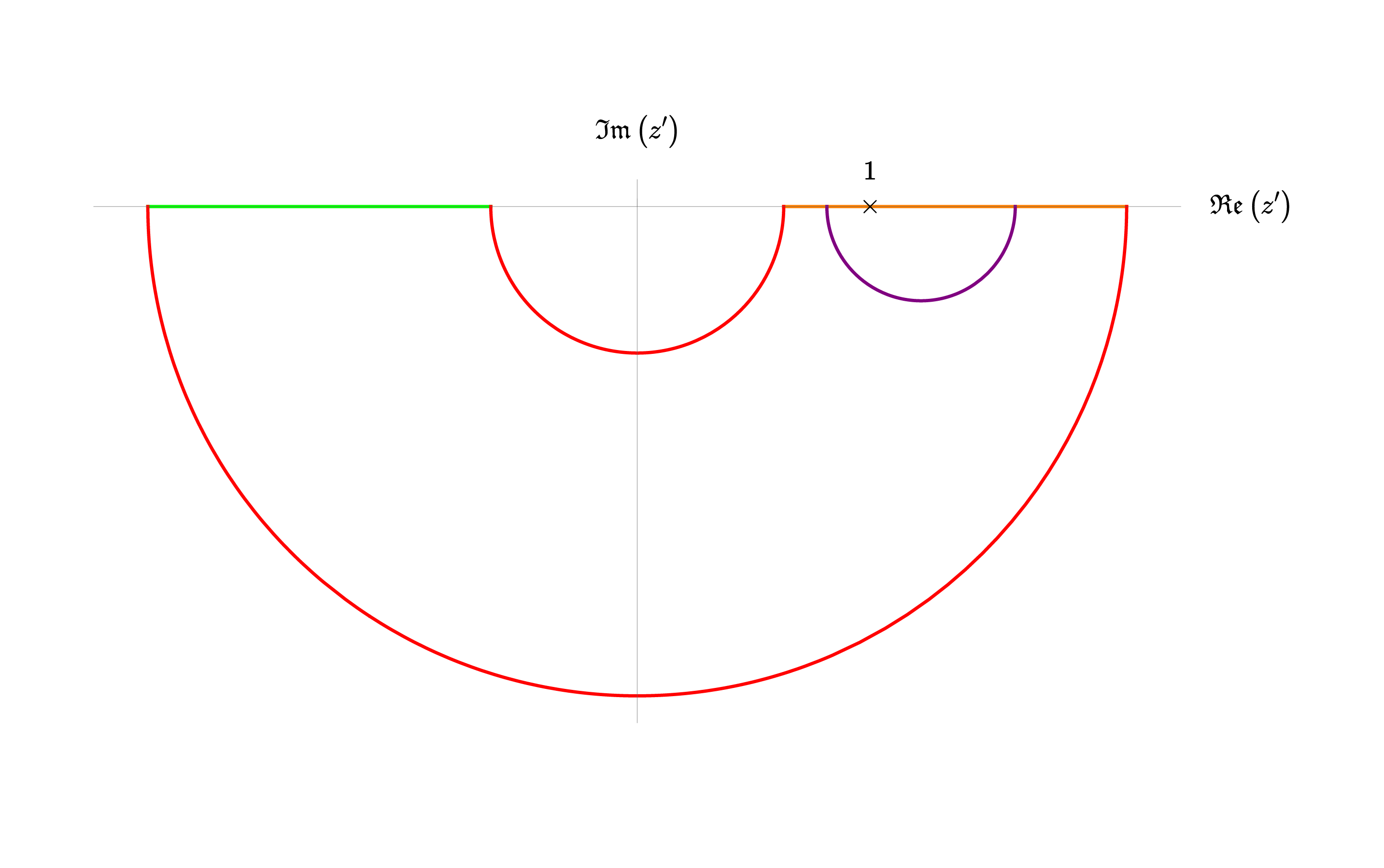}
    \caption{Annulus in the frame $z'=p^{-1}(u)$.}
    \label{fig:open2}
\end{figure}
Therefore, comparing with \eqref{annulusidentification}, we can read off how the modulus is related to the plumbing fixture parameter $q$ and the SFT data $\xi$ and $\lambda$
\begin{equation}\label{topen}
    t_{\rm o}(q, \lambda, \xi)=\frac{1}{2\pi}\log\left(\frac{\xi ^2+3 \lambda ^2 q+\sqrt{\xi ^4+9 \lambda ^4 q^2-10 \lambda ^2 \xi ^2 q}}{\xi ^2+3 \lambda ^2 q-\sqrt{\xi ^4+9 \lambda ^4 q^2-10 \lambda ^2 \xi ^2 q}}\right).
\end{equation}
Notice that by varying the plumbing fixture parameter in $[0,1]$ the modulus changes in the interval
\begin{equation}
    t_{\rm o}(q, \lambda, \xi)\in [t_{\rm max}(\lambda,\xi),\infty),
\end{equation}
in which we defined $ t_{\max}(\lambda,\xi)\coloneqq t_{\rm o}(q=1, \lambda, \xi)$.  This is the portion of moduli space covered by the non-separating open string Feynman diagram, which depends on the data $\xi$ and $\lambda$.\\
%The results obtained align perfectly with our expectations. Through the open non-separating plumbing fixture, we can cover a sub-region of the moduli space connected to the open string degeneration. The lower bound signifies the impossibility of covering the entire moduli space through this Feynman diagram. Naturally, this lower bound depends on $\lambda$ and $\xi$, the parameters used to define the maps. \\
Finally, we can easily find the local coordinate map associated to the  open string puncture on $C_{(\pi,2\pi t)}$, figure  \ref{fig:open3}
\begin{equation}
\begin{split}
    G_{\rm o}(w)&\coloneqq g_{\rm o}\circ f_{0}(w)\\&=\log \left(\frac{\xi \sqrt{\xi ^4+9 \lambda ^4 q^2-10 \lambda ^2 \xi ^2 q}+(\xi^2 -9 \lambda ^2 q) w}{\xi \sqrt{\xi ^4+9 \lambda ^4 q^2-10 \lambda ^2 \xi ^2 q}-(\xi^2 -9 \lambda ^2 q) w}\right)\\&=2\tanh^{-1}\left(\frac{1}{\xi}\sqrt{\frac{\xi^2-9\lambda^2q}{\xi^2-\lambda^2q}}w\right).
\end{split}\label{Gopen}
\end{equation}
where we defined  $g_{\rm o}(u)\coloneqq -\log \circ \,p^{-1}(u)$.\\
\begin{figure}[ht]
    \centering
    \includegraphics[scale=0.3]{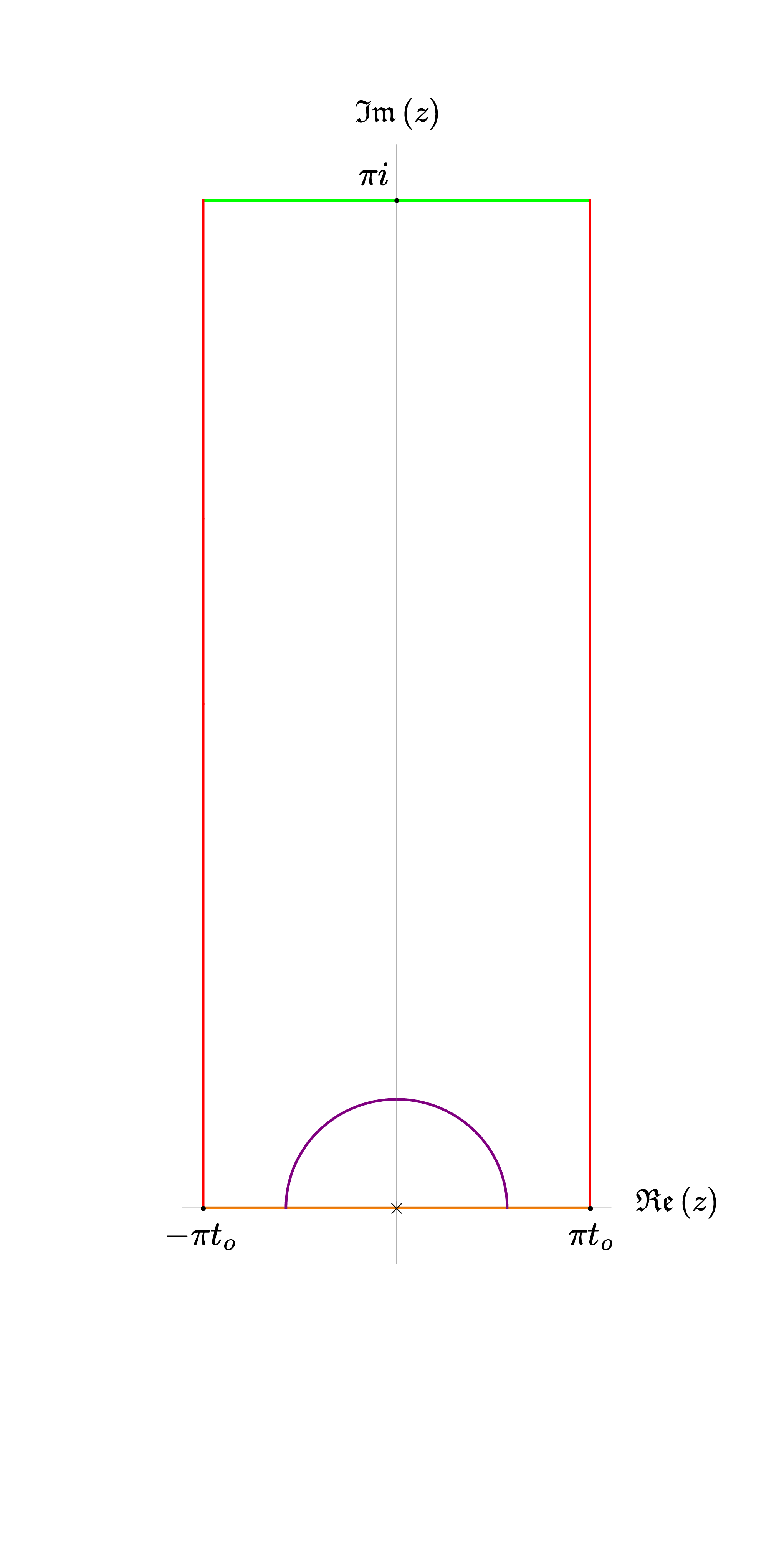}
    \caption{Annulus in the frame $z=-\log(z')$.}
    \label{fig:open3}
\end{figure}  
%%%%%%%%%%%%%%%%%%%%%%%%%%%%%%% 
%%%%%%%%%%%%%%%%%%%%%%%%%%%%%%%%
\subsection{Closed string separating plumbing fixture}
Regarding the closed channel
$$-\wo\left(\varphi_{\rm o},m_{1,0}^{(0,1)}\left(\frac{b_{0}^{+}}{L_{0}^{+}}\bar{P}_{0}^{+}l_{0,0}^{(0,1)}\right) \right),$$
 this time we have to perform the plumbing fixture on the following two fundamental vertices
\begin{equation}
    \wc\left(l_{0,0}^{(0,1)},\Phi\right)=\frac{1}{4\pi^2}\langle f_{b}\circ c_{0}^{-}\Phi(0,\bar{0}\rangle_{\rm disk},
\end{equation}
and
\begin{equation}
    \wo\left(\Psi,m_{1,0}^{(0,1)}(\Phi)\right)=\frac{(-)^{\Phi+\Psi}}{2\pi i}\langle d\circ f_{\rm o}\circ \Psi(0) d\circ f_{\rm c}\circ \Phi(0,\bar{0})\rangle_{\rm disk},
\end{equation}
where, following  \cite{cosmo}, $f_b$, $f_{\rm c}$ and $f_{\rm o}$ are the $SL(2,\mathbb{C})$ maps 
\begin{align}
    &f_{\rm b}(w_1)=\frac{w_1}{l_{\rm b}},\\
    &f_{\rm o}(w_2)=\frac{w_2}{l_{\rm o}},\\
    &f_{\rm c}(w_3)=i\frac{1+\frac{w_3}{\beta_1}}{1+\frac{w_3}{\beta_{2}}},
\end{align}
where $l_{\rm b}$ and $l_{\rm o}\in [1,\infty)$ are the fundamental closed and open stubs, respectively, and the real parameters $\beta_1$ and $\beta_2$ are chosen to be such that $\beta_2>\beta_1>1$. Additionally, imposing the non-overlap condition implies the following constraint
\begin{equation}
    l_{\rm o}>\frac{\beta_1(1-\beta_2)}{\beta_2(1-\beta_1)}.
\end{equation}
The two punctured disks are reported in figure \ref{fig:closed1}.
\begin{figure}[th]
    \centering
    \includegraphics[scale=0.25]{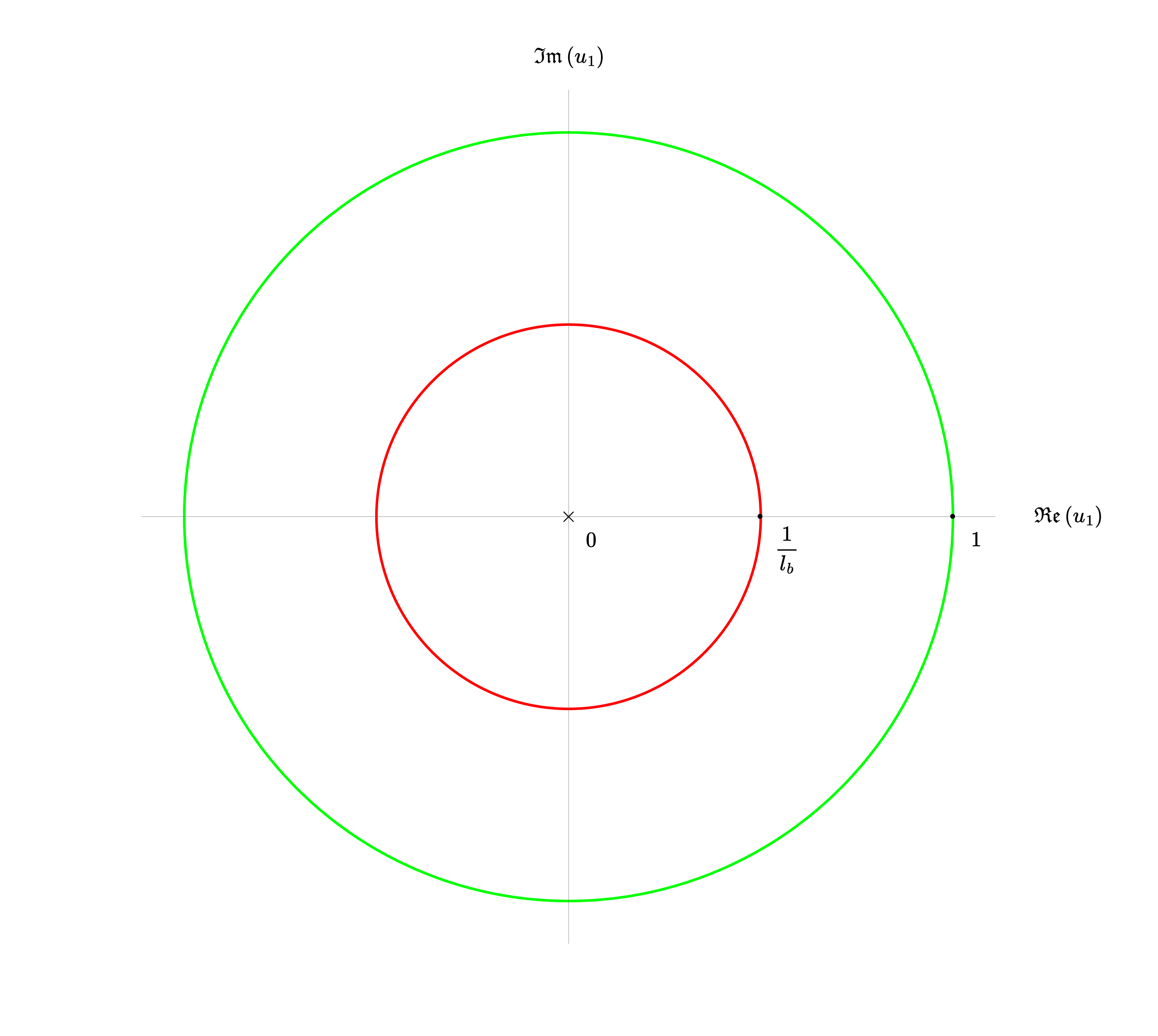}
     \includegraphics[scale=0.25]{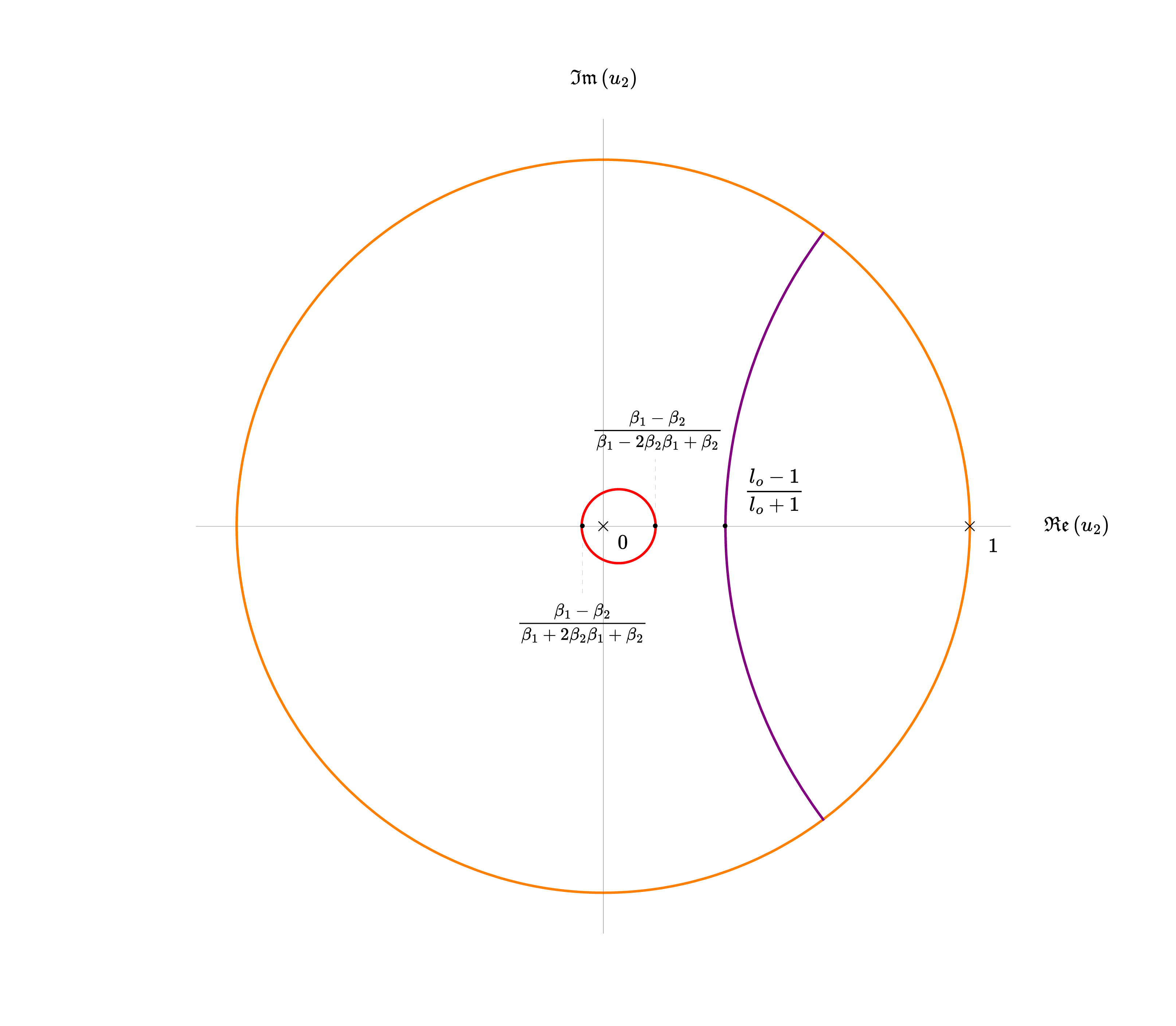}
    \caption{On the left, the unit circle with an inserction of a closed string at $0$ and local patches corresponding to a circle of radius $l_{\rm b}^{-1}$. On the right, the unit disk with a closed string puncture and an open string puncture, with local patches represented in red and purple, respectively. The colors used are consistent with the previous section. The plumbing fixture glues the two disks along the red curves, resulting in an annulus with orange and green boundaries.}
    \label{fig:closed1}
\end{figure}  
Furthermore we called $d$ the Cayley function, which maps the UHP into the unitary disk in the standard way $d(0)=1,\,d(i)=0, \,d(\infty)=-1$. Therefore the local coordinate maps on the disk are
\begin{align}
    u_1=f_{\rm b}(w_1), \qquad \qquad
    u_2=d\circ f_{\rm o}(w_2),\qquad\qquad
    u_2=d\circ f_{\rm c}(w_3),\label{fofcfb}
\end{align}
Now, let us perform the closed-string separating plumbing fixture, identifying the local coordinates associated with the closed punctures on the two disks using the relationship
\begin{equation}
    w_1\simeq\frac{q}{w_3},\label{closedplumbingfixture}
\end{equation}
where $q\in [0,1]$.\\
Inverting \eqref{fofcfb} and inserting it into \eqref{closedplumbingfixture} allows us to derive the relation between the coordinates on the two disks
\begin{equation}
    u_2=d\circ f_{\rm c}\circ I_{q}\circ f_{\rm b}^{-1}(u_1), 
\end{equation}
we will call this surface as $\Sigma_{\rm closed}^{\rm p.f.}(q)$, figure \ref{fig:closed2}.
\begin{figure}[ht]
    \centering
    \includegraphics[scale=0.4]{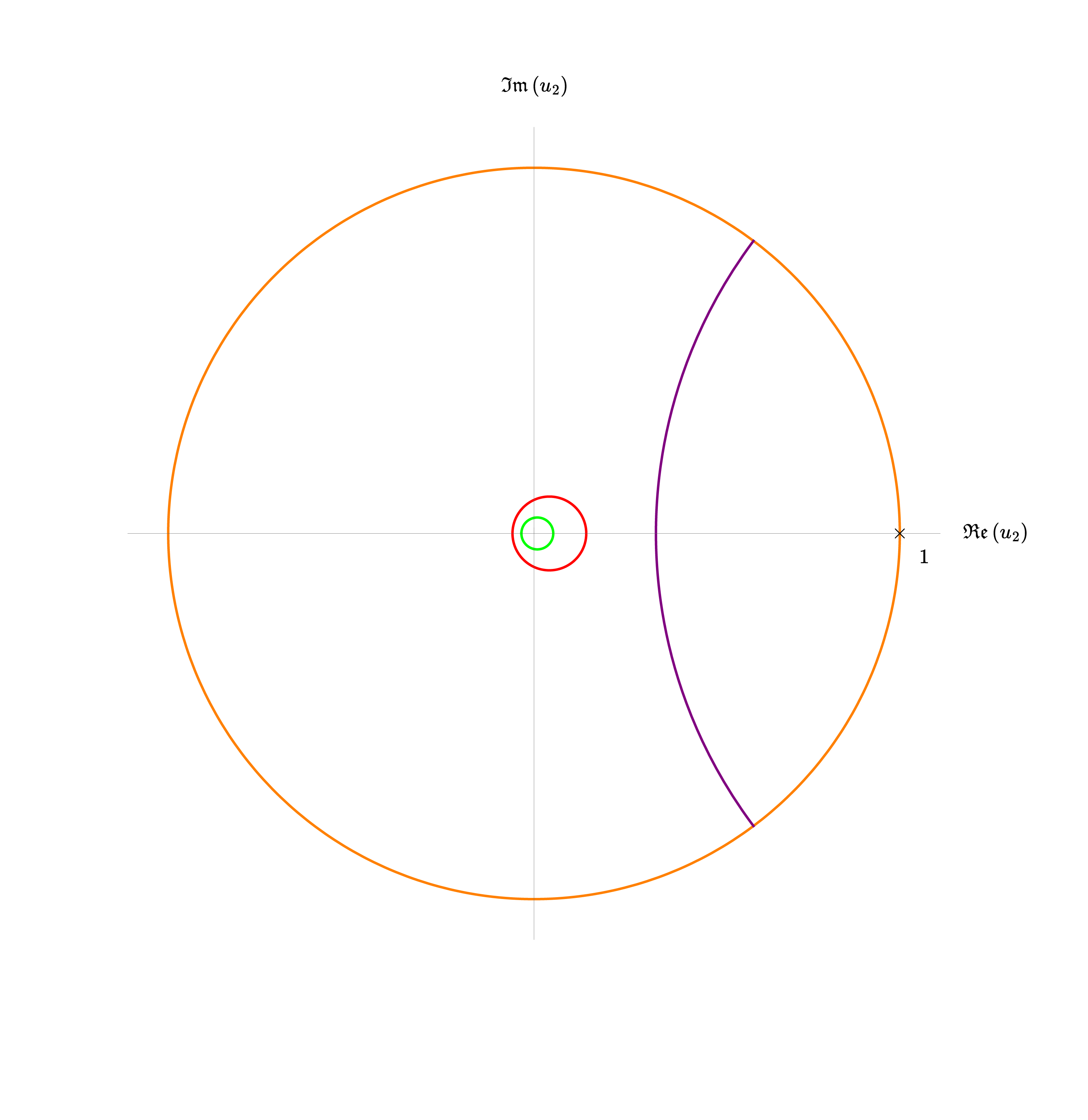}
    \caption{Boundary-punctured annulus given by the separating closed plumbing fixture}
    \label{fig:closed2}
\end{figure}  
 In terms of the $u_2$ coordinate, we are delineating an annulus with an outer boundary of unit radius with the open puncture and an inner boundary consisting of a circle with radius $R$, centered at $C$ 
\begin{align}
    &R(l_{\rm b},\beta_1,\beta_2,q)=\frac{2q \text{$\beta _1$} \text{$\beta_2 $}   (\text{$\beta_2 $}-\text{$\beta_1$})\text{$l_{\rm b} $}}{4 \text{$\beta_1 $}^2 \text{$\beta_2 $}^2 \text{$l_{\rm b}$}^2-q^2 (\text{$\beta_1 $}+\text{$\beta_2 $})^2},\\
    &C(l_{\rm b},\beta_1,\beta_2,q)=\frac{ q^2 (\text{$\beta_{2}^{2} $}-\text{$\beta_{1}^{2} $}) }{4 \text{$\beta_1 $}^2 \text{$\beta_2 $}^2 \text{$l_{\rm b} $}^2-q^2 (\text{$\beta_1 $}+\text{$\beta_2 $})^2}.
\end{align}
Therefore, to obtain the annulus in the frame where \eqref{annulusidentification} holds, it is necessary to apply an additional transformation to center both boundaries at zero. The strategy we adopt is as follows: we start with the generic real Möbius map that leaves the unit circle unchanged
\begin{equation}
    h_{\alpha}(u_2)=\frac{u_2-\alpha}{1-\alpha u_2},
\end{equation}
with $\alpha\in \mathbb{R}$, and then we impose the condition 
\begin{equation}
    h_{\alpha}\left(C+R\right)=- h_{\alpha}\left(C-R\right),
\end{equation}
thus getting
\begin{equation}
    \alpha_{\pm}=\frac{q^2 \left(\text{$\beta_1 $}^2+\text{$\beta_2 $}^2\right)\pm 2 \text{$\beta_1 $} \text{$\beta_2 $} \left(\sqrt{(q^2-\text{$\beta_{1}^{2} $} \text{$l_{b}^{2} $})  (q^2-\text{$\beta_{2}^{2} $} \text{$l_{b}^{2} $}) }\mp\text{$\beta_1 $} \text{$\beta_2 $} \text{$l_{\rm b} $}^2\right)}{q^2 (\text{$\beta_{1}^{2} $}-\text{$\beta_{2}^{2} $})},
\end{equation}
notice that both solutions are real because $1<\frac{\beta_1l_{\rm b}}{q}<\frac{\beta_2l_{\rm b}}{q}$ by assumption. These two solutions identify the two possible maps which center the boundary without the open insertion by leaving the unit circle unchanged. In particular $h_{\alpha_{+}}$ maps the interior of the unit circle into the interior of the unit circle, whereas $h_{\alpha_{-}}$ sends the interior of the unit circle into the exterior of the unit circle. We will consider $z'=h_{\alpha_{-}}(u_1)$ and thus the case in which the punctured circle becomes the inner boundary, as shown in figure \ref{fig:closed3}.
\begin{figure}[ht]
    \centering
    \includegraphics[scale=0.3]{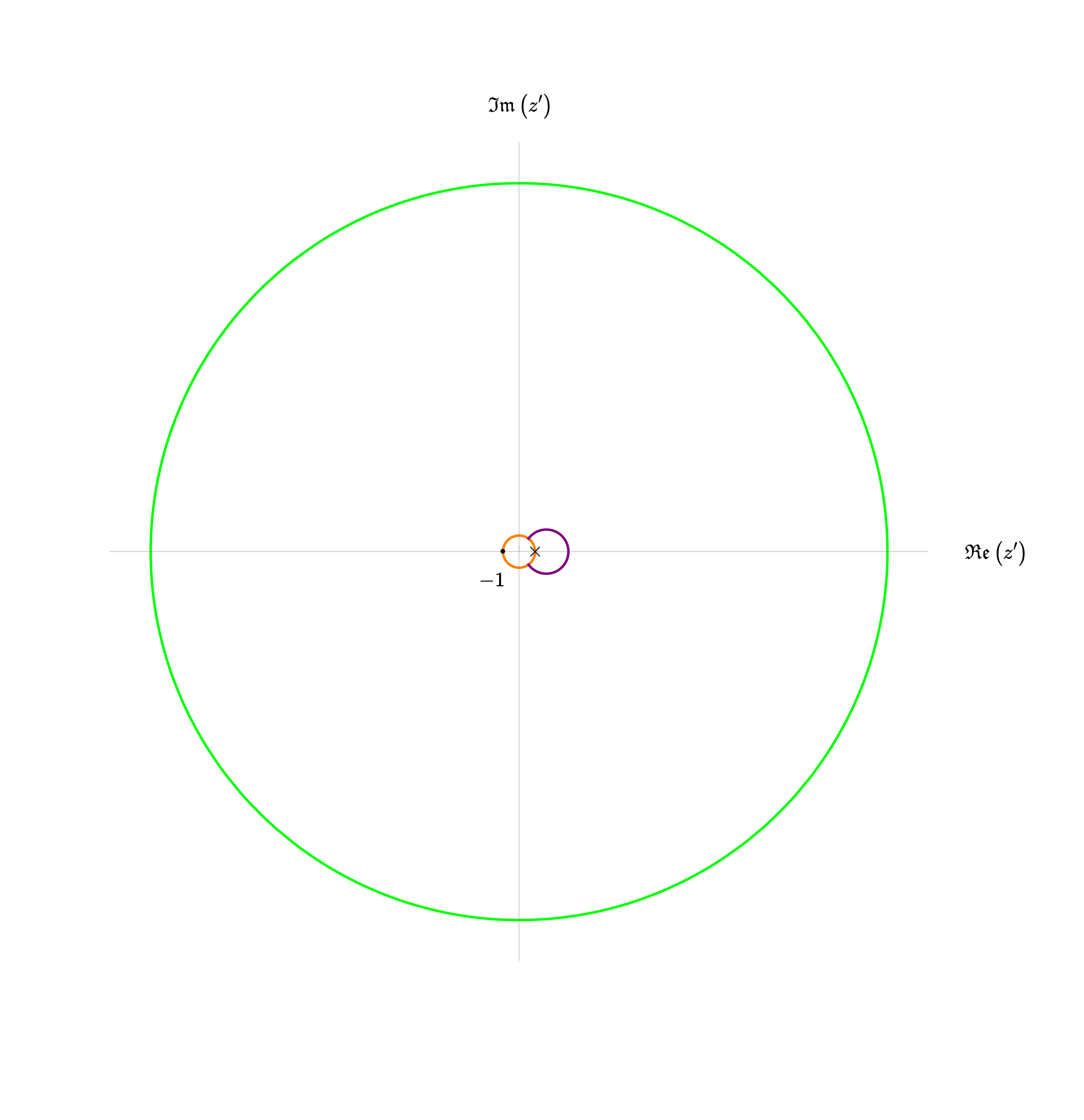}
    \caption{Annulus in the frame $z'=h_{\alpha_{-}}(u_2)$.}
    \label{fig:closed3}
\end{figure}  
Finally, by applying the logarithm, we obtain the cylinder, which, however, is rotated by $\pi/2$ and has a boundary length of $2\pi$, as reported in the first figure of \ref{fig:closed4}. To achieve the desired surface it is needed to further apply a rotation and a dilatation, second figure of \ref{fig:closed4}.
\begin{figure}[ht]
\centering
\begin{subfigure}{0.4\linewidth}
    \centering
    \includegraphics[scale=0.25]{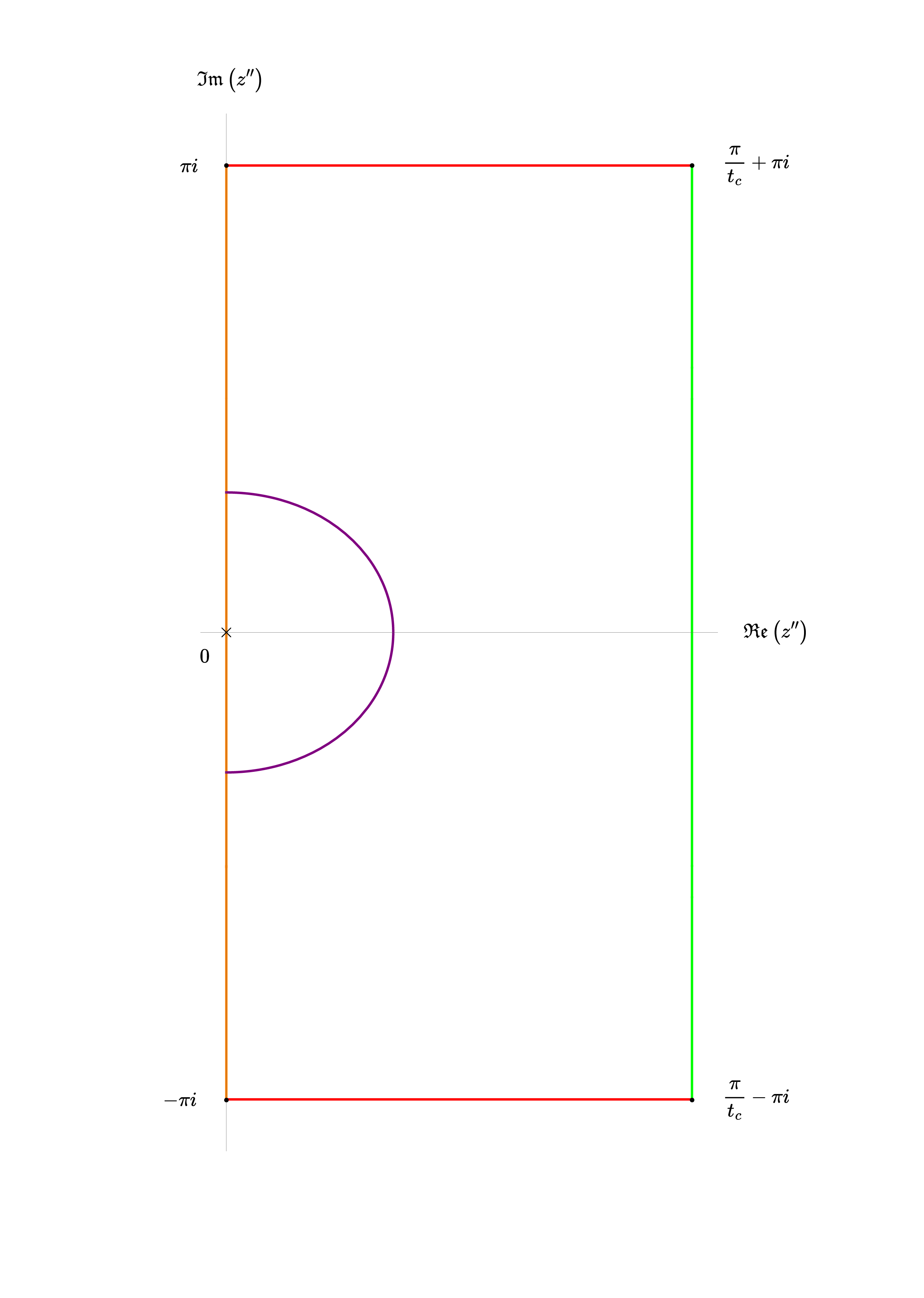}
\end{subfigure}
\begin{subfigure}{0.4\linewidth}
    \includegraphics[scale=0.25]{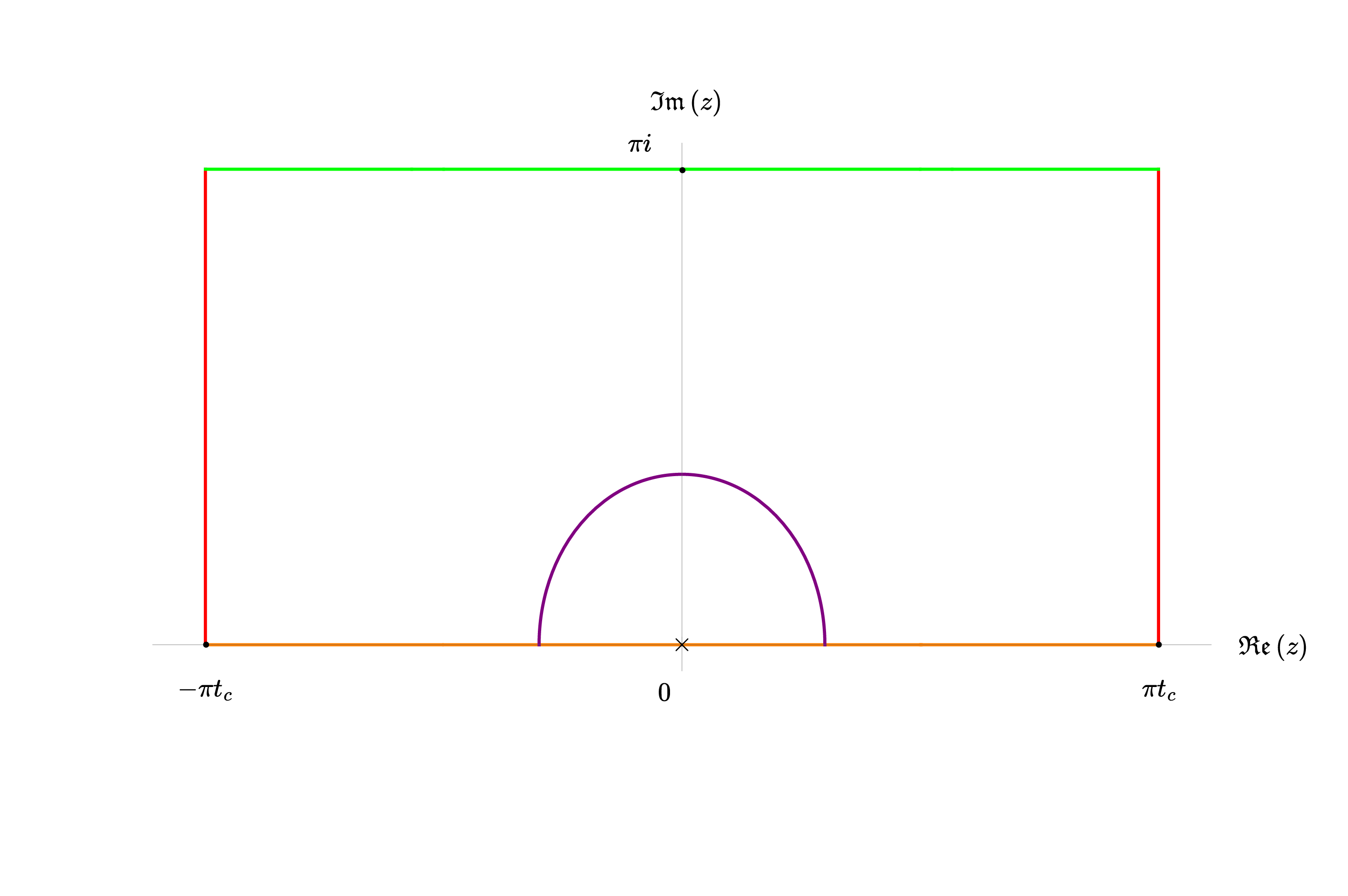}
\end{subfigure}
    \caption{Annulus in the frames $z''=\log(z')$ and $z=it_{\rm c}z''$.}
     \label{fig:closed4}
\end{figure}  
At this point, it is possible to read off the modulus by comparing with \eqref{annulusidentification}
\begin{equation}\label{tclosed}
 t_{\rm c}(q; l_{\rm b},\beta_1,\beta_2)= \pi \log \left(\frac{\text{$\beta_1 $} \text{$\beta_2 $} \text{$l_{b}^{2} $}-q^2+\sqrt{ 
 \left(q^2-\beta_{1}^{2}l_{b}^{2}\right)
 \left(q^2-\text{$\beta_{2}^{2} $} \text{$l_{b}^{2} $}\right)}}{q l_{\rm b}(\text{$\beta_2 $} -\text{$\beta_1 $})}\right)^{-1}
\end{equation}
notice that by varying the plumbing fixture parameter in [0, 1] the modulus changes as follows
\begin{equation}
    t_{\rm c}(q, l_{\rm b},\beta_1,\beta_2)\in[0,t_{\rm min}(l_{\rm b},\beta_1,\beta_2)],
\end{equation}
where we defined $t_{\rm min}(l_{\rm b},\beta_1,\beta_2)= t_{\rm c}(q=1, l_{\rm b},\beta_1,\beta_2)$.\\
As expected in the closed string channel we are able to cover a sub-region of moduli space containing the closed string degeneration $t=0$. \\
The local coordinate  associated with the open string insertion on the canonical cylinder is given by
\begin{equation}
\begin{split}
    &G_{\rm c}(w)\coloneqq g_{\rm c} \circ d\circ f_{\rm o}(w)\\&=it_{\rm c}\log\left(\frac{q^2 \left(\beta_{1}^{2}w+i\beta_{2}^{2}l_{ o}\right)-\beta_1\beta_2 (w+il_{o}) \left(\beta_1 \beta_2 l_{b}^{2}+\sqrt{(q^2-\beta_{1}^{2}l_{b}^{2})  (q^2-\beta_{2}^{2} l_{b}^{2})}\right)}
    {q^2 \left(-\beta_{1}^{2}w+i\beta_{2}^{2}l_{ o}\right)+\beta_1\beta_2 (w-il_{o}) \left(\beta_1 \beta_2 l_{b}^{2}+\sqrt{(q^2-\beta_{1}^{2}l_{b}^{2})  (q^2-\beta_{2}^{2} l_{b}^{2})}\right)} 
    \right)\\&=2t_{\rm c}\tan^{-1}\left(-\frac{1}{l_{\rm o}}\frac{q\beta_1+l_{\rm b}e^{\pi/t_{\rm c}}}{q\beta_2-l_{\rm b}e^{\pi/t_{\rm c}}}w\right).   
\end{split}\label{Gclosed}
\end{equation}
where we have introduced 
\begin{equation}
    g_{\rm c}(u_2)\coloneqq it_{\rm c}\log \circ h_{\alpha_{-}}(u_2).
\end{equation}

 %%%%%%%%%%%%%%%%%%%%%%%%%%%%%%%% 
%%%%%%%%%%%%%%%%%%%%%%%%%%%%%%%%
\subsection{Fundamental vertex and decoupling of BRST exact states}
It is important to notice the following fact. The local coordinate map in the open string channel \eqref{Gopen} can never coincide with the same quantity in the closed string channel \eqref{Gclosed}. This remains true even when the open and closed parameters are chosen such that $t_{\rm c}=t_{\rm o}$. Therefore, even if the moduli space can in principle be covered with the two Feynman diagrams containing the degenerations, this would not be sufficient off-shell. Therefore, differently from the disk amplitude with two closed strings that was analyzed in \cite{cosmo}, this time we need a fundamental vertex covering the interior of moduli space.

 In order to do so we will rewrite the amplitudes \eqref{Oannulisamplitude} in terms of surface states, by making explicit the integration over the moduli space. 
%We wil following the usual construction outlined for instance in [ZwiebachClosed], [SFTbook] and in [Erler1looptapole].
From now on, for the sake of simplicity, we will assume that the string field $\varphi_{\rm o}$ has even degree, which is true in the case of a physical tadpole amplitude.\\ Let us start with the open channel 
\begin{equation}
\begin{split}
    -\wo\left(\varphi_{\rm o},m_{0,2}^{(0,1)}\left(\boldsymbol{h}_{0}U_{\rm o}\right) \right)&=-(-)^{o^i}\wo\left(\varphi_{\rm o},m_{0,2}^{(0,1)}\left(h_0o_i,o^i\right)\right)\\
    &=-(-)^{o^i}\wo\left(o^i,m_{0,2}^{(0,1)}\left(\varphi_{\rm o},h_0o_i\right)\right)\\
    &=-\langle \Sigma_{0,3}^{(0,1)}\vert o^i\otimes \varphi_{\rm o} \otimes h_{0}o_i\\
    &=-\int_{0}^{1}\frac{dq}{q}\langle \Sigma_{0,3}^{(0,1)}\vert \left(1_{\mathcal{H}_{\rm o}}\otimes 1_{\mathcal{H}_{\rm o}}\otimes b_{0}q^{L_0}\right)\left(o^i\otimes \varphi_{\rm o} \otimes o_i\right),
    \end{split}
\end{equation}
where in the second step we used cyclicity and in the third line we introduced the surface state $\langle \Sigma_{0,3}^{(0,1)}\vert$ that is defined as follows
\begin{equation}
    \langle \Sigma_{0,3}^{(0,1)}\vert \Psi_1\otimes\Psi_2\otimes\Psi_3 \coloneqq\langle f_{-\xi}\circ\Psi_1(0)f_{0}\circ\Psi_2(0)f_{\xi}\circ\Psi_3(0)\rangle_{\rm UHP} =  (-)^{\Psi_1+\Psi_2}\wo\left(\Psi_1,m_{0,2}^{0,1}(\Psi_2,\Psi_3)\right).
\end{equation}
As for the closed string channel we have
\begin{equation}
\begin{split}
    -\wo\left(\varphi_{\rm o},m_{1,0}^{(0,1)}\left(h_{0}^{+}l_{0,0}^{(0,1)}\right) \right)&=-\frac{1}{2\pi i}\langle\Sigma_{1,1}^{(0,1)}\vert \varphi_{\rm o} \otimes h_{0}^{+}l_{0,0}^{(0,1)} \\
    &=-\frac{1}{2\pi i}\int_{0}^{1}\frac{dq}{q}\langle\Sigma_{1,1}^{(0,1)}\vert\left(1_{\mathcal{H}_{\rm o}}\otimes b_{0}^{+}q^{L_{0}^{+}}\right)\left(\varphi_{\rm o} \otimes l_{0,0}^{(0,1)}\right),
\end{split}
\end{equation}
where we have defined the surface state $\langle\Sigma_{1,1}^{(0,1)}\vert$ as follows
\begin{equation}
    \langle\Sigma_{1,1}^{(0,1)}\vert\Psi\otimes\Phi\coloneqq    \langle d\circ f_{\rm o}\circ \Psi(0) d\circ f_{\rm c}\circ \Phi(0,\bar{0})\rangle_{\rm disk}=(-)^{\Phi+\Psi}2\pi i\,\wo\left(\Psi,m_{1,0}^{(0,1)}(\Phi)\right).
\end{equation}
Now, our attention turns to the fundamental vertex
$$
\wo\left(\varphi_{\rm o},m_{0,0}^{(0,2)} \right).
$$
 Although it has not been explicitly constructed yet, we know that it corresponds to an integral over  a piece of a section of the fiber bundle having the moduli space as its base and the local coordinate map as its fiber. Clearly such subsection can be parameterized using the same variable $t$ used to characterize the moduli space. Then the region covered by $m_{0,0}^{(0,2)}$ has to align with the interval $t\in [t_{\rm min},t_{\rm max}]$, namely the portion that cannot be covered by the two previous Feynman diagrams. 
% It is worth nothing that this procedure is applicable when considering the scenario where the parameters of the local coordinate maps are chosen such that $t_{\min}\neq t_{\max}$. When $t_{\min}= t_{\max}$ a different parameterization of the subsection is required. For instance by using a variable that defines the local coordinate maps in this case the fundamental vertex shows the so-called vertical integration. Additionally, if there were a chance to equalize the local coordinate maps in this scenario, the section would reduce to a 0-dimensional surface, indicating the absence of the fundamental vertex. 
 In our specific case, based on the definitions provided in the previous section, it is actually possible to choose the parameters such that $t_{\min}=t_{\max}$. However it does not appear to be possible to match the local coordinates and therefore the fundamental vertex is needed, with our choice of lower dimensional vertices.\footnote{When $t_{\min}=t_{\max}$ the fundamental vertex would be associated to a {\it vertical integration} \cite{Sen:2014pia}
changing the local coordinate while not moving in moduli space.}\\
 In general we thus have
\begin{equation}
    \wo(\varphi_{\rm o},m_{0,0}^{(0,2)})=\int_{t_{\rm min}}^{t_{\rm max}}dt \langle \Sigma_{0,1}^{(0,2)}(t)\vert B_{t}(V_t)\varphi_{\rm o},
\end{equation}
where the surface state is defined as
\begin{equation}
   \langle \Sigma_{0,1}^{(0,2)}(t)\vert \Psi \coloneqq \langle F_t\circ \Psi(0)\rangle_{C_{(\pi,2\pi t)}}.
\end{equation}
The local coordinate map $F_t(w)$ will be fixed shortly by imposing BRST decoupling (which is equivalent to the homotopy relations). $V_t$ is the Schiffer vector and $B_{t}(V_t)$ the Beltrami form which are defined as follows
\begin{equation}
    B_{t}(V_t)=\oint_{0} \frac{dw}{2\pi i} b(w)V_{t}(w),\qquad\qquad V_t(w)=\frac{\p F^{-1}_{t}}{\p t}\left(F_{t}(w)\right),
\end{equation}
where the contour surrounds the puncture in the local coordinate frame, see \cite{Zwiebach:1992ie}  and \cite{Erbin:2021smf} for more details.\\
Crucially, the surface state satisfies
\begin{equation}\label{totalderiv}
    \frac{d}{dt}\langle \Sigma_{0,1}^{(0,2)}(t)\vert = -\langle \Sigma_{0,1}^{(0,2)}(t)\vert  B_{t}(V_t)Q_{\rm o}.
\end{equation}
With these premises the full annulus amplitude \eqref{Oannulisamplitude} becomes
\begin{equation}
\begin{split}
     A_{0;1}^{0,2}(\varphi_{\rm o})=&-\frac{1}{2\pi i}\int_{0}^{1}\frac{dq}{q}\langle\Sigma_{1,1}^{(0,1)}\vert\left(1_{\mathcal{H}_{\rm o}}\otimes b_{0}^{+}q^{L_{0}^{+}}\right)\left(\varphi_{\rm o} \otimes l_{0,0}^{(0,1)}\right)\\
    &+\int_{t_{\rm min}}^{t_{\rm max}}dt \langle \Sigma_{0,1}^{(0,2)}(t)\vert B_{t}(V_t)\varphi_{\rm o}\\
    &-\int_{0}^{1}\frac{dq}{q}\langle \Sigma_{0,3}^{(0,1)}\vert \left(1_{\mathcal{H}_{\rm o}}\otimes 1_{\mathcal{H}_{\rm o}}\otimes b_{0}q^{L_0}\right)\left(o^i\otimes \varphi_{\rm o} \otimes o_i\right).
\end{split}
\end{equation}
Now, we want to verify  if and how BRST exact states decouple. To do so let us consider $\varphi_{\rm o}=Q_{\rm o}\Lambda$
\begin{equation}
\begin{split}
     A_{0;1}^{0,2}(Q_{\rm o} \Lambda)=&-\frac{1}{2\pi i}\int_{0}^{1}\frac{dq}{q}\langle\Sigma_{1,1}^{(0,1)}\vert\left(1_{\mathcal{H}_{\rm o}}\otimes b_{0}^{+}q^{L_{0}^{+}}\right)\left(Q_{\rm o}\Lambda\otimes l_{0,0}^{(0,1)}\right)\\
    &+\int_{t_{\rm min}}^{t_{\rm max}}dt \langle \Sigma_{0,1}^{(0,2)}(t)\vert B_{t}(V_t)Q_{\rm o}\Lambda\\
    &-\int_{0}^{1}\frac{dq}{q}\langle \Sigma_{0,3}^{(0,1)}\vert \left(1_{\mathcal{H}_{\rm o}}\otimes 1_{\mathcal{H}_{\rm o}}\otimes b_{0}q^{L_0}\right)\left(o^i\otimes Q_{\rm o}\Lambda\otimes o_i\right),
\end{split}
\end{equation}
by using the fact that the surface states are BRST invariant, the property \eqref{totalderiv}, and the relations $\boldsymbol{Q}_{\rm o}\boldsymbol{U}_{\rm o}=0$ and 
$Q_{\rm c}l_{0,0}^{(0,1)}=0$ we get 
\begin{equation}
\begin{split}
    A_{0;1}^{0,2}(Q_{\rm o} \Lambda)&=+\frac{1}{2\pi i}\int_{0}^{1}\frac{dq}{q}\langle\Sigma_{1,1}^{(0,1)}\vert\left(1_{\mathcal{H}_{\rm o}}\otimes \left[b_{0}^{+}q^{L_{0}^{+}},Q_{\rm c}\right]\right)\left(\Lambda\otimes l_{0,0}^{(0,1)}\right)\\
    &\quad-\int_{t_{\rm min}}^{t_{\rm max}}dt  \frac{d}{dt}\langle\Sigma_{0,1}^{(0,2)}(t)\vert \Lambda\\
    &\quad+\int_{0}^{1}\frac{dq}{q}\langle \Sigma_{0,3}^{(0,1)}\vert \left(1_{\mathcal{H}_{\rm o}}\otimes 1_{\mathcal{H}_{\rm o}}\otimes \left[b_{0}q^{L_0},Q_{\rm o}\right]\right)\left(o^i\otimes \Lambda\otimes o_i\right)\\
    &=-\frac{1}{2\pi i}\int_{0}^{\infty}ds\frac{d}{ds}\langle\Sigma_{1,1}^{(0,1)}\vert\left(1_{\mathcal{H}_{\rm o}}\otimes e^{-sL_{0}^{+}}\right)\left(\Lambda\otimes l_{0,0}^{(0,1)}\right)\\
    &\quad+\langle\Sigma_{0,1}^{(0,2)}(t_{\rm min})\vert\Lambda-\langle\Sigma_{0,1}^{(0,2)}(t_{\rm max})\vert\Lambda\\
    &\quad-\int_{0}^{\infty}ds\frac{d}{ds}\langle \Sigma_{0,3}^{(0,1)}\vert \left(1_{\mathcal{H}_{\rm o}}\otimes 1_{\mathcal{H}_{\rm o}}\otimes e^{-sL_0}\right)\left(o^i\otimes \Lambda\otimes o_i\right)\\
    &=\frac{1}{2\pi i}\langle\Sigma_{1,1}^{(0,1)}\vert\left(\Lambda\otimes l_{0,0}^{(0,1)}\right)+\langle\Sigma_{0,1}^{(0,2)}(t_{\rm min})\vert\Lambda-\langle\Sigma_{0,1}^{(0,2)}(t_{\rm max})\vert\Lambda+\langle \Sigma_{0,3}^{(0,1)}\vert \left(o^i\otimes \Lambda\otimes o_i\right),
\end{split}
\end{equation}
where we made the change of variable $q=e^{-s}$ and we ignored the contributions at the boundary of moduli space (open and closed string degeneration). Notice that the last line is equivalent to the homotopy relation \eqref{annulushomotopyrelation}.
% inserted as the second entry in the symplectic form that has as the first entry $\Lambda$.
Therefore, the amplitude vanishes for all $\Lambda$ if we impose
\begin{align}
   &\langle \Sigma_{0,3}^{(0,1)}\vert \left(o^i\otimes 1_{\mathcal{H}_{\rm o}} \otimes o_i\right)=\langle\Sigma_{0,1}^{(0,2)}(t_{\rm max})\vert,\label{identification1}\\ 
   &-\frac{1}{2\pi i}\langle\Sigma_{1,1}^{(0,1)}\vert\left(1_{\mathcal{H}_{\rm o}}\otimes l_{0,0}^{(0,1)}\right)=\langle\Sigma_{0,1}^{(0,2)}(t_{\rm min})\vert.\label{identification2}
\end{align}
These two relations allow us to determine the local coordinate $F_{t}$ and thus to fully define the fundamental vertex. Notice that the surface states in the l.h.s are obtained respectively trough the open and closed plumbing fixture, with $q=1$. Similarly, the surface states associated to the plumbing fixture for generic $q$ can be written as
\begin{align}
    &\langle \Sigma_{\rm open}^{\rm p.f.}(q)\vert= \langle \Sigma_{0,3}^{(0,1)}\vert \left(o^i\otimes 1_{\mathcal{H}_{\rm o}} \otimes q^{L_0}o_i\right),\\
    &\langle \Sigma_{\rm closed}^{\rm p.f.}(q)\vert=-\frac{1}{2\pi i}\langle\Sigma_{1,1}^{(0,1)}\vert\left(1_{\mathcal{H}_{\rm o}}\otimes q^{L_{0}^{+}}l_{0,0}^{(0,1)}\right)
\end{align}
and we have
\begin{align}
    &\langle \Sigma_{\rm open}^{\rm p.f.}(q)\vert\Psi = \langle f_{0}\circ \Psi(0)\rangle_{\Sigma_{\rm open}^{\rm p.f.}(q)}=\langle g_{\rm o}\circ f_{0}\circ \Psi(0)\rangle_{C_{\pi,2\pi t_{\rm o}}}=\langle G_{\rm o}\circ \Psi(0)\rangle_{C_{\pi,2\pi t_{\rm o}}},\\
    &\langle \Sigma_{\rm closed}^{\rm p.f.}(q)\vert \Psi =\langle d \circ f_{\rm o}\circ \Psi(0)\rangle_{\Sigma_{\rm closed}^{\rm p.f.}(q)}=\langle g_{\rm c}\circ d\circ f_{\rm o}\circ \Psi(0)\rangle_{C_{\pi,2\pi t_{\rm c}}}=\langle G_{\rm c}\circ \Psi(0)\rangle_{C_{\pi,2\pi t_{\rm c}}}.
\end{align}
Therefore, considering the $b$-ghost insertions the amplitude becomes
\begin{equation}
\begin{split}
   A_{0;1}^{0,2}(\varphi_{\rm o})=&+\int_{0}^{1}\frac{dq}{q}\langle d\circ f_{\rm o}\circ\varphi_{\rm o}(0) d\circ f_{\rm c}\circ b_{o}\rangle_{\Sigma_{\rm closed}^{\rm p.f.}(q)}\\
    &+\int_{t_{\rm min}}^{t_{\rm max}}dt \langle  B_{t}(V_t) F_{t}\circ\varphi_{\rm o}(0)\rangle_{C_{\pi,2\pi t}}\\
    &-\int_{0}^{1}\frac{dq}{q}\langle f_{0}\circ \varphi_{\rm o}(0)f_{\xi}\circ b_{0}\rangle_{\Sigma_{\rm open}^{\rm p.f.}(q)}
    \end{split}
\end{equation}
Finally, let us focus on the local coordinate map $F_{t}$. In particular, to ensure  \eqref{identification1} and \eqref{identification2},  we need
\begin{equation}
    \begin{cases}
        &\langle\Sigma^{\rm p.f}_{\rm open}(q=1)\vert\Psi=\langle\Sigma^{(0,2)}_{0,1}(t_{\rm max})\vert\Psi\\
        &\langle\Sigma^{\rm p.f}_{\rm closed}(q=1)\vert\Psi=\langle\Sigma^{(0,2)}_{0,1}(t_{\rm min})\vert\Psi
    \end{cases}
\longrightarrow
\begin{cases}
    &G_{\rm o}(w)\vert_{q=1}=F_{t}(w)\vert_{t=t_{\rm max}}\\
     &G_{\rm c}(w)\vert_{q=1}=F_{t}(w)\vert_{t=t_{\rm min}}
\end{cases}
\end{equation}
thus $F_{t}$ can be any holomorphic function which continuously interpolates $G_{\rm c}(w)\vert_{q=1}$ and  $G_{\rm o}(w)\vert_{q=1}$ in the interval $t\in[t_{\rm min}, t_{\rm max}]$. %%%%%%%%%%%%%%%%%%%%%%%%%%%%%%% 

    \endgroup
    \end{document}